\documentclass[12pt]{article}

\usepackage{epsfig,graphicx,bm,amsmath,amsfonts}
\usepackage{epstopdf}

\newcommand{\ben}{\begin{eqnarray}\displaystyle}
\newcommand{\een}{\end{eqnarray}}
\newcommand{\be}{\begin{equation}}
\newcommand{\ee}{\end{equation}}
\newcommand{\lb}{\left (}
\newcommand{\rb}{\right )}
\newcommand{\ltb}{\left [}
\newcommand{\rtb}{\right ]}
\newcommand{\pr}{\partial}
\newcommand{\p}{\partial}
\newcommand{\ra}{\rightarrow}

\newcommand{\nn}{\nonumber\\}
\newcommand{\nt}{{1\over 16 \pi G_5}}

\newcommand{\ap}{\alpha'}
\newcommand{\ie}{{\it i.e.}\ }

\newcommand{\bi}{\begin{itemize}}
\newcommand{\ei}{\end{itemize}}
\newcommand{\ii}{\item}

\newcommand{\Ncg}{16 \pi G_5}
\newcommand{\rl}{{r^2 \over L^2}}
\newcommand{\mn}{\mu \nu}

\newcommand{\ga}{\gamma}
 \newcommand{\CD}{\mathcal{D}}

\newcommand{\hhp}{\hspace{.15cm}}

\newcommand{\ep}{\epsilon}
\newcommand{\om}{\omega}
\newcommand{\cA}{{\cal A}}
\newcommand{\cB}{{\cal B}}
\newcommand{\ph}{\phi(r,k)}
\newcommand{\php}{\phi'(r,k)}
\newcommand{\cph}{\phi(r,-k)}
\newcommand{\cphp}{\phi'(r,-k)}
\newcommand{\phpp}{\phi''(r,k)}
\newcommand{\cphpp}{\phi''(r,-k)}
\newcommand{\intk}{\int {d^4 k \over (2 \pi)^4}}

\newcommand{\bc}{\begin{center}}
\newcommand{\ec}{\end{center}}

\newcommand{\app}{\alpha'}
\newcommand{\<}{\langle}
\renewcommand{\>}{\rangle}

\newcommand{\mt}[1]{\textrm{\tiny #1}}
\newcommand{\seff}{S_\mt{eff}}

\newcommand{\higho}{{\cal O}(q\omega^2, \omega q^2, q^3, \omega^3)}

\def\lfig#1#2#3#4{
 \begin{figure}[h]
 \refstepcounter{figure}
 \label{#4}
 \addtocounter{figure}{-1}
 \epsfxsize=#3
 \centerline{\epsfbox{#2}}
 {\bf \caption{{\rm #1}}}
 \end{figure}
}

\begin{document}

\setlength{\unitlength}{1mm}

\begin{titlepage}

\begin{flushright}

\end{flushright}
\vspace{2cm}

\begin{center}
{\bf \Large Holographic hydrodynamics: models and methods}\\
\vspace*{1cm}
\end{center}

\begin{center}
\bf{Nabamita Banerjee$^{a}$,}
\bf{Suvankar Dutta$^{b}$}

\vspace{.5cm}

{\small \it $^{a}$ITF, Utrecht University, Utrecht, The Netherlands\\
and\\
Tata Institute of Fundamental Research, Mumbai, India}\\
\vspace{2mm}
{\small \it $^{b}$ Department of Physics, \\
Indian Institute of Science Education and Research(IISER)\\
Bhopal, India}\\[.5em]

{Email: \small {\tt N.Banerjee@uu.nl, \ suvankar@iiserbhopal.ac.in}}

\end{center}

\vspace{1cm}

\begin{abstract}
We review recent developments in holographic hydrodynamics. 
We start from very basic discussion on hydrodynamic systems and 
motivate why string theory is an essential tool to deal with 
these systems when they are strongly coupled. 
The main purpose of this review article is to understand 
different holographic techniques to compute transport 
coefficients (first order and higher order) and their corrections 
in presence of higher derivative terms in the bulk Lagrangian.
We also mention some open challenges in this subject.

\end{abstract}

\vspace{2cm}


\end{titlepage}

\tableofcontents

\section{Introduction and motivation}

In this article we shall study the characteristics of hydrodynamic
system from string theory perspective.

Hydrodynamics is an effective theory, describing the dynamics of  some
field theory at large distances and time-scales. A fluid system is
considered to be a continuous medium. When we talk about a small
volume element (or "fluid particle") in fluid it has to be remembered
that the small volume element consists of a large number of molecules
(specifically the size of the fluid particle is much much greater
than the mean free path of the system). The equations of
hydrodynamics assume that the fluid is in local thermodynamic
equilibrium at each point in space and time, even though different
thermodynamic quantities like fluid velocity ($\vec v(\vec x, t)$),
energy ($e(\vec x, t)$), pressure  ($p(\vec x, t)$), fluid density
($\rho(\vec x, t)$) $etc$ may vary. Fluid mechanics applies only when
the length scales of variation of thermodynamic variables are large
compared to equilibration length scale of the fluid, namely mean
free path\cite{landau}.

Hydrodynamic description does not follow from the action principle
rather it is normally formulated in the language of equations of
motion. The reason for this is the presence of dissipation in thermal
media. Due to internal friction called $viscosity$, a dissipative
fluid loses its energy over time as it propagates. The fluid without
any viscous drag is called $ideal\ fluid$. In the simplest case, the
hydrodynamic equations are just the laws of conservation of energy and
momentum (we are considering the fluid without any global charge
or current),
\be \label{conserveqn}
\nabla_{\mu}T^{\mu\nu}=0\ .
\ee
At any space time point the fluid is characterized by $d+1$  variables
(in $d$ dimensions): velocities $u^{\mu}(x)$ (of fluid particles)
and its temperature $T(x)$\footnote{Other thermodynamic variables can
be expressed as a function of temperature.}. All these variables are
functions of space and time. Since velocities are time like they
follow $u_{\mu}u^{\mu}=-1$, therefore total number of variables
describing the fluid is $d$ which is equal to the number of equations
of motion as in equation (\ref{conserveqn}).

In hydrodynamics we express $T_{\mu\nu}$ in terms of $T(x)$ and
$u^{\mu}(x)$. Following the standard procedure of effective field
theories, we expand energy momentum tensor in powers of spatial
derivatives. As we have already explained local thermodynamic
quantities (velocity and temperature) vary very slowly over space-time
therefore their derivatives are very small. Thus it is legitimate
to express the energy-momentum tensor in powers of derivatives of
local quantities $i.e.$, $\pr s \ll (\pr s)^2, \pr^2 s \ll(\pr s)^3,
\pr^3 s, \cdots \ll \cdots $ where, $s$ stands for any local
quantity.

At the zero$^{th}$ order, $T_{\mu\nu}$ is given by the familiar
formula
for
ideal fluids,
\be
T_{\mu \nu} = (e(x)+p(x)) u_{\mu}(x) u_{\nu}(x)+ p(x) g_{\mu \nu}(x)
\ee
where $e(x)$ is energy density and $p(x)$ is pressure\footnote{Next
time onwards we shall drop the functional dependence of local
variables i.e. we shall write $u_{\nu}(x)$ as $u_{\mu}$ and similarly
for other variables.}. $g_{\mu\nu}$ is the metric tensor of background
spacetime. If we also consider the fluid system has conformal
invariance (a system which is scale invariant) then its stress tensor
becomes traceless $i.e.$ $T^{\mu}_{\mu}=0$. This implies that
$p=\frac{1}{d-1}e$. This is the thermodynamic equation of state.

At the next order in derivative expansion fluid energy-momentum tensor
is given by,
\ben\displaystyle
\label{emt1st}
T_{\mu \nu}&=& (e+p) u_{\mu} u_{\nu}+ p g_{\mu \nu} - 2
\sigma_{\mu\nu}\nn
\sigma_{\mu \nu}&=&\eta { P^{\alpha}_{\mu} P^{\beta}_{\nu} \over 2}
\bigg[\nabla_{ \alpha}u_{\beta}+ \nabla_{ \beta}u_{\alpha}
-{2 \over d-1} g_{\alpha \beta} \nabla \cdot u \bigg] + \zeta\
g_{\mu\nu}\nabla \cdot u, \qquad P^{\mu\nu}=
g^{\mu\nu}+u^{\mu}u^{\nu}
\een
where $\sigma^{\mu\nu}$ is proportional to derivatives of
$u^{\mu}(x)$ and is termed as dissipative part of $T^{\mu\nu}(x)$.
$\sigma_{\mu\nu}$ is a symmetric tensor. It has been divided into a
traceless part and a trace part. The coefficient of traceless part
is denoted by $\eta$ and that of trace part is denoted by $\zeta$.
The coefficient $\eta$ is called $shear$ viscosity coefficient and
$\zeta$ is called $bulk$ viscosity coefficient. Shear viscosity
coefficient describes fluid's reaction against applied shear stress
where as bulk viscosity coefficient measures reaction against volume
stress. $\eta$ and $\zeta$ are called the first order transport
coefficients as they appear at the first order in derivative expansion
of energy-momentum tensor.
For conformal fluid bulk viscosity coefficient vanishes. Since
trace of stress tensor is proportional to $\zeta $ (from equation
(\ref{emt1st})) $i.e.$ ,
$$T^{\mu}_{\mu} \sim \zeta \nabla \cdot u.$$

In a similar fashion one can write the expression for stress tensor
to the second order in derivative expansion. We shall come to the
second order hydrodynamics later in section \ref{prob2}.

Before we go ahead let us first motivate why we use string theory to
study different properties of fluid dynamics.

\subsection{Why string theory ?}

The hydrodynamic
behavior of a system is characterized by a set of transport
coefficients, like shear viscosity, bulk viscosity $etc$ as mentioned
above.  After Relativistic Heavy Ion Collider (RHIC)
experiments\cite{rhic}, the study of shear viscosity to entropy
density ratio of gauge theory plasma has developed lots of attention.
The QGP (Quark-Gluon Plasma)
produced at RHIC behaves like viscous fluid with very small shear
viscosity coefficient (near-perfect fluid). Such a low ratio of shear
viscosity to entropy density is very hard to describe with
conventional methods. The temperature of the gas of quarks and gluons
produced at RHIC is approximately $170 \ MeV$ which is very close to
the confinement temperature of QCD. Therefore, at this high
temperature they are not in the weakly coupled regime of QCD. In fact
near the transition temperature the gas of quarks and gluons belongs
to the non-perturbative realm of QCD. Thus usual perturbative
gauge theory computations are not applicable to explain RHIC results.
On the other hand in Lattice gauge theory (a technique to study the
properties of strongly coupled system) it is difficult to compute
real time correlators of energy momentum tensor as the theory is
formulated on Euclidean lattice\footnote{Please look at
\cite{lattice} where the author have found a lattice technique to
compute the ratio. But their result is far from universality(
$\eta/s =1/4\pi$).}. The AdS/CFT correspondence, at this point,
appears to be a technically powerful tool to deal with strongly
coupled (conformal) field theory in terms of weakly coupled
(super)-gravity theory in AdS space. Holographic techniques (motivated
from the AdS/CFT correspondence) to compute hydrodynamic transport
coefficients exhibit a remarkable quantitative agreement with those
arising from numerical fits to RHIC data.

This motivates us to study properties of hydrodynamic systems from the 
point of view of string theory, in particular, the AdS/CFT correspondence.

Now, we pause our discussion on fluid system a little and discuss 
the basic idea and working principle of the AdS/CFT correspondence. 


\section{The AdS/CFT conjecture  and fluid/gravity correspondence}

In this section we briefly review the AdS/CFT conjecture
\cite{maldacena}. The
correspondence is itself a vast subject and
there are lots of good review articles on this\footnote{See
\cite{maldarev} for example}. Therefore,
here we shall briefly discuss the important points of this conjecture
to introduce ourselves to the notations which will be used through out
this article.

The original conjecture states the {\it equivalence} between two
seemingly unrelated theories: Type IIB string theory on $AdS_5 \times
S^5$, where both $AdS_5$ and $S^5$ has radius $b$, with a
five form field strength $F_5$, which has integer flux $N$ over $S^5$,
and complex string coupling $\tau_S=a+i e^{-\phi}$  where $a$ is axion
and $\phi$ is dilaton field and ${\cal N} =4$ SYM theory in 4
dimension, with gauge group $SU(N)$, Yang-Mills coupling $g_{YM}$ and
instanton angle $\theta_I$ (together define a complex coupling
$\tau_{YM} = {\theta_I \over 2 \pi} + {4 \pi i \over g_{YM}^2}$) in
its superconformal phase, with $g_S = {g^2_{YM} \over 4 \pi}$,
$a= {\theta_I \over 2\pi}$ and $b^4 = 4 \pi g_S N (\alpha')^2$.

The best understood examples of the AdS/CFT correspondence relate the
strongly coupled dynamics of certain conformal field theories (CFT) to
the dynamics of gravitational systems in AdS spaces.The
description is holographic, as the CFT is living on the boundary of
the bulk AdS space and thus has one lower dimension than the full bulk
spacetime. All the characteristics or information of a lower
dimensional theory is captured in a theory living in one higher
dimension and $vice\ versa$. The description is also called a duality
because when the field theory is strongly coupled the gravity theory
is weakly coupled and $vice\ versa$.

These boundary theories are quantum field theories of a particular
kind. They pertain to strongly coupled systems which are difficult if
not impossible to study perturbatively. The AdS/CFT correspondence
seems to be a powerful tool to deal with these strongly coupled
(conformal) field theory. Since our main target is to
understand the properties of strongly coupled plasma produced at
RHIC\footnote{In general, QCD does not have any conformal
invariance ($\beta$ function is not zero).  However, at high enough
temperature, the QCD plasma is well described by some conformal field
theory and thus AdS/CFT correspondence can approximate it by a gravity
dual. With this assumption we apply this correspondence to explain the
properties of strongly coupled QGP produced at RHIC.}, lets us first
understand how such field theory can be dealt with
via weak gravity computations.

\subsection{The Large {\textquoteleft t} Hooft coupling ($\lambda$)
Limit}

 The large {\textquoteleft
t} Hooft coupling limit corresponds to
taking $\lambda = g_{YM}^2 N = g_S N \ra \infty$, while $N \ra \infty$
also.This is the weakest form of the conjecture. In this large $N$
limit, the super Yang-Mills theory becomes nonperturbative $i.e.$ a
strongly coupled super Yang-Mills theory, but the string theory side
becomes very much tractable. This can be realized as follows: from the 
relation $b^4 = 4 \pi g_S N (\alpha')^2$, one can easily check that large 
$\lambda$ limit corresponds to the radius of curvature $b$ being much much larger than
the string length, $b^2\gg\ap$. Now, the classical string theory effective
action is given by,
\be
{\cal L} \sim a_1\ap {\cal R} + a_2 \ap^2 {\cal R}^{(2)} + a_3 \ap^3
{\cal R}^{(3)} + \ \cdot \ \cdot
\ee
where, ${\cal R}^{(n)}$ is a $2n$ derivative term constructed out of
Ricci tensor or Riemann tensor or Ricci scalar.  Since the near
horizon geometry of $D3$ brane spacetime is asymptotically AdS, with
both the $AdS_5$ and $S_5$ has radius of curvature $b$, the scale of
Riemann tensor is set by,
\be
{\cal R} \sim {1\over b^2} \sim {1\over \sqrt{g_SN} \ap } \sim {1\over
  \sqrt{\lambda} \ap}.
\ee
Therefore the expansion of effective Lagrangian in powers of $\ap$
effectively becomes an expansion in powers of $(\lambda)^{-{1\over
    2}}$,
\be\label{genstringacn}
{\cal L} \sim a_1 (\lambda)^{-{1\over
    2}} + a_2 (\lambda)^{-1} + a_3 (\lambda)^{-{3\over
    2}} + \ \cdot \ \cdot .
\ee
Thus taking $\lambda \ra \infty$ limit corresponds to considering the
classical supergravity action instead of full classical string theory.

\subsection{The Field Operator Mapping}\label{field-op-map}

Another important machinery in the AdS/CFT correspondence is the
mapping between bulk fields and boundary operators. In \cite{witten}
Witten proposed that the boundary values of supergravity fields acts
as a source to the corresponding operator in the field theory
side. To be more explicit, let us define the string theory partition
function as,
\be
Z_{S} = \int \prod_i \ltb d \chi_i \rtb \exp\ltb -S [\chi_i] \rtb
\ee
where we collectively denote all string theory fields by $\chi_i$.
Since AdS is a space with a boundary, we need to specify the
boundary values for these fields to compute the partition function.
Let us denote the boundary value of the field $\chi_i$ by $\chi_i^B$.
Therefore the partition function  is, in general a function of the
boundary values $\chi_i^B$'s of the fields,
\be
Z_{S}[\chi_i^B] = \int \ltb \prod_i \ltb d \chi_i \rtb \exp\ltb -S
[\chi_i] \rtb \rtb_{\chi_i \ra \chi_i^B}.
\ee

On the other hand in the filed theory side the correlation functions
of any operators ${\cal O}_i$ are given by,
\be \label{npointfunc}
\langle {\cal O}_{i_1}(x_1) {\cal O}_{i_2}(x_2) \cdot \cdot \cdot
{\cal O}_{i_n}(x_n \rangle = {\delta^n Z_{CFT}[\{J_i\}] \over \delta
J_{i_1} (x_1) \delta J_{i_2} (x_2) \cdot \cdot \cdot \delta J_{i_n}
(x_n) }|_{J_i=0}
\ee
where $Z_{CFT}[\{J_i\}]$ is given by,
\be
Z_{CFT}[\{J_i\}] = \int \prod_i \ltb d \phi_i \rtb \exp\ltb -S
[\phi_i]  + \sum_i \int [d^4x] \sqrt{-g_{CFT}} J_{i}(x) {\cal
    O}_i(x) \rtb.
\ee
According to \cite{witten} there is an one to one correspondence
between $\chi_i$'s and ${\cal O}_i$'s such that
\be
Z_{CFT}[\{J_i\}] = Z_{S}[\chi_i^B], \ \ \ with \ \ \chi_i^B=J_i.
\ee
For example, if we want to compute correlation function of boundary
stress tensor $T_{\mu\nu}$ then the corresponding bulk field is metric
$g_{\mu\nu}$. We need to first calculate the bulk partition function
(string theory/gravity partition function). Then following equation
(\ref{npointfunc}) we can find the n-point correlation function of
boundary operators by taking functional derivative of the bulk
partition function with respect to the boundary values of the
corresponding bulk fields. We shall use these results later in this
article.

\subsection{Fluid/gravity correspondence}

The power of AdS/CFT is not confined to characterizing only the
thermodynamic properties of boundary field theories. If we consider a
black object with translation invariant horizon, for example black
$D3$ brane geometry, one can also discuss hydrodynamics - long wave
length deviation (low frequency fluctuation) from thermal equilibrium.
In addition to the thermodynamic quantities the black brane is also
characterized by the hydrodynamic parameters like viscosity, diffusion
constant, etc.. The black $D3$ brane geometry with low energy
fluctuations (i.e. with hydrodynamic  behavior) is dual to some finite
temperature gauge theory plasma living on boundary with hydrodynamic
fluctuations. Therefore studying the hydrodynamic properties of
strongly coupled gauge theory plasma using the AdS/CFT duality is an
interesting subject of current research.

The first attempt to study hydrodynamics via AdS/CFT was
in \cite{son1}, where authors related the shear viscosity coefficient
$\eta$ of strongly coupled ${\cal N}=4$ gauge theory plasma in large
$N$ limit with the absorption cross-section of low energy gravitons by
black $D3$ brane. Other hydrodynamic quantities like speed of sound,
diffusion coefficients, drag force on quarks $etc$ can also be
computed in the context of AdS/CFT.


\section{Goal of this article}

In this article we explain how to compute different transport
coefficients of strongly coupled gauge theory plasma using
gauge/gravity correspondence.

The most famous transport coefficient is shear viscosity coefficient.
We plan to study a systematic procedure to compute this particular
transport coefficient of dual plasma in presence of possible $stringy$
correction to the bulk Lagrangian. We
first
discuss about the famous Kubo formula in section \ref{kubo-section}.
In section \ref{prob1} we show how the shear viscosity coefficient of
strongly coupled boundary gauge theory plasma depends on horizon value
of the effective coupling of transverse graviton moving in black
brane background. However to completely specify the boundary plasma,
it is necessary to understand its higher order transport coefficients
(coefficients which appear in second order in derivative expansion).
In section \ref{prob2} we study the second order hydrodynamics. We
define a response function in bulk and show that the response function
flows non-trivially with the radial direction and depends on the full
black hole geometry. We study the radial flow of this response
function (of energy-momentum tensor of dual gauge theory) in presence
of generic higher derivative terms in bulk Lagrangian and solve these
flow equations analytically to obtain second order transport
coefficients of boundary plasma. We consider explicit example to
explain this formalism in section \ref{r4sec2nd}. Finally we
end this article with some concluding remarks in section
\ref{conclusion}.


\section{Holographic hydrodynamics}
\label{kubo-section}

As we have already mentioned, at present the best available analytical
tool one has to study the properties of strong coupling
hydrodynamics is precisely the AdS/CFT correspondence. In this section
we briefly discuss how one can apply the knowledge of the
correspondence to capture the hydrodynamic properties of the plasma.
The goal of this section is to introduce $Kubo \ formula$ and use this
formula in the context of gauge/gravity duality to compute different
transport coefficients of strongly coupled plasma.

\subsection{Kubo formula}

The Kubo formula relates transport coefficients of plasma with their
thermal correlators. Let us consider the response of the
fluid to small metric fluctuation $g_{xy}=\eta_{xy}+h_{xy}$ and
consider $h_{xy}=h_{xy}(t,x_3)$. The expression for energy momentum
tensor in fluid's rest frame is given by (from equation (\ref{emt1st})
and also considering conformal fluid),
\be\label{Txy}
T_{xy}= p h_{xy} + \eta \frac{\pr h_{xy}}{\pr t}+ {\cal O}(\pr^2)\ .
\ee
At the level of linear response theory the one-point function of
$T_{xy}$ is linear in $h_{xy}$ and when expressed in Fourier space the
proportionality constant is simply the thermal retarded correlator
$G^{(R)}_{xy,xy}$ (Green's function),
\be\label{lin-resp}
\langle T_{xy}\rangle = G_{xy,xy}^{(R)} h_{xy} \ .
\ee
Therefore writing the expression of equation (\ref{Txy}) in momentum
space we get,
\be\label{kubo-formula}
G_{xy,xy}^{(R)} = p - i\omega \eta +  {\cal O}(\omega^2).
\ee
Thus the shear viscosity coefficient is given by\footnote{Similarly
other transport coefficients are also captured in $G^{(R)}$. We shall
consider them in section \ref{prob2}. },
\be\label{kubo}
\eta = \lim_{\omega \ra 0} -\frac{\Im G_{xy,xy}^{(R)}}{\omega}.
\ee
Hence, to compute shear viscosity coefficient we need to find the
real time retarded Green's function of EM tensor. Now, the usual
prescription of AdS/CFT to compute boundary correlator is Euclidean.
 In principle, some real time Green's
function can be obtained by analytic continuation of
the corresponding Euclidean ones. However, in many cases it is
actually very difficult to get. In particular the low frequency low
momentum limit Green's function (which is interesting for
hydrodynamics) is difficult to obtain from analytic continuation of
Euclidean one. The difficulty here is, we need to analytically
continue from a discrete set of points in Euclidean frequencies (the
Matsubra frequencies) $\omega= 2 \pi i n$ (n integer) to real values
of $\omega$. The smallest value of the Matsubra frequency is quite
large. Hence to get information in small $\omega$ limit that we are
interested in, is quite difficult. The authors of
\cite{son2} have done a detailed analysis of this
difficulty and have given a prescription to compute the real time
correlator. This is a well-defined holographic method to compute
thermal correlators of boundary theory and capture the values of
different transport coefficients (for example shear viscosity in first
order hydrodynamics). However we will follow a different path to
compute boundary thermal correlator. In the next two sections we try
to elaborate on this new technique to compute transport coefficients
holographically.

In the next section(s), we use this formula holographically to extract 
the values of different transport coefficients for strongly coupled
boundary fluid system.

\section{First order hydrodynamics}
\label{prob1}

We plan to discuss the properties of hydrodynamic system order by order 
in derivative expansion. As we have already mentioned at the first order 
we encounter shear viscosity, the only non-trivial transport coefficient, 
for conformal fluid without any other conserved current. 
This section is dedicated to discuss the holographic techniques to compute
this transport coefficient in generic higher derivative gravity.

\subsection{Hydrodynamic limit in AdS/CFT and  Membrane paradigm}

In this subsection we briefly review the proposal described in
\cite{liu} relating the hydrodynamic limit of AdS/CFT to the membrane
paradigm. This proposal relates a generic transport coefficient of
the boundary theory to some geometric quantities evaluated at the
black hole horizon in the bulk. We will concentrate on the shear
viscosity coefficient of the boundary fluid and confine ourselves at
the level of linear response and low frequency limit of the strongly
coupled gauge theory.\\

\bc
{\underline{\bf Membrane side}}
\ec

Let us start with classical black hole membrane paradigm, which says
that the black hole has a fictitious fluid on its horizon. In general,
the black hole action can be expressed as,
\begin{equation}
S_{eff}=S_{out}+S_{surf} , \nonumber \\
\end{equation}
where $S_{out}$ contains integration over space time out side the
horizon and $S_{surf}$ is the boundary term on the horizon.
Physically, $S_{surf}$ represents the effect of the horizon fluid on
the spacetime. Let us consider a general black hole background,
\begin{equation} \label{genmet}
dS^2= \ \ g_{MN}dx^M dx^N\ \ =\ \ g_{rr}dr^2+ g_{\mu \nu}dx^{\mu} dx^{\nu}
\end{equation}
where ${M,N}$ runs over $d+1$ dimensional bulk spacetime and  ${\mu,
\nu}$ runs over $d$ dimensional boundary spacetime. This black hole
has a horizon at $r_h$ and asymptotic boundary (where the dual gauge
theory sits) at $r_b$. We assume $SO(3)$ invariance in the boundary
spatial directions, that is all the metric components and the
couplings in the theory are only functions of $r$. We consider a small
perturbation $h_{xy}$ in the $SO(3)$ tensor sector of this metric. We
use $\phi(r,x^{\mu})= h^x_y$ as the off diagonal component of
graviton and in the Fourier space, the perturbation looks like,
\begin{equation} \label{fr}
\phi (r, k_{\mu})= \int {d^d x \over (2 \pi)^d} \, \phi (r, x^{\mu})
\, e^{i
k_{\mu} x^{\mu}}, \qquad k_{\mu} = (-\omega, \vec k) \ .
\end{equation}
The action for this massless perturbation can be written as,
\begin{eqnarray}\displaystyle \label{memch}
S_{out} = - \int_{r > r_h} d^{d+1}x \sqrt{-g}\, {1 \over q (r)} \,
(\nabla\phi)^2,\quad
S_{surf} = \int_{\Sigma} d^{d}x \sqrt{-\gamma}
\left(\frac{\Pi(r_h,x)}{\sqrt
{-\gamma}}\right) \phi(r_h,x)
\end{eqnarray}
here, $\Pi$ is the conjugate momentum to $\phi$ for the $r-$foliation
and $\gamma$ is the induced metric on the horizon. We can interpret
$S_{surf}$ as the effect of the membrane fluid on the spacetime and
$\Pi_{mb}= ({\Pi(r_h) \over \sqrt{\gamma}})$ as the ``membrane
$\phi-$charge''.

Now, following membrane paradigm, the horizon is a regular place for
the in-falling observer, hence, any physical deformation of the system
has to satisfy the in-falling boundary condition. The
in-falling boundary condition implies that near the horizon $r_h$,
\begin{itemize}
\item
 the
deformation should behave as $\phi \sim (r-r_h)^{i \omega \beta}$,
for some constant $\beta$ and
\item
the solution should be a function of the non singular
``Eddington-Finklestein'' co-ordinate $v$ defined as,
\begin{equation} \label{vcoord}
dv= dt + \sqrt{{g_{rr} \over g_{tt}}} dr.
\end{equation}
\end{itemize}
This implies near the horizon $r_h$, the deformation satisfies,
\begin{equation} \label{ifbc}
\partial_r \phi= \sqrt{{g_{rr} \over g_{tt}}} \partial_t \phi\ .
\end{equation}
The above equation (\ref{ifbc}) puts constraint on the constant
$\beta$ as,
\begin{equation} \label{beta}
\beta= \sqrt{{g_{rr}(r-r_h)^2 \over g_{tt}}}\bigg |_{r_h}.
\end{equation}
With equations (\ref{fr},\ref{ifbc}) and some redefinition of the
time, we can
also express the membrane charge as,
\begin{equation}
\Pi_{mb} = -\frac{1}{q(r_h)}\partial_{\hat{t}}\phi(r_h).
\end{equation}

As per our interpretation, $\Pi_{mb}$ is the response of the membrane
fluid induced by $\phi$ and using equation (\ref{fr}), in linear
response, we define a shear viscosity coefficient for the membrane
fluid as,
\begin{eqnarray}\displaystyle \label{memvis}
\Pi_{mb}= i \omega \eta_{mb}\phi
\een
and thus we get,
\ben
\eta_{mb}={1 \over q(r_h)}.
\end{eqnarray}

\bc
{\underline{\bf Boundary Side}}
\ec

With this much of analysis of the membrane fluid, we concentrate on
the boundary side where the gauge theory lives. This is a interacting
theory at finite temperature and behaves as a
fluid at sufficiently long length scale or low energy. The real time
(Lorentzian signature) finite temperature version of AdS/CFT
correspondence allows to compute various hydrodynamic quantities of
this gauge theory at strong coupling by doing some supergravity
calculations in the AdS space. Now using the Kubo formula, the shear
viscosity of the boundary fluid is given in terms of retarded Green's
function, the response of graviton to the boundary stress tensor. In
\cite{son2}, the
authors have given a simple prescription to compute the boundary
correlator using the bulk field $\phi$, the
off diagonal
component of the graviton. Their prescription requires to find a
solution for the graviton which is in-falling at the horizon and
constant at the boundary. Then one computes the on-shell action with
this solution and the retarded Green's function is related to the
surface term of the on-shell action at the boundary. Taking
$\phi(k_{\mu},r)=f(k_{\mu},r) \phi_0(k_{\mu})$ with normalization
$f(k_{\mu},r_b) \rightarrow 1$,
\begin{eqnarray}\displaystyle \label{rgf}
S &=& - \sum_{r=r_h,r_b} \int
\frac{d^{d}k}{(2\pi)^{d}}\,\phi_0(k_{\mu})\,G(k_{\mu},r)\,\phi_0(-k_{
\mu}) \nn
G^R(k_{\mu})&=& \lim_{r\rightarrow r_b}2 G(k_{\mu},r) =
\lim_{r\rightarrow r_b}2 {\sqrt{-g} g^{rr}
  \over q(r)} \partial_rf(k_{\mu},r).
\end{eqnarray}
For this profile of the graviton, we can evaluate its conjugate
momenta $\Pi$, it readily gives us the relation,
\begin{eqnarray}\displaystyle
G^R(k_{\mu})= -\lim_{r \rightarrow r_b}{\Pi (k_{\mu},r) \over
\phi(k_{\mu},r)}\ .
\end{eqnarray}
Hence, the shear viscosity coefficient $\eta$ can be written as,
\begin{equation}
\eta= \lim_{k_{\mu} \rightarrow 0} \lim_{r \rightarrow r_b} {\Pi
(k_{\mu},r) \over i \omega \phi(k_{\mu},r)}.
\end{equation}
Important point to note is that, in the low frequency limit $(k_{\mu}
\rightarrow 0,  \ with  \  \Pi, \ \omega \phi \,\,fixed)$, the flow
of $\Pi$ and $\omega
\phi$ in the $r-$ direction are trivial. Hence, we can actually
compute the
shear viscosity coefficient of the boundary fluid at any constant
$r-$slice
and its value would be same. We compute it at the horizon and get,
\be \label{visco}
\eta= {1 \over q(r_h)} \sqrt{\frac{-g} {g_{rr}g_{tt}}} \biggr|_{r_h} =
{1 \over q (r_h)} {A \over V}
\ee
Comparing equations (\ref{visco}) and (\ref{memvis}), we see that the
viscosity coefficient of the boundary fluid is related to that of the
membrane fluid and more importantly they are given as just the value
of the inverse effective coupling of the transverse graviton evaluated
at the horizon. To emphasize, equation (\ref{ifbc}) plays a crucial
role
in this equivalence. The AdS/CFT response of the graviton is almost
same as that of the membrane except that the membrane now has to sit
in the boundary. The in-falling boundary condition of the graviton
field in AdS/CFT is precisely the regularity condition given by
equation (\ref{ifbc}) of
the membrane paradigm. In the low frequency limit, we can place a
fictitious membrane at each constant $r$ and define the transport
coefficient as $\eta(r)$. Since the flow is trivial in this limit,
$\eta(r)$ actually comes out to be a constant, ${1 \over q(r_h)}$.

\subsection{Example: Two derivative gravity}
\label{rev}

In this section explain how to calculate the shear viscosity
coefficient of the boundary fluid from the effective coupling
constant of transverse graviton in Einstein-Hilbert gravity.

We  first fix the background spacetime. We start with the
following Einstein-Hilbert action in AdS space.
\begin{equation}\label{EHacn}
I=\nt \int d^5x \sqrt{-g} \lb R + 12
\rb\ .
\end{equation}
Here we have taken the radius of the AdS space to be 1. The background
spacetime is given by the following metric\footnote{We are working
in a coordinate frame where
  asymptotic boundary is at $r \ra 0$.}
\begin{equation}\label{metric0}
ds^2=-h_t(r)dt^2 + {dr^2 \over h_r(r)} + {1\over r} d{\vec x}^2, \quad
h_t(r)={1-r^2 \over r}, \quad h_r(r)=4r^2 (1-r^2).
\end{equation}
The black hole has horizon at $r_0=1$ and the temperature is
given by,
$
T={1 \over \pi} \ .
$

We
consider the following metric perturbation,
\begin{equation}\label{petmet}
g_{xy}=g^{(0)}_{xy}+ h_{xy}(r,x)=g^{(0)}_{xy}(1+\ep \Phi(r,x))
\end{equation}
where $\epsilon$ is an order counting parameter. We  consider terms
up to order $\epsilon^2$ in the action of $\Phi(r,x)$. The action (in
momentum space) is given by,(up to some total derivative terms in the
action)\footnote{Though throughout this article we have written the
four
vector $k$, but in practice we have worked in ${\vec k}\rightarrow 0$
limit. In all the expressions we have dropped the terms proportional
to ${\vec k}$ or its power.},
\begin{equation} \label{acn02}
S=\nt \int {d\om d^3{\vec k}\over (2 \pi)^4}dr \lb \cA^{(0)}_1(r)
\cphp \php
+ \cA^{(0)}_0(r,k) \ph \cph \rb
\end{equation}
where,
\begin{equation}
\cA_1^{(0)}(r) = {r^2 -1 \over r},\quad \cA_0^{(0)}(r,k)= {\omega^2
\over 4r^2(1-r^2)}\ .
\end{equation}
This can be viewed as an action for minimally coupled scalar field
$\ph$ with
effective coupling given by,
\begin{equation}\label{effcoup}
K_{{\rm eff}}(r)=\nt {\cA^{(0)}_{1(r)} \over \sqrt{-g^{(0)}} g^{rr}} .
\end{equation}
Therefore according to \cite{liu} the effective coupling
$K_{{\rm eff}}$ calculated at the \index{horizon}horizon $r_0$ gives
the shear viscosity coefficient of boundary
\index{fluid!boundary}fluid,
\begin{eqnarray}\displaystyle
\eta &=& r_0^{-{3\over 2}} (-2 K_{{\rm eff}}(r_0)) = \nt \ .
\end{eqnarray}
However, when we add higher derivative terms (effect of string theory)
in bulk action the situation becomes complicated. In the next two
subsections we generalize this idea to higher derivative gravity theories.

\subsection{The Effective Action}\label{gendis}

Having understood the above procedure of determining the shear viscosity
 coefficient from the effective coupling of transverse graviton it is
tempting to generalize this method for any higher derivative gravity.
The first problem one faces is that the action for transverse graviton
no more has the canonical form as in equation (\ref{acn02}). For generic
'n'
derivative gravity theory the action can have terms with (and up to)
`n' derivatives of $\Phi(r,x)$. Therefore, from that action it is not
very clear how to determine the effective coupling. In this section we
try to address this issue.

We construct an effective action which is of form given
by equation (\ref{acn02}) with
different coefficients capturing higher derivative effects. We
determine these two coefficients by claiming that the equation of
motion for $\ph$ coming from these two actions (general action and
effective action) are same up to first order in perturbation expansion
(in coefficient of higher derivative term). Once the
effective action for transverse graviton is obtained in canonical form then one can
extract the effective coupling from the coefficient of $\php \cphp$
term in the action.  Needless to say, our method is perturbatively
correct.

\subsubsection {The General Action and Equation of Motion}

Let us start with a generic '$n$' derivative term in the action with
 coefficient $\mu$. We study this system perturbatively and all our
 expressions are valid up to order $\mu$. The action is given by,
\begin{equation} \label{gacnhd}
S=\nt \int d^5x \lb R + 12 + \mu  \ {\cal R}^{(n)}
\rb
\end{equation}
where, ${\cal R}^{(n)}$ is any $n$ derivative Lagrangian. $\mu$ is
the perturbation parameter.
The metric in general is given by (assuming planar symmetry),
\begin{equation}
ds^2=-(h_t(r)+\mu\ h_t^{(n)}(r))dt^2 + {dr^2 \over h_r(r) +  \mu\
h_t^{(n)}(r)}
+ {1\over r}
(1+ \mu\
h_s^{(n)}(r))d{\vec x}^2 \
\end{equation}
where $h_t^{(n)},h_r^{(n)}$ and $h_s^{(n)}$ are higher derivative
corrections to the  metric.

Substituting the background metric with fluctuations in the action
given in equation
(\ref{gacnhd}) (we call it general action or original action)
for the scalar
field $\ph$ we get,
\begin{equation} \label{ghdacnphi}
S=\nt \int {d^4 k \over (2 \pi)^4} dr \sum_{p,q=0}^{n} \cA_{p,q}(r,k)
\phi^{(p)}(r,-k) \phi^{(q)}(r,k)
\end{equation}
where, $\phi^{(p)}(r,k)$ denotes the $p^{th}$ derivative of the field
$\ph$ with respect to $r$ and $p+q\leq n$. The coefficients
$\cA_{p,q}(r,k)$ in general depends on the coupling constant $\mu$.
$\cA_{p,q}$ with $p+q \ge 3$ are proportional to $\mu$ and vanishes in
$\mu \rightarrow 0$ limit , since the terms $\phi^{(p)} \phi^{(q)}$
with $p+q\ge 3$ appears as an effect of higher derivative terms in
equation (\ref{gacnhd}).  Up to some total derivative terms, the
general
action in equation
(\ref{ghdacnphi}) can also be written as,
\begin{eqnarray}\displaystyle \label{ghdacnphi2}
S&=&\nt \int {d^4 k \over (2 \pi)^4} dr \sum_{p=0}^{n/2}\cA_{p}(r,k)
\phi^{(p)}(r,-k) \phi^{(p)}(r,k), \hspace{.88cm} n \quad {\rm even}
 \nonumber \\
&=&\nt \int {d^4 k \over (2 \pi)^4} dr
\sum_{p=0}^{(n-1)/2}\cA_{p}(r,k)
\phi^{(p)}(r,-k) \phi^{(p)}(r,k), \hspace{.4cm} n \quad {\rm odd} \ .
\nn
\end{eqnarray}
The equation of motion for the scalar field $\ph$ is given by,
\begin{eqnarray}\displaystyle
\sum_{p=0}^{n/2} \lb - {d \over dr} \rb ^{p}{\p {\cal
L}(\{\phi^{(m)}\})
\over  \p
  \phi^{(p)}(r,k)} =0 (n \ {\rm even}), \quad
\sum_{p=0}^{(n-1)/2} \lb - {d \over dr} \rb ^{p}{\p  {\cal
L}(\{\phi^{(m)}\})
\over  \p
  \phi^{(p)}(r,k)} =0,(n \ {\rm odd})
\end{eqnarray}
where ${\cal L}(\{\phi^{(m)}\})$ is given as
\begin{equation}
{\cal L}(\{\phi^{(m)}\})=\sum_{p}\cA_{p}(r,k)
\phi^{(p)}(r,-k) \phi^{(p)}(r,k)  \ .
\end{equation}
We  analyze the general action for the scalar field $\ph$ and their
equation of motion perturbatively and write an effective action for
the field $\ph$.

The generic form of the equation of motion (varying the general
action) up to order $\mu$
is given by,
\begin{equation}\label{geom}
\cA_0(r,k) \ph - \cA_1^{'}(r,k) \php -\cA_1(r,k) \phpp = \mu \ {\hat
{\cal
    F}}(\{\phi^{(p)}\}) + {\cal O}(\mu^2)
\end{equation}
where ${\hat {\cal
    F}}(\{\phi^{(p)}\})$ is some linear function of double and higher
derivatives of
$\ph$, coming from two or higher
derivative terms in equation (\ref{ghdacnphi}).
The zero$^{th}$ order ($\mu \rightarrow 0$) equation of motion is
given
by,
\begin{equation} \label{eom0}
\cA_0^{(0)}(r,k) \ph - \cA_1^{'(0)}(r,k) \php -\cA_1^{(0)}(r,k) \phpp
=0
\end{equation}
where, $\cA_p^{(0)}$ is the value of $\cA_p$ at $\mu \rightarrow 0$.
From this equation we can write $\phpp$ in terms of $\php$ and $\ph$
in $\mu \rightarrow 0$
limit.
\begin{equation} \label{phi20}
\phpp= {\cA_0^{(0)}(r,k)\over  \cA_1^{(0)}(r,k)} \ph -
{\cA_1^{'(0)}(r,k)
  \over \cA_1^{(0)}(r,k)} \php \ .
\end{equation}
Then the full equation of motion can be written in the following way,
\begin{eqnarray}\displaystyle \label{eomg}
\cA_0^{(0)}(r,k) \ph - \cA_1^{'(0)}(r,k) \php -\cA_1^{(0)}(r,k)
\phpp = \mu\
   {\tilde {\cal F}}(\ph,\php,\phpp,...)  + {\cal O}(\mu^2) \ .
\end{eqnarray}
Since the right hand side of equation (\ref{eomg}) is proportional to
$\mu$, we can replace the $\phpp$ and other higher (greater than 2)
derivatives of $\ph$ by its leading order value in equation
(\ref{phi20}).
Therefore up to order $\mu$ the equation of motion for $\phi$ is given
by,
\begin{eqnarray}\displaystyle \label{eomg2}
\cA_0^{(0)}(r,k) \ph - \cA_1^{'(0)}(r,k) \php -&\cA_1^{(0)}(r,k)
\phpp = \mu\
   {\cal F}(\ph,\php) + {\cal O}(\mu^2) \nonumber \\
&=  \mu ({\cal F}_1 \php + {\cal F}_0 \ph) + {\cal O}(\mu^2) \
\end{eqnarray}
where ${\cal F}_0$ and ${\cal F}_1$ are some function of $r$.
This is the perturbative equation of motion for the scalar field $\ph$
obtained from the general action in equation (\ref{ghdacnphi}).

\subsubsection{Strategy to Find The Effective Action}
\label{gendis1}

In this subsection we describe the strategy to write an effective
action for the field $\ph$ whose has equation of motion has the form 
of (\ref{acn02}) but with different functions. The prescription is
following:

{\bf (a)} We demand the equation of motion for $\ph$ obtained from the
original action and the effective action are same up to order $\mu$.
This will
fix the coefficients of $\phi^{'2}$ and $\phi^2$ terms in effective
action.

Let us start with the following form of the effective action.
\begin{eqnarray}\displaystyle \label{acneff}
\seff&=& {1 \over 16 \pi G_5}
\int {d\om d^3{\vec k}\over (2 \pi)^4}dr \bigg[ (\cA^{(0)}_1(r,k) +
\mu \cB_1(r,k)) \cphp \php \nn
&& \qquad \qquad + (\cA^{(0)}_0(r,k)+\mu \cB_0(r,k)) \ph \cph \bigg ]\
.
\end{eqnarray}
The functions $\cB_0$ and $\cB_1$ are yet to be determined. We
determine these functions by claiming that the equation of motion for
the scalar field $\ph$ obtained from this effective action is same as
equation (\ref{eomg2}) up to order $\mu$. The equation of motion for
$\ph$ from
the effective action is given by,
\begin{eqnarray}\displaystyle
\cA_0^{(0)}(r,k) \ph &- \cA_1^{'(0)}(r,k) \php -\cA_1^{(0)}(r,k) \phpp
= \mu \lb \cB_1^{'}(r,k) - {\cA_1^{'(0)}(r,k)
  \over \cA_1^{(0)}(r,k)} \cB_1(r,k) \rb \php  \nonumber\\
& + \mu \lb \cB_1(r,k) {\cA_0^{(0)}(r,k)\over
  \cA_1^{(0)}(r,k)} - \cB_0(r,k) \rb \ph + {\cal O}(\mu^2) \ .
\end{eqnarray}
Therefore comparing with equation (\ref{eomg2}) we get,
\begin{equation}
\cB_1'(r,k)  - {\cA_1^{'(0)}(r,k)
  \over \cA_1^{(0)}(r,k)} \cB_1(r,k) - {\cal F}_1(r,k)  =0
\end{equation}
and
\begin{equation}\label{cb00}
\cB_0(r,k)= \cB_1(r,k) {\cA_0^{(0)}(r,k)\over
  \cA_1^{(0)}(r,k)} - {\cal F}_0(r,k) \ .
\end{equation}
The solutions are given by,
\begin{eqnarray}\displaystyle \label{cb1}
\cB_1(r,k) &=& \cA_1^{(0)}(r,k)\int dr {{\cal F}_1(r,k)
\over \cA_1^{(0)}(r,k)} +
\kappa \cA_1^{(0)}(r,k)
={\tilde {\cal B}}_1(r,k) + \kappa  \cA_1^{(0)}(r,k)
\end{eqnarray}
and
\begin{equation}\label{cb0}
\cB_0= {\tilde \cB}_0(r,k) + \kappa \cA_0^{(0)}\
\end{equation}
for some constant $\kappa$. We need to fix this constant.

{\bf (b)} Condition {\bf (a)}
can not fix the overall normalization
factor of the
effective action.
In particular we can multiply it by $(1 +
\mu \Gamma)$ (for some constant $\Gamma$) and still get the same
equation of motion. Considering this normalization, the effective
action is given by,
\begin{eqnarray}\displaystyle
\seff&=& {\ 1 + \mu \ \Gamma \ \over 16 \pi G_5}
\int {d\om d^3{\vec k}\over (2 \pi)^4}dr \bigg[ (\cA^{(0)}_1(r,k) +
\mu \cB_1(r,k)) \cphp \php \nonumber \\
&&\hspace{2cm} + (\cA^{(0)}_0(r,k)+\mu \cB_0(r,k)) \ph \cph \bigg ]\ .
\end{eqnarray}
Substituting the values of $\cB$'s in equations (\ref{cb1}) and
(\ref{cb0}) we get,
\begin{eqnarray}\displaystyle
\seff &=& (1+ \mu (\Gamma+\kappa)) S^{(0)} + \mu\int dr \bigg [\tilde
{\cal
    B}_1(r,k) \cphp \php  + {\tilde \cB}_0(r,k) \cph \ph \bigg]
\end{eqnarray}
where $S^{(0)}$ is the effective action at $\mu \rightarrow 0$ limit.
This implies that the integration constant $\kappa$ can be absorbed in
the overall normalization constant $\Gamma$. Henceforth we will denote
this combination as $\Gamma$.

Our prescription is to take $\Gamma$ to be ${\bf zero}$ from the
following
observation.
\begin{itemize}
\item
The shear viscosity coefficient of boundary fluid is related to the
imaginary part of retarded Green function in low frequency limit. The
retarded Green function $G^R_{xy,xy}(k)$ is defined in the following
way: the on-shell action for graviton can be written as a surface
term,
\begin{eqnarray}\displaystyle
\label{rgfn}
S \sim  \intk
\,\phi_0(k)\,{\cal G}_{xy,xy}(k,r)\,\phi_0(-k)
\end{eqnarray}
where $\phi_0(k)$ is the boundary value of $\ph$
and $G^{R}_{xy,xy}$ is given by,
\begin{eqnarray}\displaystyle
G^R_{xy,xy}(k)=\lim_{r\rightarrow 0}2 {\cal G}_{xy,xy}(k,r)
\end{eqnarray}
and shear viscosity coefficient is given by\footnote{To calculate
this number   one has to know the exact solution, \ie the form of
$\xi$ and the value of   $\beta$ in equation (\ref{phiform}).},
\begin{equation}
\eta = \lim_{\omega \rightarrow 0} \bigg[ {1\over \omega} {\rm Im}
G^R_{xy,xy}(k)
  \bigg ] \qquad (computed \ \ on-shell)\ .
\end{equation}
\item
Now it turns out that the imaginary part of this retarded Green
function obtained from the original action and effective action are
same up to the normalization constant $\Gamma$ in presence of generic
higher derivative terms in the bulk action. Therefore it is quite
natural to take $\Gamma$ to be $zero$ as it ensures that starting from
the effective action also one can get same shear viscosity using Kubo
machinery. To show that the above statement is true we do not need to
know the full solution for $\phi$, in other words to find the
difference between the two Green functions one does not need to
calculate the Green functions explicitly. Assuming the following
general form of solution for $\phi$
\begin{equation} \label{phiform}
\phi \sim (1-r^2)^{-i \omega \beta} \lb 1 + i \omega \beta \mu \xi(r)
\rb
\end{equation}
it can be shown generically.  For a detailed proof see \cite{ns1}
\item
Because of the canonical form of the effective action, it follows
from the argument in \cite{liu} and the statement above, that the
shear viscosity coefficient of boundary \index{fluid!boundary}fluid is
given by the \index{horizon}horizon value of the effective coupling
obtained from the effective action in presence of any higher
derivative terms in the bulk action.
\end{itemize}
{\bf (c)} After getting the effective action for
$\ph$, the effective coupling is given by,
\begin{equation}\label{gkeff}
K_{{\rm eff}}(r)=  {1\over 16 \pi G_5}\
{\cA_1^{(0)}(r,k) + \mu \cB_1(r,k) \over
\sqrt{-g}
  g^{rr}}
\end{equation}
where $g^{rr}$ is the '$rr$' component of the inverse perturbed
metric and $\sqrt{-g}$ is the determinant of the perturbed metric.
Hence the shear viscosity coefficient is given by,
\begin{equation}
\eta = r_0^{-{3\over 2}} (-2 K_{{\rm eff}}(r=r_0))
\end{equation}
where $r_0$ is the corrected \index{horizon}horizon radius.
To summaries, we have obtained a well defined procedure to find the
correction (up to order $\mu$) to the coefficient of shear viscosity
of the boundary \index{fluid!boundary}fluid in presence of
general higher derivative terms in the action.

\subsection{Membrane fluid in higher derivative gravity}

Let us also consider the effect of the higher derivative terms on 
the membrane fluid. In higher derivative gravity, since the canonical 
form of the action
(\ref{acn02}) breaks down, it is not very obvious how to define the
membrane
charge $\Pi_{{\rm mb}}$. Instead of the original action if we consider
the
effective action (\ref{acneff}) for graviton then it is possible to
write the
membrane action perturbatively and define the membrane charge
($\Pi_{{\rm
mb}}$) in higher derivative gravity. As if the membrane
\index{fluid}fluid is
sensitive to the effective action $\seff$ in higher derivative
gravity.

Following \cite{liu} we can write the membrane action and charge in
the
following way (in momentum space)
\begin{equation}
S_{{\rm mb}}= \int_{\Sigma} {d^4k \over (2 \pi)^4}
 \sqrt{-\sigma} \lb{\Pi(r_0,k) \over
  \sqrt{-\sigma}} \phi(r_0,-k) \rb
\end{equation}
where $\sigma_{\mu\nu}$ is the induced metric on the membrane and
$\Pi(r,k)$ is conjugate momentum of $\phi$ with respect to $r$
foliation where $\Pi(r,k)=(\cA_1^{(0)}(r,k) + \mu \cB_1(r,k))\phi'(r,-k)$.
The membrane charge is then given by,
\begin{equation}
\Pi_{{\rm mb}} = {\Pi(r_0,k) \over \sqrt{-\sigma}}=- {\tilde
  K}_{{\rm eff}}(r_0) \sqrt{g^{(0)^{rr}}} \p_r
\ph \big |_{r_0} \ .
\end{equation}
Imposing the in-falling wave boundary condition on $\phi$, it can be
shown
that the membrane charge $\Pi_{{\rm mb}}$ is the response of the
\index{horizon}horizon
\index{fluid}fluid to the bulk graviton excitation and the membrane
\index{fluid}fluid transport
coefficient is given by,
\begin{equation}
\eta_{{\rm mb}}= {\tilde K}_{{\rm eff}}(r_0) \ .
\end{equation}

Hence, we see that even in higher derivative gravity the shear
viscosity coefficient of boundary fluid is captured by the
membrane fluid.

Motivated by the fact that like entropy of boundary theory, shear viscosity
coefficient is also determined in terms of data specified on the black 
hole horizon, we can think that if we know the near-horizon geometry of 
an asymptotically AdS black hole space time, it would be enough to compute 
the ratio $\eta/s$. In the next section we elaborate this particular issue
with an example.


\section{Near-horizon analysis of $\eta/s$}
\label{nhsection}

As we know that it is possible to find the entropy of boundary
field theory by only specifying the complete near horizon geometry of
its dual black hole spacetime, one can also ask in the same spirit,
whether different hydrodynamics coefficients of the boundary plasma
can be found from the knowledge of bulk near-horizon geometry.
There are enough hints from the above analysis which seems to
point to an affirmative answer to this question. 

In last section we have shown that in the low frequency limit 
($\omega \ra 0$) the shear
viscosity coefficient $\eta$ is related to transport coefficient of
membrane fluid, which in turn is given by the effective coupling
constant of graviton ($h_{xy}$) evaluated on the horizon. We have also 
described how to generalize this prescription has been generalized for any higher
derivative gravity. All these approaches
indicate that there can be an "IR" description of transport
coefficients $\eta$. However, it is not quite clear how other
transport coefficients like $\tau_{\pi}$ etc.
which appear in next
order in $\omega$ can be calculated only in terms of horizon data. The
derivation of \cite{liu,ns1} was strictly valid only in
$\omega \ra 0$ limit.

In this section, we
observe that only with the knowledge of near horizon geometry of a
black hole spacetime one can easily calculate the thermodynamic and
hydrodynamic quantities, namely entropy and shear viscosity of
boundary fluid and their ratio. One does not need to know the full
analytic solutions of Einstein equations. This observation is helpful
when bulk Lagrangian is very complicated. For example when gravity is
coupled to various matter fields (gauge field, scalars etc.) in
non-trivial way, it may not be always possible to find an analytic
solution\cite{ABD}. But if a black hole solution exists then it is
possible to
find the near horizon geometry, i.e. how the metric and other fields
behave in near horizon limit. The method is very simple. We do not
need to solve any differential equation to find the near horizon
geometry. First we write the field equations. Then take a suitable
near horizon ansatz for different fields. In general different
components of the metric goes like $g_{tt} \sim a_1 (r-r_h) + a_2
(r-r_h)^2+ \cdots, \ \ g_{rr}\sim {b_1 \over r-r_h}(1+ b_2
(r-r_h)+\cdots$)(for non-extremal case), scalar fields behaves as
$\varphi \sim \varphi_h + \varphi_1 (r-r_h)+\cdots$ and gauge fields
goes as $A_t = q(r-r_h)+\cdots$. Substituting these in the field
equations we find the coefficients consistently order by order in
$(r-r_h)$. Once we find the near-horizon structure of the spacetime we
find entropy of the system using Wald's formula.
which says we only need to calculate the horizon values of some
quantities. To calculate shear viscosity we adopt the method proposed
in last section\cite{liu,ns1}. We write the effective action for transverse
graviton in black hole near-horizon region and find the coupling
constant. For non-extremal black hole usually (in presence of higher
derivative terms) the effective coupling depends on radial coordinate
and we need to take $r\ra r_h$ limit at the end to find the horizon
value of the coupling constant. See \cite{nsnh} for detailed
discussion and examples.

However we would like to clarify the following
$important$ $points:$

Using the AdS/CFT conjecture we are studying the low frequency
hydrodynamic properties of field theory plasma which has a gravity
dual.
Our observation is certainly valid for those fluids whose dual gravity
satisfies the asymptotic AdS boundary condition. The presence of a
cosmological constant term in the bulk action ensures that the black
hole solution must be asymptotically AdS. We found near-horizon
geometry by solving the background equations of motion in presence of
the cosmological constant term. This ensures that we are solving the
near horizon geometry of an asymptotically AdS black hole space time.

One can also consider Lagrangian with different higher
derivative terms.
Higher derivative corrections typically make energy conditions e.g.
weak energy conditions in general relativity hard to interpret, as a
result of which the underlying solutions could be singular. This means
that it is not necessarily true that the solutions are asymptotically
AdS.
However in many
cases, the higher derivative terms of
$R^{(n)}$ kind (contraction of Riemann, Ricci or scalar) can be
treated perturbatively. Usually if the zeroth order black
hole solution is non singular then these higher derivative terms (when
treated perturbatively) do not create any potential hazard in the bulk
theory (blowing up of some invariant quantity between horizon and
infinity) and one gets solutions which are asymptotically AdS
\cite{ns1},\cite{ns2},\cite{Kats}. For more generic situations, if
the higher derivative terms are treated exactly, one has to take
special care to show that there exists a non-singular solution
everywhere between the horizon and the boundary which is
asymptotically AdS.

 The asymptotic AdS boundary condition is true for all the gravity
models we studied in this chapter. For generic models (where we
think our result is most useful) of realistic boundary plasmas such as
QGP or more generically QCD, one should encounter a dual gravity
theory with non-trivial matter coupling. A complete analytic solution
for such a background is almost impossible. For these cases, the
existence of the asymptotic AdS solution can be checked numerically,
and the corresponding near-horizon geometry can be obtained
analytically. In these scenarios, our observation will be useful to
extract information about the coefficient $\eta$ for the plasma.

Now we will present a particular model where the above observation
will be useful and we will compute the shear viscosity of the
boundary fluid dual to this system just from the knowledge of the
near-horizon geometry.

\subsection{The model} \label{model}
We will focus on a 5-dimensional theory of gravity coupled
to a massless scalar and an abelian electromagnetic field
whose action is\footnote{We also have to add counterterms
to regularize the action --- for (\ref{action1}), the
counterterms are given in \cite{Taylor:2000xw} (see, also,
the nice review \cite{Skenderis:2002wp} and references therein
for a more detailed discussion).}
\be
\label{action1}
S_{{\rm EM}} = \frac{1}{16 \pi G_5} \int d^5x \sqrt{-g}
\left( R + {12} - e^{\alpha \varphi(r)} F_{\mu\nu} F^{\mu\nu}
-\partial_{\mu}\varphi \partial^{\mu}\varphi \right) + S_{{\rm CS}},
\ee
\be
\label{yb}
S_{{\rm CS}} = {{\zeta\over 3} \over 16\pi G_5} \int  d^5x\
\epsilon^{\mu\nu\rho\sigma\gamma}A_{\mu} F_{\nu\rho} F_{\sigma\gamma}.
\ee
In this section, we consider a constant moduli potential,
$V(\phi)=2\Lambda=-12/l^2$, and also fix the radius of $AdS$ to be
$l=1$.

Since the equations of motion for the gauge field simplify, we choose
the constant in front of the Chern-Simons term to be $\zeta/3$. The
action
(\ref{action1}) with various values for $\alpha$ resembles (truncated)
actions obtained in string compactifications \cite{Liu:2010sa}. The
coupling $\zeta$
captures the strength of the anomaly of the boundary current.

The equations of motion for the metric, scalar, and electromagnetic
field ($F_{\mu\nu}=\partial_{\mu}A_{\nu}-\partial_{\nu}A_{\mu}$) are
\be
\label{einstein}
R_{\mu\nu}+4 g_{\mu\nu}+e^{\alpha \varphi(r)}\lb {1\over 3}
F^2 g_{\mu\nu}-2
F_{\mu\rho}F_{\nu\sigma}g^{\rho\sigma}\rb-\partial_{\mu}\varphi
\partial_{\nu}\varphi=0\ ,
\ee
\be
\label{scalar}
\frac{1}{\sqrt{-g}}\partial_{\mu}(\sqrt{-g}\partial^{\mu}\varphi)
  = -\frac{1}{2}\alpha e^{\alpha \varphi(r)}F_{\mu\nu} F^{\mu\nu}\ ,
\ee
\be
\label{gauge}
e^{\alpha\varphi(r)}\partial_{\nu}\lb \sqrt{-g}F^{\nu\mu}\rb +
{\zeta\over 4}
\epsilon^{\mu\rho\sigma\gamma\delta}F_{\rho\sigma}F_{\gamma\delta}=0\
,
\ee
where we have varied the scalar and the electromagnetic field
independently. The Bianchi identities for the gauge field are
$F_{[\mu\nu;\lambda]}=0$.

Since we are interested in a theory for which the Chern-Simons term
has a non-trivial contribution, we consider the following ansatz for
the
gauge field:\footnote{Please note that our $P'(r)$ is the same as
$P(r)$ in
\cite{D'Hoker:2009bc} and our $Z(r)$ in the ansatz of the metric
(\ref{anz2}) is their $C(r)$.}
\be
A=E(r) dt - {B\over 2} y dx + {B\over 2} x dy - P(r) dz\ .
\ee
The magnetic field, $B$, is fixed to be constant by the Bianchi
identities. Thus, the field strength is \footnote{Our convention
for the coordinates is $(r,t,x,y,z)$.}
\be
\label{Fansatz}
F=\left(
\begin{array}{ccccc}
 0 & E'(r) & 0 & 0 & -P'(r) \\
 -E'(r) & 0 & 0 & 0 & 0 \\
 0 & 0 & 0 & B & 0 \\
 0 & 0 & -B & 0 & 0 \\
 P'(r) & 0 & 0 & 0 & 0
\end{array}
\right)\ ,
\ee
where $'$ denotes derivatives with respect to $r$.

Our analysis is on time-independent black hole solutions and so we
consider
the following ansatz for the metric
\be
\label{anz2}
ds^2 = {dr^2 \over U(r)} - U(r) dt^2 + e^{2V(r)} \left ( dx^2 + dy^2
\right  )
+ e^{2 W(r)} \left ( dz + Z (r) dt \right )^2\ ,
\ee
which is compatible with the symmetries of the problem.

In this case, the horizon is located at (the biggest
root of) $U(r_h)=0$ and the temperature can be easily computed on the
Euclidean section --- we obtain
\be\label{temp}
T= {U'(r_h)\over 4\pi}\ .
\ee

By using the metric ansatz (\ref{anz2}), we can rewrite the Maxwell
equations as
\be
\label{ME1}
[Q(r) e^{2 V(r)+W(r)+\alpha  \varphi (r)}]' - 2  \zeta B P'(r)=0,
\ee
\be
\label{ME2}
[e^{2 V(r)+W(r)+\alpha  \varphi (r)} \left(U(r) e^{-2 W(r)}
   P'(r)-Q(r) Z(r)\right)]' - 2 \zeta B E'(r)=0
\ee
where
\be
Q(r) = E'(r) + Z(r) P'(r).
\ee

It is easier to work with combinations of Einstein equations rather
than using directly (\ref{einstein}). First, we extract the
expressions
of second derivatives of the functions that characterize the metric in
the following way: we obtain $W''(r)$ from $(rr)$- , $V''(r)$ from
$(xx)$- , and $U''(r)$ from $(zz)$-component of Einstein equations.

Let us now consider the $(tz)$-component of Einstein equations in
which
we replace $W''(r), V''(r),$ and $U''(r)$ --- we obtain
\be
\label{E4}
e^{2 W(r)} \left[2 V'(r) Z'(r)+3 W'(r) Z'(r)+Z''(r)\right]-4
   Q(r) P'(r) e^{\alpha  \varphi (r)}=0.
\ee
%
then
considering the

An important observation, which can be drawn by studying the system of
equations
(\ref{ME1})--(\ref{E4}), is that a non-zero magnetic field is not
compatible
with a constant function $Z(r)$ (and, also, $P(r)$).

The other (independent) combinations of Einstein equations
are obtained as follows:
by replacing $U''(r)$ in the $(rr)$-component of Einstein equations we
get
\ben\label{E1}
&& 2 B^2 e^{\alpha  \varphi (r)-4 V(r)}+2 Q(r)^2 e^{\alpha  \varphi
   (r)}+U(r) [2 V'(r)+W'(r)]^2\nn
&&+[U(r) \left(2
   V'(r)+W'(r)\right)]'+\frac{1}{2} e^{2 W(r)} Z'(r)^2-12=0
\een
by replacing $W''(r)$ in the $(zz)$-component of Einstein equations
\ben\label{E3}
&&-4 Q(r)^2 e^{\alpha  \varphi (r)}+U''(r)+U'(r) [2
   V'(r)-W'(r)]-2 e^{2 W(r)} Z'(r)^2\nn
&&+2 U(r) [2 V''(r)-2 V'(r) W'(r)+2
   V'(r)^2+\varphi '(r)^2]=0
\een
and the last one is, in fact, the $(xx)$-component of Einstein
equations
\ben\label{E2}
&&e^{2 W(r)} [-4 B^2 e^{\alpha  \varphi (r)}-3 e^{4 V(r)}
   U'(r) V'(r)+12 e^{4 V(r)}]-2 Q(r)^2
   e^{4 V(r)+2 W(r)+\alpha  \varphi (r)}\nn
&&+U(r) e^{4 V(r)} [2
   P'(r)^2 e^{\alpha  \varphi (r)}-3 e^{2 W(r)}
   \left(V''(r)+V'(r) W'(r)+2 V'(r)^2\right)]=0.
\een
We also use the ansatz of the metric in the equation of motion for
the scalar (\ref{scalar}) and so this equation becomes
\be
\label{SC}
[U(r) e^{2 V(r)+W(r)} \varphi '(r)]'+ \alpha  e^{2 V(r)+W(r)+\alpha
\varphi (r)} [B^2 e^{-4
   V(r)}+U(r) e^{-2 W(r)} P'(r)^2-Q(r)^2]=0.
\ee

Due to the non-trivial coupling between the scalar and gauge fields,
the equation (\ref{scalar}) has a non-trivial right hand side. The
non-vanishing electromagnetic field may also be understood as a source
for the scalar field. Thus, the scalar charge is determined by the
electric and magnetic charges and so it is not an independent
parameter
that characterizes the system --- this charge plays an important
role when the asymptotic value of the scalar is not fixed (see
\cite{Gibbons:1996af}).

We conclude this section with an observation on the
Hamiltonian constraint. A vanishing Hamiltonian is a characteristic
feature of any theory that is invariant under arbitrary coordinate
transformations --- for our system, we can obtain a first order
differential equation by replacing $W''(r), V''(r),$ and $U''(r)$
in the $(tt)$-component of Einstein equations.

The Hamiltonian constraint, which can be enforced as an initial
condition, has the following expression:
\ben\label{CON}
&& 2 B^2 e^{\alpha  \varphi (r)-4 V(r)}+U(r)[-2 P'(r)^2
   e^{\alpha  \varphi (r)-2 W(r)}+4 V'(r) W'(r)+2
   V'(r)^2-  \nn
&& -\varphi '(r)^2]+2 Q(r)^2 e^{\alpha  \varphi
   (r)}+2 U'(r) V'(r)+U'(r) W'(r)+\frac{1}{2} e^{2 W(r)}
   Z'(r)^2-12=0 \ .
\een

\subsection{Non-extremal Near horizon geometry}\label{nhsec}

In this section, we will first find the near-horizon geometry of
the non-extremal black hole, which we will need to compute
the shear viscosity to entropy density ratio. In
the extremal limit, due to the attractor mechanism,
the near horizon geometry is universal regardless of the asymptotic
values of the scalars. We will not present a detailed analysis of the
extremal branches of solutions here, those can be found in \cite{ABD}.
Here we only present the near-horizon geometry of the non-extremal
case.


As in \cite{D'Hoker:2009bc}, we work with a coordinate system in which
the solution takes the canonical form at the horizon. That is, the
field
strength $F_H$ and the metric $ds_H^2$ are
\ben
\label{nearh}
F_H & = & q \, dr\wedge dt + B \, dx \wedge dy -p \ dr\wedge dz ,
\nn
ds_H^2 & = & r_H^2(dx^2 + dy^2 + dz^2) ,
\een
where $q$ and $B$ are the charge density (of the black brane) and
the magnetic field at the horizon, respectively. In this way, the
gauge freedom is removed and the initial conditions are
\ben
U(r_h)=Z(r_h)=P(r_h)=0\,\,\, , \,\,\,\,\,\,\,\,\,
V(r_h)=W(r_h)=\ln(r_h).
\een
A similar analysis (and numerical solutions) in the presence of
the Gauss-Bonnet term but without the Chern-Simons term was presented
in \cite{Astefanesei:2008wz}.

The generic solutions have a non-degenerate horizon. Near the event
horizon, they admit a power series expansion of the form (using the
definition of the temperature (\ref{temp}) in the expression of $U$):
\ben
\label{nonextremal}
U(r)&=& 4 \pi T (r-r_h)+ u_2 (r-r_h)^2 + \cdots , \nn
V(r)&=& \ln(r_h)+ v_1 (r-r_h)+ v_2 (r-r_h)^2 + \cdots ,\nn
W(r)&=& \ln(r_h)+ w_1 (r-r_h)+ w_2 (r-r_h)^2 + \cdots ,\nn
Z(r)&=& z_1 (r-r_h)+ z_2 (r-r_h)^2 + \cdots ,\nn
E(r)&=& q (r-r_h)+ q_1 (r-r_h)^2 + q_2 (r-r_h)^3 +\cdots ,\nn
P(r)&=& p (r-r_h) + p_1 (r-r_h)^2 + \cdots ,\nn
\varphi(r)&=& \varphi_h + \varphi_1 (r-r_h)+\varphi_2 (r-r_h)^2 +
\cdots
\een
It is important to emphasize that, what is generally called
near horizon geometry for a non-extremal black hole is just a
truncation of the above series expansion. To compute the shear
viscosity, though, we need also some data at the order $(r-r_h)^2$.

Another observation is that, in principle, one can use a boost
transformation in $z$ direction to set $p=0$ (see
\cite{D'Hoker:2009bc}).
However, the boost transformation is singular at some point
outside the black hole horizon. In our analysis we keep the value of
$p$ non-zero and determine
it in terms of other horizon data. We will see in section \ref{etabys} that
the expressions for the entropy, shear viscosity, and
their ratio remain unchanged if we set the horizon value of $P'(r)$
to be zero (in other words, they do not depend of $p$). This is
expected due to the fact that the physical quantities should
be invariant under the boost transformations.

By substituting the ansatz (\ref{nonextremal}) in the field equations,
we get the following expressions for the coefficients at the order
$(r-r_h)$:\footnote{We obtain the results as functions of the
coefficients
($T, \ q, \ B, \ \varphi_h$, $r_h$, and $z_1$) --- this will simplify
the
computations of the shear viscosity.}
\ben
\label{AA}
v_1&=&-\frac{2 B^2 e^{\alpha  \varphi _h}+r_h^4 \left(q^2 e^{\alpha
   \varphi _h}-6\right)}{6 \pi  T r_h^4},
\nn
w_1&=&\frac{4 B^2 e^{\alpha  \varphi _h}-r_h^4 \left(4 q^2 e^{\alpha
   \varphi _h}+3 z_1^2 r_h^2-24\right)}{24 \pi  T r_h^4},
\nn
p&=& \frac{q \left(2 B \zeta  e^{-\alpha  \varphi _h}+z_1
   r_h^3\right)}{4 \pi  T r_h},\nn
\varphi_1 &=& \frac{\alpha  e^{\alpha  \varphi _h} \left(q^2
   r_h^4-B^2\right)}{4 \pi  T r_h^4},
\nn
q_1&=& {e^{-2 \alpha  \varphi _h} \over 16 \pi
   T r_h^4}\bigg( 2 B^2 [\left(\alpha
   ^2+2\right) e^{3 \alpha  \varphi _h} +4 \zeta ^2]\nn
&& -r_h^4
   e^{2 \alpha  \varphi _h} [2 q^2 \left(\alpha ^2-2\right)
   e^{\alpha  \varphi _h}+z_1^2 r_h^2+24]\bigg).
\een

However, in higher derivative gravity theories the only
coefficient at the order $(r-r_h)^2$, which we need for viscosity
bound computation, is $u_2$. But, for completeness, we
present the expressions of all the other coefficients
that appear at order $(r-r_h)^2$, in Appendix \ref{hocoeff}.

A non-extremal charged scalar black hole is characterized by
four independent parameters: the mass, electric charge, magnetic
field, and also the value of the scalar at the horizon, $\varphi_h$.
In this case, the horizon radius (and so the entropy) and the
horizon value of the scalar depend of the asymptotic boundary
data ($\varphi_\infty$). We will see in the next subsection that
this is in contrast with the extremal case for which we obtain
an attractor behavior at the horizon.

At first sight, it may seem surprising that the data
(\ref{AA}, \ref{BB}) we need to compute the entropy and
shear viscosity depend also on $z_1$, a
coefficient that we do not compute explicitly. However, we will see
in the next subsection that the final values of
the physical quantities depend in fact just on four independent
parameters, namely $(q,B,r_h,\varphi_h)$ --- we `trade' the mass
for the horizon radius and so the independent parameters that
completely characterize the black hole are the ones mentioned above.

\subsection{Shear Viscosity to Entropy Density Ratio}
\label{etabys}

Interestingly, as has already been observed that only
with the knowledge of the near horizon geometry one can easily
calculate the shear viscosity of boundary fluid. One does not need to
know the full analytic solutions of Einstein equations --- this method
is especially useful in higher derivative AdS gravity theories.

At first sight, it seems that the dual gravitational mode $(4.1)$ does
not generally decouple. Interestingly enough, the decoupling occurs
when the momentum vanishes and this is what we need for the
computation
of the viscosity in the hydrodynamics limit --- a detailed
derivation of this claim has been provided in Appendix \ref{hxy}.

By plugging (\ref{petmet}) in the action and keeping the terms at
order
$\epsilon^2$ (at the first order in $\epsilon$, we obtain
the equations of motion for gravitons), we get the effective
action for the perturbation of the form equation (\ref{ghdacnphi})
:\footnote{The terms that contain
the derivatives with respect to the spatial coordinates, $\vec{x}$,
combine in terms whose coefficient is proportional with
$\vec{p}^2$. Since we work in the hydrodynamic approximation
$\vec{p}=0$, these terms do not play any role in our analysis.}

Next, we integrate by parts to obtain the bulk action for the graviton
in the following form (up to some total derivative terms):
\be
S=\nt \int {d^4 k \over (2 \pi)^4} dr [{\cal A}_1(r,k)
\phi'(r,k)\phi'(r,-k) +{\cal A}_0(r,k) \phi(r,k)\phi(r,-k)] ,
\ee
where
\be
{\cal A}_1(r,k)=-{1\over 2}e^{2V(r)+W(r)} U(r), \ \ \ \ \ {\cal
A}_0(r,k)={e^{2V(r)+W(r)}\omega^2 \over 2U(r)}.
\ee
At this point, it is important to emphasize that there are many
total derivatives in this action that do not affect the equations
of motion for the graviton. For the computation of the
{\it imaginary part} of the two-point function, the coefficient of
the term $\phi'\phi'$ in the bulk action is important. The other
total derivatives in the bulk action and the Gibbons-Hawking surface
term contribution exactly cancel on the boundary \cite{cai1}. It was
also shown \cite{cai1} (straightforward to check 
for present case) that, in the case of Einstein gravity, the ratio of
viscosity and entropy density is not affected when the matter fields
are minimally coupled. The effective coupling \cite{ramy1,
cai1} is
\be
K_{eff}={1\over 16 \pi G_5}{{\cal A}_1(r,k) \over \sqrt{-g}g^{rr}}=
-{1\over 32 \pi G_5}
\ee
and so the viscosity coefficient of the boundary fluid stress tensor
is
\be
\eta = e^{2V(r_h)+W(r_h)}(-2 K_{eff}(r_h)).
\ee

In this case, the shear viscosity to entropy density ratio turns out
to be universal, namely
\be
{\eta\over s}=\frac{1}{4 \pi }.
\ee

\subsubsection{Four derivative action}

Let us now consider the action (\ref{action1}) supplemented
with the most general four-derivative interactions \cite{mps}:
\begin{eqnarray}\label{hdaction}
S_{{\rm HD}}&=&S_{EM}+ \frac{\alpha'}{16 \pi G_5}\int d^5x\sqrt{-g}
\bigg [c_1 R_{abcd}R^{abcd}
+c_2 R_{abcd}F^{ab}F^{cd} +c_3
(F^2)^2 \\ \nn
&& \ \ \ \ \ \ \ \ \ \ \ \ \ \ \ \ \ \ \ \ \ \ \ \ \ \ \ \ \ \ \ \ \ \
\ \
+ c_4\,F^4 +c_5\, \epsilon^{abcde} A_a R_{bcfg} R_{de}{}^{fg}\bigg ]\,
.
 \nonumber
\end{eqnarray}

Since in supergravity actions the gauge kinetic terms couple to
various scalars, it will be interesting to understand the role
of the moduli in computing the viscosity bound. Unlike \cite{mps},
our action contains a scalar, $\varphi$, and the coefficients $c_i$
depend on the value of $\varphi$. This resembles the four-derivative
supergravity action \cite{Hanaki:2006pj}.

We treat the higher derivative terms perturbatively and apply the effective action
method of section \ref{gendis} to compute the shear viscosity coefficient of
the boundary fluid. However, to obtain the viscosity bound we also
need the entropy
density. We start by using the Noether charge formalism of Wald
\cite{Wald:1993nt} (see, also, \cite{dg, Astefanesei:2008wz} for a
discussion
in AdS) to compute the entropy density --- we will need just the
data in Section $3.1$ and Appendix B.

When we add higher derivative corrections to the action, the entropy
is no longer given by the area law --- instead, we use a general
formula proposed by Wald
\be
s=-2\pi \int_{{\cal
H}} {\partial L \over \partial R_{abcd}}\epsilon_{ab}\epsilon_{cd} ,
\label{eq:wald2}
\ee
where $\epsilon_{ab}$ is the binormal to the surface ${\cal H}$.

By using (\ref{eq:wald2}), we obtain the following expression for
the entropy density:
\be
\label{s}
s=\frac{r_h^3}{4G_5}+\frac{\alpha'}{4G_5}\,r_h^3\,[
c_1(3z_1^2r_h^2-4u_2)-2c_2q^2]+{\cal}O(\alpha'^2).
\ee
We use the expression of $u_2$ given in Appendix B to rewrite this
expression as
\be
s=\frac{r_h^3}{4 G_5}-{\alpha' \ r_h^3\over G_5}\ltb {c_1\over 3}
\lb {5B^2\over r_h^4} e^{\alpha \varphi _h} + 7 q^2 e^{\alpha \varphi
_h}-6\rb
+{c_2\over 2}q^2 \rtb +{\cal}O(\alpha'^2).
\ee
As expected, the entropy density depends on four independent
parameters,
namely $(r_h,q,B,\varphi_h)$. Since in Wald formula only the four
derivative
interactions that involve the curvature tensor are important, the
entropy
only depends on $c_1$ and $c_2$ ($c_5$ does not appear because
the binormal has just $rt$ components and the contribution from this
term vanishes).

To compute the four derivative corrections to the shear viscosity
coefficient, we have to find the quadratic action for the
transverse graviton moving in the background spacetime. As in previous
section, we consider again the following metric perturbation
\be
g_{xy}=g^{(0)}_{xy}+ h_{xy}(r,x)=g^{(0)}_{xy}[1+\ep \Phi(r,x)] ,
\ee
where $\epsilon$ is an order counting parameter. We find the effective
action for transverse graviton following the prescription described in previous
section. From that effective action one can extract the form of effective coupling.

Evaluating the effective coupling in the near horizon we obtain the
shear viscosity coefficient
\ben
\eta&=&\frac{1}{16 \pi  G_5} \nn
 && +\frac{\alpha' \left(c_1 r_h^4 \left(8 q^2 e^{\alpha  \varphi _h}
+3 z_1^2 r_h^2-32 \pi  T v_1-36 \pi  T w_1-10 u_2+48\right)-B^2
\left(c_2-8 c_1
   e^{\alpha  \varphi _h}\right)\right)}{8 \pi  G_5 r_h^4}\nn
&& + {\cal O}(\alpha'^2)
\een
which can be rewritten as (we use the results in Appendix \ref{hocoeff})
\be
\eta=\frac{r_h^3}{16 \pi  G_5}-{\alpha' \ r_h^3\over 2 \pi G_5}  \ltb
c_1 \lb q^2
+ {B^2\over r_h^4}\rb e^{\alpha \varphi _h} + {c_2\over 4}{B^2\over
r_h^4}\rtb
+ {\cal O}(\alpha'^2).
\ee
The ratio of the shear viscosity and entropy density turns out to be
\be
{\eta\over s}=\frac{1}{4 \pi }+\frac{\alpha'}{\pi} \ltb {c_1\over
3}\lb (q^2
- \frac{B^2}{r_h^4}) e^{\alpha \varphi _h} -6\rb +{c_2\over 2}(q^2
- \frac{B^2}{r_h^4}) \rtb+ {\cal O}(\alpha'^2).
\ee
In $B\ra 0$ limit this result matches with \cite{mps}.\footnote{Note
that our
`$q$' is different than `$q$' of \cite{mps}. In \cite{mps}, $q$ is the
physical
charge (up to some normalization). In our case, the physical charge is
$\sim r_h^6 q$.}

Let us end up this section with a discussion of the extremal limit. In
the absence of the moduli, the extremality condition is
$2B^2  +r_h^4\lb q^2 -6\rb = 0$ and so the shear viscosity to entropy
density ratio becomes
\be
{\eta\over s}=\frac{1}{4 \pi }+\frac{\alpha'}{\pi} \ltb -c_1 {B^2\over
r_h^4}
+ {3 c_2 \over 2} \lb 2- {B^2 \over r_h^4}\rb \rtb + {\cal
O}(\alpha'^2).
\ee

Therefore, there is a drastic change when the magnetic field is turned
on. That is,
unlike the electrically charged solution studied in \cite{mps}, the
leading
correction of $\eta/s$ in the extremal limit depends on both, $c_1$
and $c_2$.
As expected, in $B\ra 0$ limit our result matches with the one of
\cite{mps}.

Thus we see, with the knowledge of near-horizon geometry we have been able to
compute the ratio $\eta/s$ of a boundary system which is dual to some non-trivial
bulk theory, whose analytic solution is not known completely.
 
In the next section we will discuss about the transport coefficients, 
which appear at the second order of derivative expansion in energy momentum tensor. 
Other examples of higher derivative correction to first order transport coefficients
have been discussed in section \ref{r4sec2nd}.

\section{Second order Hydrodynamics}\label{prob2}

In the introduction section we have discussed that hydrodynamics is
an effective description of field theory in terms of derivative
expansions of its local variables. One can write the energy-momentum
tensor of the system in terms of derivative expansion. We have written
the form of energy-momentum tensor up to first order in derivative
expansion in equation (\ref{emt1st}). In this section we discuss about
the
second order expansion of energy-momentum tensor. At the second order
we
encounter many more transport coefficients. Before discussing the
holographic method to compute these transport coefficients we must
mention that second order hydrodynamic description is necessary in a
relativistic theory, as the first order description is not consistent
with causality issues \cite{causa1,causa2,causa3}. For example, consider
a fluid system with some conserved current ${\cal J}^{\mu}$. The current
satisfies the conservation equation and diffusion equation,
\be\label{ficks}
\partial_{\mu}{\cal J}^{\mu}=0, \qquad {\cal J}_{i}=-D \partial_i \rho,
\ee
where $D$ is the diffusion constant and $\rho$ is corresponding charge
density. Using these two equations one can write,
\be
\partial_t \rho -D \nabla^2 \rho =0.
\ee
This is a parabolic equation (first order in $t$ and second order in $x_i$'s), and
does not satisfy causality (see \cite{causa1,causa2,causa3} for details). In order to
restore causality one needs to make the diffusion equation hyperbolic. The only possible
way to do that is to modify the Fick's law, i.e. to introduce a new term in the
second equation of (\ref{ficks}),
\be\label{modficks}
{\cal J}_{i}=-D \partial_i \rho- \tau \partial_t J_i.
\ee
The coefficient $\tau$ is called the relaxation time. Now using equation (\ref{modficks}) and
conservation equation we can write,
\be
\tau \partial_t^2 \rho + \partial_t \rho -D \nabla^2 \rho =0.
\ee
This is hyperbolic equation. A propagating solution $\rho = \exp[i \omega t + i q z]$
leads to dispersion relation
\be
-\tau \omega^2 - i \omega +D q^2 =0.
\ee
Therefore in $q\ra \infty$ limit, the wavefront velocity turns out to be
$v \sim \sqrt{D/\tau}$, which is consistent as long as $v <c$. Thus, the
second order terms in diffusion equation or dispersion relation saves
the causality problem which appear at the first order hydrodynamics
(i.e. if we consider first order terms only).

Like the conserved current ${\cal J}^{\mu}$ the first order energy
momentum tensor also faces the same problem. Therefore
one has to add higher derivative terms to energy-momentum tensor of a
relativistic fluid. From the conformal symmetry of the system the form of the
second order energy momentum tensor can be determined (like first
order). In this article we will not discuss about the derivation of
the second order EM tensor rather, we start with the expression. The derivation 
can be found in\footnote{In appendix
  \ref{revshiraz} we have briefed the method of \cite{causa1}.}
\cite{loga,causa1,causa2}. In \cite{loga}, the author has developed a 
Weyl-covariant formalism which simplifies the study of conformal hydrodynamics. 

The energy-momentum tensor up to second order in derivative expansion
is given by,
\ben \label{T2nd}
T^{\mu\nu}&=&(\epsilon + P)u^\mu u^\nu + Pg^{\mu\nu} -\sigma^{\mu\nu} +
 \Theta^{\mu\nu}\nn
  \Theta^{\mu\nu} & =&
    \eta \tau_\Pi \left[ {}^\<D\sigma^{\mu\nu}{}^\> + \frac1{d-1}
\sigma^{\mu\nu}
    (\nabla{\cdot}u) \right]
 +\kappa\left[R^{\<\mu\nu\>}-(d-2) u_\alpha R^{\alpha\<\mu\nu\>\beta}
     u_\beta\right]\nn
  &&\quad + \lambda_1 {\sigma^{\<\mu}}_\lambda \sigma^{\nu\>\lambda}
  + \lambda_2 {\sigma^{\<\mu}}_\lambda \Omega^{\nu\>\lambda}
  + \lambda_3 {\Omega^{\<\mu}}_\lambda \Omega^{\nu\>\lambda}\, .
\een
The notations are following,
\ben
\Omega^{\mu\nu} &=& \frac12 \Delta^{\mu\alpha}\Delta^{\nu\beta}\lb
\nabla_{\alpha}u_{\beta}-\nabla_{\beta}u_{\alpha} \rb\ ,\nn
\Delta^{\mu\nu} &=& u^{\mu}u^{\nu}+g^{\mu\nu}\ ,\nn
D&=& u^{\mu}\nabla_{\mu}
\een
and for a second rank tensor tensor $A^{\mu\nu}$, 
\ben
^{<}A^{\mu\nu>} \equiv \frac12 \Delta^{\mu\alpha}\Delta^{\nu\beta} 
(A_{\alpha\beta}+A_{\beta\alpha}) -\frac{1}{d-1} \Delta^{\mu\nu}
\Delta^{\alpha\beta} A_{\alpha\beta} \equiv A^{<\mu\nu>}\ .
\een
It is an easy exercise to check that $^{<}A^{\mu\nu>}$ is transverse 
and traceless i.e. $u_{\mu}^{<}A^{\mu\nu>}=0$ and $
g_{\mu\nu} ^{<}A^{\mu\nu>}=0$. Thus $T^{\mu\nu}$ is also transverse 
and traceless up to second order in derivative expansion.

We see
that, we encounter five second order transport coefficients which are
named $\tau_{\Pi}, \kappa, \lambda_{1,2,3}$ after
\cite{causa2}. For
non-conformal fluid there are eight other transport coefficients appear
in the second order. 

Note that if we consider the
hydrodynamic system in flat space, then $\kappa$  term does not
appear, as the Ricci and Riemann tensor vanishe for flat
metric. However, $\kappa$ contributes to the two-point Green’s
function of the stress-energy tensor (please look at equation
(\ref{Gdef})). The $\lambda_{1,2,3}$ terms are nonlinear in
velocity. They also do not appear when we consider small
perturbations and thus do not appear in Green's function.  The
parameter $\tau_{\Pi}$ has dimension of time and can be thought of as
the relaxation time.  This $\tau_{\Pi}$ term is enough to get rid of
the causality problem.  

The main goal of this section is to discuss a holographic technique
to compute the 
second order transport coefficients. If we follow the analysis given
in section \ref{kubo-section}, we find that the expression for
Green's function up to second order in frequencies is given by,
\be\label{Gdef}
G_R^{xy,xy}(\omega, k) = p - i\eta\omega + \eta\tau_\Pi \omega^2
      -\frac\kappa 2 [(d-3)\omega^2+k^2]+{\cal O}(k^3)\, .
\ee
One important thing to note here (also mentioned earlier) that other three transport
coefficients do not appear in the expression of Green's function.
Therefore we can not compute those three transport coefficients
calculating retarded Green's function. One needs to adopt different
holographic methods \cite{causa1,causa2}.

In this section we study the radial evolution of response function
defined by equation (\ref{barchi}). The retarded Green's function of
boundary
fluid is given by the asymptotic value of this response function.
We show that, at low frequencies, the evolution of this response
function is independent of the radial direction and hence it can be
computed either at the horizon or at the boundary. Thus, computing the
response function at horizon, the shear-viscosity coefficient, which
is a first-order transport coefficient of the boundary plasma, can be
obtained from the characteristics of the membrane fluid. In order to
completely specify the boundary plasma, it is necessary to understand
its higher-order transport coefficients and one needs to move away
from the low frequency limit. In this case, the response function
flows non-trivially with the radial direction and depends on the full
black hole geometry. Hence, although the boundary plasma and the
membrane fluid have the same shear-viscosity coefficients, other
transport coefficients (higher order) can differ and it is not clear
how the two are related. One possibility is that the higher-order
coefficients of the two are related by renormalization-group flow
equations.

We again look at leading Einstein-Hilbert (E-H) action with a negative
cosmological constant in 4+1 dimensions and study the motion of a
transverse graviton in this background\footnote{Here we restrict
ourselves to five space-time dimensions, but the discussions are quite
generic and can be extended to arbitrary dimensions.}. The action and
the solution is given by, equations (\ref{EHacn}) and (\ref{metric0}).

We study the graviton's fluctuation in this background like equation
(\ref{petmet}). Here we use Fourier transform to work in the momentum
space $k=\{-\omega,\vec{k}\}$.From this action, we can find the
conjugate momentum $\Pi(r,k_\mu)$ of the transverse graviton (for
r-foliation) and the equation of motion from action given in equation
(\ref{acn02})\footnote{The factor $1/16\pi G_5$ has been absorbed in
${\cal A}_i^{(0)}$},
\be\label{eom}
\Pi(r,k_\mu)= 2 {\cal A}^{(0)}_1(r,k) \phi'(r,k),  \quad
\Pi'(r,k_{\mu})-2\
{\cal A}^{(0)}_0(r,k) \phi(r,k)=0 \ .
\ee
The on-shell action reduces to the following surface
term\footnote{Please look at \cite{nsflow} for detailed discussion on
other boundary terms.}
\be\label{boundacn}
S= \sum_{r=0,1}\int {d^4 k \over (2 \pi)^4 }({\cal A}^{(0)}_1(r,k)
\phi'(r,k)\phi(r,-k)).
\ee
Following the AdS/CFT prescription given in \cite{son2} (also look
at
section \ref{prob1}), the boundary retarded Green's function is given
as,
\be
G_R(k_\mu)=\lim_{r \rightarrow 0}\frac{2{\cal A}^{(0)}_1(r,k)
\phi'(r,k)\phi(r,-k)}{\phi_0(k) \phi_0(-k)},
\ee
where, $\phi_0(k_\mu)$ is the value of the graviton fluctuation at
boundary. Full solution of the graviton can be written as
$\phi(r,k_\mu)= \phi_0(k_\mu)F(r,k_\mu)$, where $F(r,k_\mu)$
goes to identity at the boundary.
We can rewrite the boundary retarded Green's function as,
\ben
G_R(k_\mu)&=& \lim_{r \rightarrow 0}{\Pi(r,k_\mu)\over \phi(r,k_\mu)}.
\een
Let us define a response function of the boundary theory
as\footnote{We set the zero frequency part of G to zero, as it gives
contact terms},
\be\label{barchi}
\bar \chi(k_\mu,r)= {\Pi(r,k_\mu)\over  i \omega \phi(r,k_\mu)}
\ee
where $\omega=k_{0}$.
This function is defined for all $r$ and $k_\mu$. Therefor the
boundary Green's function is given by,
\be
G_R(k_\mu)=\lim_{r \rightarrow 0} i \omega \bar \chi(k_\mu,r).
\ee

We will study the radial evolution of the response function $\bar
\chi(k_\mu)$
from horizon to boundary. Differentiating equation (\ref{barchi}) and
using
the equations of motion (\ref{eom}) we get,
\be\label{flow}
\partial_r \bar \chi(k_{\mu},r)= i \omega \sqrt{- {g_{rr} \over
g_{tt}}}
\Bigg[{\bar \chi(k_{\mu},r)^2 \over \Sigma(r)}- {\Upsilon(r) \over
\omega^2}\Bigg],
\ee
where we define
\ben
\Sigma(r)= - 2 {\cal A}^{(0)}_1(r,k_\mu)\sqrt{-{g_{rr} \over g_{tt}}},
\quad
\Upsilon(r)= 2 {\cal A}^{(0)}_0(r,k_\mu)\sqrt{-{g_{tt} \over g_{rr}}}.
\een

As mentioned earlier, the flow equation (\ref{flow}) is
valid for any value of momentum. This is a first
order differential equation and we need to specify one
boundary condition to solve this equation. That naturally
comes from the behavior of the equation at the horizon.
 Demanding the solution to be regular at the horizon,
 we get the following condition,
\be\label{bocon}
\bar \chi({k_{\mu}},r)^2\Bigg|_{r=1}= {\Sigma(r) \Upsilon(r) \over
\omega^2}\Bigg|_{r=1}.
\ee
For two derivative gravity this boundary condition implies
that\footnote{We choose the negative brunch. The sign
of the boundary condition in equation (\ref{bocon}) depends on the
choice of
coordinate. In our coordinate the boundary is at
$r\ra 0$ hence we need to choose the negative branch.},
\be\label{bocon2}
\bar \chi({k_{\mu}},1)=- \sqrt {{\Sigma(1) \Upsilon(1) \over
\omega^2}}=- \nt
\ee
which is independent of $k_{\mu}$. Therefore the full momentum
response at the
horizon corresponds to only to the zero momentum limit of
boundary response, $\bar\chi({k_{\mu}},1)=\bar\chi(k_{\mu}\ra 0,r \ra
0)$ \footnote{ However,
in higher derivative gravity we will see that the
$\bar\chi(k_{\mu},1)$ depends
on spatial momentum.}.

With this boundary condition, one can integrate out the
differential equation (\ref{flow}) from horizon to
asymptotic boundary and obtain the AdS/CFT response
for all momentum $k_{\mu}$. In particular, it is
trivial to see that at $(\omega,k_{i}) \rightarrow 0$ limit,
the flow is trivial $\partial_r \bar \chi(k_{\mu},r)=0$
and using the boundary condition in equation (\ref{bocon}) we get  the
first order
transport
coefficient of boundary fluid, i.e.
the shear viscosity coefficient coefficients turns out to be
$\eta=\nt$.

In this section, we will go away from $(\omega,k_i \rightarrow 0)$
limit. As we have already mentioned, it is possible to integrate
the flow equation for any momentum (perturbatively) and we can easily
find the higher order transport coefficients. The usual Kubo approach
to compute these coefficients requires the full profile of
the transverse graviton in black hole background background (solving a
second order
differential equation), where as, using the
flow equation, one can get these transport coefficients without
explicit knowledge of the
 graviton's profile.


\subsection{A renormalized response function}\label{renormalization}

When we solve the flow equation (\ref{flow}) to get the boundary
response function
in general it involves divergence at the boundary ($r\ra 0$). These
are usual $UV$ divergences and to remove them we need to re-normalize
the response function properly.

We follow the holographic renormalization prescription of
\cite{skende,Skenderis:2002wp}.
As the graviton is massless, we only need to add the following
counterterm
to the graviton's action,
\be\label{CT}
S_{C}= {1\over 16 \pi G_5} \int_{r=\delta} d^4 x \sqrt{-\gamma}
\frac{1}{4}\Phi(\epsilon, x)\Box\Phi(\epsilon, x).
\ee
In momentum space,
\be
S_{C}= {1\over 64 \pi G_5} \int_{r=\delta} {d^4 k\over (2\pi)^4}
\sqrt{-\gamma} \phi(\delta, k)(g^{tt}\omega^2+k_ik^i)\phi(\delta, -k).
\ee
Therefore the renormalized Green's function is given by,
\be
{\cal G}_{R}= \lim_{r \rightarrow 0}\ltb{\Pi(r,k_\mu)\over
\phi(r,k_\mu)}
+ \frac{\sqrt{-\gamma}}{32 \pi G_5} (g^{tt}\omega^2+k_ik^i)\rtb.
\ee

However we will study the flow of un-renormalized response function
defined
in equation (\ref{barchi}) and we define our renormalized response
function as,
\be
\bar{\chi}^{Ren}(r,k_{\mu}) =\bar{\chi}(r,k_{\mu})+\frac{1}{i \omega}
{ \sqrt{-\gamma} (g^{tt}\omega^2+k_ik^i)\over 32 \pi G_5}.
\ee
The counter term will cancel the UV divergences appearing
in the expression of $\bar{\chi}$ and we will get a finite result at
the boundary, i.e.
$ \lim_{r \rightarrow 0}\bar{\chi}^{Ren}(r,k_{\mu})$ will be finite.
From the above analysis, we understand that one can get rid of the UV
divergences appearing in the response function by following the
holographic renormalization technique. But, an important
observation is, this counter term does not add any finite
contribution to the result, it
only cancels out the divergences. Thus, one can study the flow
of the un-renormalized response function and
ignore the divergences piece to get the finite contribution
at the boundary.


\subsection{Second order transport coefficients from flow equation}

In this subsection, we compute the higher order transport coefficients
by
solving the flow equation (\ref{flow}) perturbatively up to
order $\omega^2$ and $k_i^2$. This is a non-linear first order
differential equation. Now, the right hand side of this equation is
proportional to $\omega$. Hence, to solve $\bar {\chi}$ to order
$\omega^2$, we can replace the leading order
 solution for $\bar{\chi}$ in the right hand side of equation
(\ref{flow}).
 This simplifies the situation a lot as the
 non-linear equation becomes linear.
 Now, to
leading order, $\bar{\chi}=-\eta=-{1 \over 16 \pi G_5}$.
Therefore up to order $\omega^2$, we get,
\be
\partial_r \bar \chi(k_\mu,r)= i \omega \sqrt{- {g_{rr} \over g_{tt}}}
\Bigg[{\eta^2 \over \Sigma(r)}- {\Upsilon(r) \over \omega^2}\Bigg] +
{\cal O}(\omega^2,k_i^2).
\ee
The integration constant for the equation can be fixed form the
boundary
condition in equation (\ref{bocon}). Putting the value of the
constant, the solution takes the form,
\ben\label{chi}
i \omega \bar \chi(k_{\mu},0)&=&\lim_{r\ra 0}-{1\over 96 \pi G_5
r}\bigg[3 q^2 (r-1)
+\omega  (3 \omega  +r (\omega  (\log (8)-3)+6 i))\bigg] + \higho
\nn
&=& -i\omega \lb \nt \rb
+ \omega^2 \ltb {1\over 2}(1-\ln 2) \lb \nt \rb  \rtb
 - {q^2 \over 2} \lb \nt \rb +{\cal O}({1\over r}).
\een
Here we have chosen the four momentum to be $k= \{\omega,0,0,q\}$.

This expression has divergence as $r\ra 0$ (UV divergence) and
can be removed by adding suitable counter term (as explained in the
last section).

Comparing the finite piece of equation (\ref{chi}) at $r\ra 0$ with
the
generic expansion of the retarded Green's function in
equation (\ref{Gdef})\footnote{The overall sign depends on
the choice of coordinate.},
we get,
\ben\label{htf}
\eta = {T^3 \pi^3\over 16 \pi G_5},\quad
\tau_{\pi}={2-\ln2 \over 2 \pi T},\ \ \ \
 \kappa = {\eta\over \pi T}\ .
\een
Here $T={1\over \pi}$.
These results are in agreement with \cite{causa1,causa2,causa3}.
Thus, we see that,
studying the flow equation of the response function we can
compute the higher order transport coefficients perturbatively.
Here, we present the results for the second order transport
coefficients, but,
in general it is possible to go beyond second order.

At this point, it is not clear why only considering the
boundary term from action in equation (\ref{ghdacnphi}) is enough to
get the correct results. In usual Kubo approach,
one needs to take into account the Gibbons-Hawking term also. But
as the action in equation (\ref{ghdacnphi}) has well defined
variational
principle, one does not need to add any Gibbons-Hawking term with it.
It has been explained in great details in \cite{nsflow} that the boundary terms coming
from the original action and the corresponding
Gibbons-Hawking action are exactly same as the boundary
terms coming from the action in equation (\ref{ghdacnphi}) up to terms
proportional to $\phi^2$ and pure divergence
terms. The $\phi^2$ terms do not contribute to any
transport coefficients\footnote{They only contribute to pressure of
the boundary theory.}.
The divergent terms will get
canceled by the proper counterterms and hence are
not important for finding the transport coefficients.
Thus it is clear that the effective action will give
us the correct transport coefficients for the boundary plasma. This
observation also holds
for higher derivative gravity theory\footnote{In \cite{nsflow}, it
has been proved explicitly
for $R^{(n)}$ gravity theory.}.

%

\subsection{Higher derivative correction to flow
equation}\label{modelhd}

So far we have discussed the flow equation of two point correlation
function of energy-momentum tensor of boundary theory whose gravity
dual is given by Einstein-Hilbert action (two derivative action).
But it is not obvious how to generalize this for higher derivative
case.  The construction in the previous section was based on the canonical form
of graviton's action given in equation (\ref{ghdacnphi}). AS discussed 
in section \ref{gendis}, in presence
of arbitrary higher
derivative terms in the bulk, the general action for the perturbation
$h_{xy}$
does not have the above form as equation (\ref{ghdacnphi}). Rather it
will have more than
two derivative (with respect to $r$) terms like $\phi'\phi''$,
$\phi''^2$ $etc$.
In presence of these terms it is not possible to bring this action
into a
canonical form (up to some total derivative terms). In this section we
consider generic higher derivatives terms in the bulk
Lagrangian.  We follow the prescription of section \ref{gendis1} to construct an
effective action
$"S_{\rm eff}"$ for transverse graviton in canonical form in presence
of
generic higher derivative terms in the bulk. The effective action and
original action
give same equation of motion perturbatively in the coupling of the
higher derivative terms.

Let us consider a gravity set-up with $n$ derivative action
in equation (\ref{gacnhd}). We write an effective action for
transverse graviton in canonical form,
\ben
S_{\rm eff}&=&\nt \int {d^4 k \over (2 \pi)^4} dr \bigg[ {\cal
A}_1^{\rm HD}(r,k)
 \phi'(r,k)\phi'(r,-k) + {\cal A}_0^{\rm HD}(r,k)
\phi(r,k)\phi(r,-k)\bigg]
\een
with some unknown function ${\cal A}_1^{HD}$ and ${\cal A}_0^{HD}$. We
fix these functions by
demanding that the equations of motion obtained from the effective
action and the original action
are same perturbatively in $\mu$ (see section \ref{gendis1}).

The generalized canonical momentum and equation of motion
are given by,
\ben
\Pi^{\rm HD}(r,k)&=& 2 {\cal A}_1^{\rm HD}(r,k) \phi'(r,k), \quad \lb
\Pi^{\rm HD}(r,k)\rb'= 2 {\cal A}_0^{\rm HD}(r,k) \phi(r,k).
\een
Once we find the effective action for the graviton,
we follow the procedure in the previous section to
obtain the flow equation for the boundary Green's
function in generic higher derivative gravity.

The boundary Green's function is given by,
\be
G_R^{\rm HD}(k_\mu)=\lim_{r \rightarrow 0}\frac{2{\cal A}_1^{\rm
HD}(r,k) \phi'(r,k)\phi(r,-k)}{\phi_0(k) \phi_0(-k)},
\ee
which can be written using the definition of canonical momentum as,
\be
G_R^{\rm HD}(k_\mu)=\lim_{r \rightarrow 0}{\Pi^{\rm HD}(r,k_\mu)\over
\phi(r,k_\mu)}.
\ee
Let us define a response function of the boundary theory in higher
derivative theory as,
\be
\bar \chi^{\rm HD}(k_\mu,r)= {\Pi^{\rm HD}(r,k_\mu)\over  i \omega
\phi(r,k_\mu)}.
\ee
Therefore the flow equation is given by,
\be\label{hdflow}
\partial_r \bar \chi^{\rm HD}(k_{\mu},r)= i \omega \sqrt{- {g_{rr}
\over g_{tt}}}
\Bigg[{\bar \chi^{\rm HD}(k_{\mu},r)^2 \over \Sigma^{\rm HD}(r,k)}-
{\Upsilon^{\rm HD}(r,k) \over \omega^2}\Bigg],
\ee
where we define
\ben
\Sigma^{\rm HD}(r,k)= - 2 {\cal A}_1^{\rm HD}(r,k_\mu)\sqrt{-{g_{rr}
\over g_{tt}}},\quad
\Upsilon^{\rm HD}(r,k)= 2 {\cal A}_0^{\rm HD}(r,k_\mu)\sqrt{-{g_{tt}
\over g_{rr}}}.
\een
This is the flow equation for two point correlation function of
energy-momentum tensor in presence of generic higher derivative
term in the bulk action. Therefore integrating this equation from
horizon to asymptotic boundary one can find the higher derivative
correction to the transport coefficients at any order in
frequency/momentum.

Like two derivative case here also we need to provide
a boundary condition to solve this equation. The response function
$\bar \chi^{\rm HD}(k_{\mu},r)$
should be well-defined at horizon. This implies,
\be
\bar \chi^{\rm HD}({k_{\mu}},r)\Bigg|_{r=r_h}= \sqrt {{\Sigma^{\rm
HD}(r) \Upsilon^{\rm HD}(r)} \over \omega^2}\Bigg|_{r=r_h}
\ee
here the horizon is located at $r=r_h$.

One important point to mention here is that unlike two derivative
gravity where $\bar \chi({k_{\mu}},r_h)$ was independent of $k_{\mu}$,
$\bar \chi^{\rm HD}(k_{\mu},r_h)$ can in general depend on $k_{\mu}$.
We will see this explicitly in the next section. Therefore the full
momentum response
at the horizon may not be able to correspond only to the zero momentum
limit of boundary response in higher derivative theory.

Here we have solved the equation (\ref{hdflow}) numerically and plotted the
function in fig. \ref{respnplot2}. We can see that for non-zero $k$ the
horizon value of real part of the response function is different than that
for zero momentum.

\lfig{Flow of response function for Higher derivative bulk action.}
{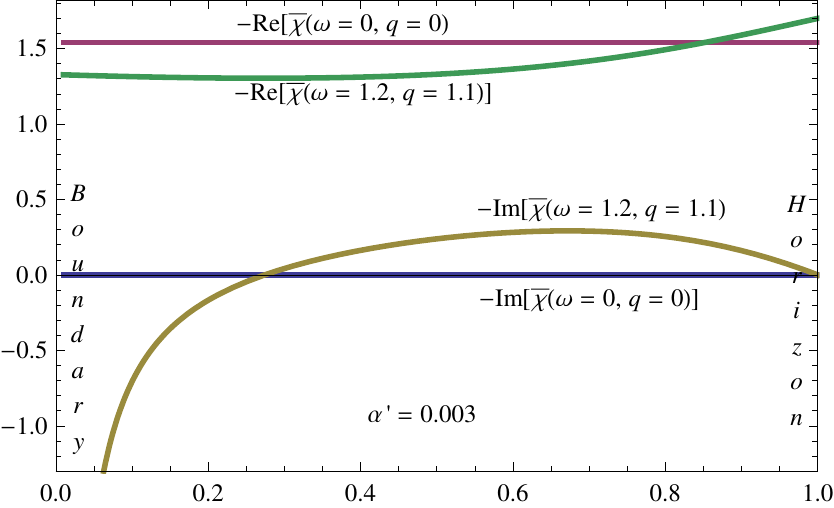}{6.5cm}{respnplot2}

Like two derivative case, the response function
in higher-derivative gravity theory also contains UV divergences. We
need
to add proper counter term following the holographic renormalization
procedure to cancel these divergences. A little more thinking also
says
that in presence of any higher-derivative term in the action the
structure
of the counterterm remains same as equation (\ref{CT}). Only the
overall normalization
 constant depends on higher-derivative coupling. Thus, similar
to the leading gravity, the counterterm in higher derivative
gravity also cancels out the divergence and does not add any finite
contribution to the boundary response function. One can study the flow
equation of the un-renormalized response function and read off the
transport coefficients from its finite piece.

After learning the generic techniques to compute the first and second 
order transport coefficients of strongly coupled boundary fluid  in 
section \ref{prob1} and \ref{prob2},
we would like to discuss some examples where we can apply these methods
to compute them. In the next section we plan to discuss some 
examples of higher derivative gravity, motivated from string theory.

\section{Examples of 2nd Order Transport Coefficients 
for Higher Derivative Theories} \label{r4sec2nd}

\subsection{Eight derivative correction}

In this section we  apply the effective action approach for eight
derivative terms in the Lagrangian. We  consider the well known
$Weyl^4$ term. This term appears in type II \index{string theory!type
II}string theory. Adding this term in the bulk action corresponds to
${1\over \lambda^{3/2}}$ correction in dual large $N$ theory. The five
dimensional bulk action is given by,
\begin{equation} \label{w4acn}
S=\nt \int d^5x \sqrt{-g} \lb R + 12 + \mu W^{(4)} \rb
\end{equation}
where,
\begin{equation}
 W^{(4)}=C^{hmnk}C_{pmnq}C_h^{\hhp rsp}C^q_{\hhp rsk}+{1\over
2}C^{hkmn}C_{pqmn}C_h^{\hhp rsp}C^q_{\hhp rsk}
\end{equation}
and the weyl tensors $C_{abcd}$ are given by,
\begin{equation}
C_{abcd}=R_{abcd} + {1 \over 3}
(g_{ad}R_{cb}+g_{bc}R_{ad}-g_{ac}R_{db}-g_{bd}R_{ca})+{1 \over
  12}(g_{ac}g_{bd}-g_{ad}g_{cb})R\ .
\end{equation}
The background metric is given by \cite{gks} (with horizon at
$r_0=1$),
\begin{eqnarray}\displaystyle\label{w4met}
ds^2 &=& -\frac {(1-r^2)}{r}  \left (1 + 45 \mu  r^6 - 75 \mu  r^4 -
     75 \mu
         r^2 \right) dt^2 \nonumber \\
&&+ {1\over 4 (1-r^2) r^2 } \lb 1-285 \mu  r^6 + 75 \mu  r^4 + 75 \mu
r^2
   \rb  dr^2 + {1\over r} d{\vec x}^2\ .
\end{eqnarray}
The temperature of this \index{black hole temperature}black hole is
given by $T= {1 \over \pi} \lb 1 + 15 \mu \rb \ .$

\subsection{ The General Action}

Putting the perturbed metric in equation (\ref{w4acn}) we get the
general  action for the scalar field $\ph$. As mentioned in section
\ref{gendis}, we see that in presence of higher derivative terms the
general action does not have canonical form. The action for $\phi$ is
given by,
\begin{eqnarray}\displaystyle
S&=&\nt \int {d^4 k \over (2 \pi)^4} dr \bigg [
A_1^W(r,k) \ph \cph + A^W_2(r,k) \php
\cphp \nonumber \\
&& \hspace{2cm}+ A_3^W(r,k) \phpp \cphpp + A^W_4(r,k) \ph \cphp
 \nonumber
\\
&&\hspace{2cm}+ A^W_5(r,k) \ph \cphpp + A^W_6(r,k)
\php \cphpp \bigg ] \ .
\end{eqnarray}
Expressions for $A_i^W$ can be found in \cite{ns1}.
Up to some total
derivative terms this action can be written as,
\begin{eqnarray}\displaystyle
S &=& \nt \int {d^4 k \over (2 \pi)^4} dr \bigg [ \cA_0^{W} \ph \cph
+ \cA_1^{W}\php \cphp + \cA_2^{W} \phpp \cphpp \bigg] \nonumber
\end{eqnarray}
where,
\begin{eqnarray}\displaystyle
\cA_0^{W} &=& A^W_1(r,k) -{A^{'W}_4(r,k) \over 2} + {A_5^{''W}(r,k)
\over 2}
\nonumber \\
\cA_1^{W} &=& A_2^W(r,k) - A_5^W(r,k) -{A^{'W}_6(r,k) \over 2}, \quad
\cA_2^{W} = A_3^W(r,k) \ .
\end{eqnarray}
With this form of action it is not possible to define the effective
coupling constant of transverse graviton or a response function
(following equations (\ref{effcoup}) and (\ref{barchi}) respectively).
Therefore we
find the effective action for transverse graviton $\phi$ in presence
of $W^4$ term in the action.

\subsection{ The Effective Action and Shear Viscosity}

Following the general discussion in section \ref{gendis1} we find
the effective action for graviton in presence of $W^4$ term. We write
the effective action for the scalar field in the following way,
\begin{eqnarray}\displaystyle
\label{effacnw4}
\seff^{W} &=& {(1+ \Gamma \mu) \over 16 \pi G_5} \intk  \bigg[
 (\cA^{(0)}_1(r,k) +
\mu \cB_1^{W}(r,k)) \cphp \php \nn
&& \hspace{2cm}+ (\cA^{(0)}_0(r,k)+\mu \cB_0^{W}(r,k)) \ph \cph
\bigg]\ .
\end{eqnarray}
The functions $\cB_0^W$ and $\cB_1^W$ are given by,
\begin{eqnarray}
\cB_0^W(r,k)=-\frac {\omega ^2 \left (663 r^6 - 573 r^4 + 75 r^2
     \right)} {4 r^2
       \left (r^2 - 1 \right)},\
\cB_1^W(r,k)=\frac {\left (r^2 - 1 \right) \left (129 r^6 + 141 r^4 -
75 r^2
\right)} {r} \ .
\end{eqnarray}
We set the normalization constant $\Gamma=0$ (see \cite{ns1} for
detailed discussion). Then the effective coupling constant is given by
equation (\ref{gkeff}),
\begin{eqnarray}\displaystyle
K_{{\rm eff}}(r) &=& \nt {\cA_1^{(0)}(r,k) + \mu \ \cB_1^W(r,k)
\over \sqrt{-g}
  g^{rr}}
=\nt \lb -{1\over 2} \lb 1 + 36 \mu \ r^4 (6 -r^2) \rb \rb \ .
\end{eqnarray}
Therefore the shear viscosity is given by,
\begin{eqnarray}\displaystyle\label{w4eta}
\eta &=&  r_0^{-\frac{3}{2}} (-2 K_{{\rm eff}}(r_0))
=\nt \lb 1 + 180 \ \mu \rb, \ \ \ \ (r_0=1) \
\end{eqnarray}
and shear viscosity to
entropy density ratio
\begin{equation}
{\eta \over s}={1\over 4 \pi }\lb 1 + 120 \ \mu \rb
\end{equation}
where entropy density $s$ is given by $s={1\over 4G_5}
\lb 1 + 60 \ \mu \rb$ \cite{gks}.
These results agree with the one in the literature.

\subsection{String theory correction to flow equation}

Now let us compute string theory corrections to the second order
transport coefficients following the general discussion given in
section \ref{modelhd}.

From the effective action for transverse graviton, computed in
equation (\ref{effacnw4}), the flow equation is given by,
\be\label{W4flow}
\partial_r \bar \chi^{W^4}(k_{\mu},r)= i \omega \sqrt{- {g_{rr} \over
g_{tt}}}
\Bigg[{\bar \chi^{W^4}(k_{\mu},r)^2 \over \Sigma^{W^4}(r,k)}-
{\Upsilon^{W^4}(r,k) \over \omega^2}\Bigg],
\ee
where,
\ben
\Sigma^{W^4}(r,k)&=& - 2 {\cal A}_1^{W^4}(r,k_\mu)\sqrt{-{g_{rr} \over
g_{tt}}}, \ \
\Upsilon^{W^4}(r,k)= 2 {\cal A}_0^{W^4}(r,k_\mu)\sqrt{-{g_{tt} \over
g_{rr}}}
\een
where,
\ben
{\cal A}_1^{W^4}={\cal A}_1^{0}+\mu {\cal B}_1^W,\quad {\cal
A}_0^{W^4}={\cal A}_0^{0}+\mu {\cal B}_0^W\ .
\een
The explicit expressions for $\Sigma$ and $\Upsilon$ can be obtained
by using ${\cal A}_0^{W^4}$ and ${\cal A}_1^{W^4}$.
From the regularity of $\bar \chi^{W^4}(k_{\mu},r)$ at horizon we get,
\ben\label{bcW4}
\bar \chi^{W^4}(k_{\mu},1) &=& {\sqrt {\Sigma^{W^4}(1)
\Upsilon^{W^4}(1) \over \omega^2}}
= {r_0^3\over 16 \pi G_5} +  {\mu  r_0\over 4 \pi G_5} \left(45
r_0^2+11 q^2\right)\ .
\een
Here, we see that unlike the two-derivative gravity, the horizon value
of the response function depends on spatial momenta $q$.
With this boundary condition we solve the flow equation up to order
$\omega^2$ and $q^2$ (ignoring ${\cal O}(\omega q^2)$ term). Here we
write the final
result\footnote{$k=\{\omega,0,0,q\}$ and we ignore the UV divergence
piece.}.
\ben\label{chiw4}
i \omega \ \bar \chi^{W^4}(k_{\mu},0)&=& -i (1+180 \mu ) {r_0^3
\over 16 \pi G_5} \omega
 + \bigg[ \frac{1}{2} (1-\log
   (2))  +\frac{5}{4} \mu  (199-66 \log (2))\bigg] {r_0^2\over
16 \pi G_5} \omega^2\nn
&&\qquad  - {1 \over 2}(1+20 \mu)
{r_0^2\over 16 \pi G_5} q^2 + \higho \ .
\een
Comparing this result with equation (\ref{Gdef}) we get
\ben
{\eta \over \pi^3 T^3} &=& 1+ 135 \mu+ {\cal O}(\mu^2),\quad
\kappa = {\eta \over \pi T}\lb 1- 145 \mu\rb + {\cal
O}(\mu^2)\nonumber \\
\tau_{\pi} T &=& \frac{2-\log (2)}{2 \pi } + \frac{375 \mu }{4 \pi
}+ {\cal O}(\mu^2)\ .
\een
These results are in agreement with \cite{buchel-paulos1} who applied
usual Kubo formula to obtain these results. The agreement provides
a non-trivial check to this approach of obtaining higher order
transport coefficients from the flow equation (\ref{flow}).

\subsection{Four derivative correction}

Next, we will concentrate on the generic four derivative
corrections to Einstein-Hilbert action. These terms arise in the
effective
action for the heterotic string theory. In fact, the complete
super-symmetrized
$R^2$ correction to effective Heterotic string theory is known and one
way to
obtain it is the super-symmetrization of the Lorentz Chern-Simons
terms \cite{Berg,panda}.
This terms also arises in the context of Type IIB string theory
\cite{Kats,Blau},
where the theory is on $AdS_5 \times X^5$, the compact space $X^5$
being $S^5/Z_2$.
The dual theory is ${\cal N}=2 Sp( N)$gauge theory with 4 fundamental
and 1 antisymmetric
traceless hyper-multiplets. This super-conformal theories arises in
the context
of $N$D3-branes sitting inside 8 D7-branes coincident on an
orientifold 7-plane.
In this case, generic four derivative $R^2$ correction comes form the
DBI action of the branes.

Here we compute the generic four derivative correction to the second
order transport
coefficients, the relaxation time $\tau_{\pi}$ and $\kappa$. We can
choose
the coefficients of the higher derivative terms to be the
four-dimensional Euler density
and get pure Gauss-Bonnet correction to these coefficients.

The action
\ben \label{6dacnreqg}
{\cal I} &=& \nt \int d^5x \sqrt{-g} \bigg [ R +12 +
 \app  \bigg (  \beta_1 R^2  + \beta_2
R_{\mu\nu\rho\sigma} R^{\mu\nu\rho\sigma} + \beta_3
R_{\mu\nu}R^{\mu\nu}  \bigg ) \bigg].
\een

In particular for Gauss-Bonnet correction,
$\beta_1=1, \beta_2=1, \beta_3=-4.$ One can get rid
of the $Ricci^2$ and $Scalar^2$ terms by a field
redefinition and therefore all physical quantities
should depend on the coefficient $\beta_2$ only. Here,
we prefer to work with the generic case as it would be
easier for us the get the results for pure Gauss-Bonnet
combination at every step.

The background solution is given by \cite{BD2},
\be \label{6derimet}
ds^2 =  f(r) dt^2 + {g(r)\over 4 r^3} dr^2 + {1\over r} d\vec{x}^2
\ee
where $f(r)$ and $g(r)$ are given by,
\ben
f(r)&=&r - \frac{1}{r} -
 2 r\big (r^2 - 1 \big)\beta_2 \app
\een
and
\ben
g(r) &=&  \frac {r} {1- r^2} + \frac {2 r \big (
      10 \beta_1 + (1- 3 r^2) \beta_2 +
      2 \beta_3 \big) \app} { 3(r^2 - 1)}
      \ .
\een
This is the background metric corrected up to order $\app$.
We have fixed the integration constant such that the
boundary metric is Minkowskian and the
horizon is located at $r=1$. The temperature of the
black brane is given by,
\ben
T&=& {1 \over \pi} +{ 10 \beta_1-5 \beta_2+ 2 \beta_3 \over 3 \pi}
\app.
\een

Similar to the $Weyl^4$ case, we can write the following effective
action for this model,
\ben
S_{\rm eff} &=& \nt \int {d^4 k \over (2 \pi)^4} dr \bigg[ {\cal
A}_1^{GB}(r,k)
 \phi'(r,k)\phi'(r,-k)  + {\cal A}_0^{GB}(r,k)
\phi(r,k)\phi(r,-k)\bigg]
\een
where, ${\cal A}_1^{GB}$ and ${\cal A}_0^{GB}$ are given in appendix
\ref{gbapp}.
 Now, it is straightforward to write the corresponding flow equation
(\ref{flow}) in this case,
\be\label{GBflow}
\partial_r \bar \chi^{\rm GB}(k_{\mu},r)= i \omega \sqrt{- {g_{rr}
\over g_{tt}}}
\Bigg[{\bar \chi^{\rm GB}(k_{\mu},r)^2 \over \Sigma^{\rm GB}(r,k)}-
{\Upsilon^{\rm GB}(r,k) \over \omega^2}\Bigg],
\ee
where we define
\ben
\Sigma^{\rm GB}(r,k)&=& - 2 {\cal A}_1^{\rm GB}(r,k_\mu)\sqrt{-{g_{rr}
\over g_{tt}}}\\
\Upsilon^{\rm GB}(r,k)&=& 2 {\cal A}_0^{\rm GB}(r,k_\mu)\sqrt{-{g_{tt}
\over g_{rr}}}.
\een
Now, the boundary condition (\ref{bocon}) takes the following form,
\be\label{GBbc}
\bar \chi^{GB}(k_{\mu},1)=\nt \ltb 1+\left(\left(q^2-8\right) \beta
_3-40 \beta _1\right) \alpha '\rtb\ .
\ee
As mentioned earlier, we see that even in this case, the boundary
condition
depends on spatial momenta $q$ through the coefficient $\beta_3$.
With this boundary condition, one can solve the flow equation
(\ref{GBflow}) and the solution is given by,
\ben\label{chigb}
i \omega \ \bar \chi^{GB}(k_{\mu},0)&=& \nt\bigg[-i(1-  \left(40 \beta
_1+8 \beta _3\right) \alpha ')\omega +\frac{\omega^2}{2}
   \bigg[(1-\log2)
 +\frac{\app}{6}(130 \beta _1 (\log2-1)\nn
&&-\beta _2 (5
   \log2-2)
   +26 \beta _3 (\log2-1))\bigg]-\frac{q^2}{2}\bigg[1 -
\frac{1}{3} (130 \beta _1+25 \beta _2+26 \beta _3)
   \alpha'\bigg]\bigg]\nn
   && + \higho \ .
\een
From this expression we get the following transport coefficients,
\be
\eta= \nt \lb 1-  8\left(5 \beta _1+ \beta _3\right) \alpha '\rb +
{\cal O}(\app^2).
\ee
This matches with results in \cite{Kats,sd,myers-shenker}. Now, we find 
the higher order coefficients,
\ben\label{rgbg}
\kappa &=& {\eta \over \pi T}\lb 1- 10 \beta_2 \app\rb + {\cal
O}(\app^2)\nn
\tau_{\pi} T &=& {2-\ln2 \over 2 \pi}- {11 \beta_2 \over 2\pi}\app +
{\cal O}(\app^2) .
\een

As we can see, the physical quantities $\eta/s, \kappa,\tau_{\pi} T$
only
depend on the coefficient $\beta_2$. In particular to Gauss-Bonnet
combination, the corrections are,
\ben\label{GBTC}
\kappa &=& {\eta \over \pi T}\lb 1- 10 \app\rb + {\cal O}(\app^2)\nn
\\
\tau_{\pi} T &=& {2-\ln2 \over 2 \pi}- {11 \over 2 \pi}\app+ {\cal
O}(\app^2).
\een
 Thus, we show that studying the flow of the response function 
(constructed from the effective action), we can find out all higher 
order transport coefficients systematically.

\subsubsection{Exact result for Gauss-Bonnet black hole}

As we have done the above computation perturbatively, the above
 expressions are valid only at order $\alpha'$. But, one can
  consider the Gauss-Bonnet term exactly in coupling. For pure
  Gauss-Bonnet combination the equations of motion remain second order
  differential equation and hence it is easy to solve exactly to find
  the background space-time.
  We solve the flow equation in this background exactly in coupling
constant,
  and find the exact expressions
  for relaxation time $\tau_{\pi}$ and $\kappa$. In this section we
briefly
  outline the result.

  The action and the solution is given by,
\ben
{\cal I}_{GB}&=&\nt\int d^5x \sqrt{-g}\bigg [ R +12 +
{\lambda_{gb}\over 2}  \bigg (R^2 +
R_{\mu\nu\rho\sigma} R^{\mu\nu\rho\sigma}  -4 R_{\mu\nu}R^{\mu\nu}
\bigg ) \bigg]\nn
ds^2&=& r^2\lb -{f(r)\over f_{\infty}} dt^2 + d\vec x^2\rb + {dr^2
\over r^2 f(r)}
\een
where,
\be
f_{\pm}(r) = {1\over 2 \lambda_{gb}}\ltb 1 \pm  \sqrt{1-4\lambda_{gb}\lb
1- {r_0^4\over r^4}\rb}\rtb
\ee
and
\be
f_{\infty}=\lim_{r\ra \infty} f(r) = {1-\sqrt{1-4 \lambda_{gb}}\over 2
\lambda_{gb}}\ .
\ee

In this coordinate the boundary metric is $\eta$. We also consider
only the $'-'$ branch of $f_{\pm}$ which
corresponds to a non-singular black hole solution with non-degenerate
horizon.

The black hole temperature is given by,
\be
T={r_0\over \pi f_{\infty}}\ .
\ee

With exact GB, the effective action for fluctuation has a canonical
form.
Therefore we derive the flow equation for the response function (as we
did
in section \ref{prob2}) and solving this equation we get,
\ben\label{exact}
\eta &=& \nt (1- 4 \lambda_{gb}), \ \ \
\kappa = \frac{2 \lambda _{gb} \left(8 \lambda
   _{gb}-1\right)}{\left(1-\sqrt{1-4 \lambda
   _{gb}}\right) \left(4 \lambda _{gb}-1\right)}\nn
\tau_{\pi}T &=& {1\over 4\pi (-1 + 4 \lambda_{gb})}\bigg[-8
\lambda_{gb}^2+12 \sqrt{1-4 \lambda_{gb}}
   \lambda_{gb}+10 \lambda_{gb}-2 \sqrt{1-4
   \lambda_{gb}}
    -4 \log (2) \lambda _{gb}\nn
    && +\left(1-4
   \lambda_{gb}\right) \log \left(-4 \lambda_{gb}+\sqrt{1-4
\lambda_{gb}}+1\right)+\left(4
   \lambda_{gb}-1\right) \log \left(1-4 \lambda_{gb}\right)-2+\log
(2)\bigg].\nn
\een
One can easily check that up to first order in $\lambda_{gb}$, the
results in
(\ref{exact}) reduces to the one in (\ref{GBTC}). In \cite{myers1} the
authors
obtained the relation between second order transport coefficients and
$\lambda_{gb}$
numerically, however we are able to present the result exactly.\\

\noindent
{\underline {\bf Violation of bound on shear viscosity to entropy density ratio}}\\

In our units the entropy density $s$ turns out to be $s = \frac{1}{4G_5}$ and
hence shear viscosity to entropy density ratio turns out to be
\be
\frac{\eta}{s}= \frac{1}{4\pi}(1-4\lambda_{gb}),
\ee
which violates the famous $KSS$ bound for $\lambda_{gb} >0$ \cite{myers-shenker,Kats}.
In \cite{kss} the
authors conjectured a
bound on shear viscosity to entropy density ratio for gauge theory plasmas which
have a holographic dual, i.e.,
\be
\frac{\eta}{s} \geq \frac{1}{4\pi}.
\ee
In \cite{myers-shenker} the authors argue that when $\lambda_{gb} > \frac{9}{100}$
the theory violates microcausality and is inconsistent. Therefore,
for (3+1)-dimensional CFT duals of (4+1)-dimensional
Gauss-Bonnet gravity, consistency of the theory requires
\be
\frac{\eta}{s} \geq \frac{1}{4\pi}\cdot {\frac{16}{25}}.
\ee

As we have mentioned in introduction that the flow equation is a
first order non-linear differential equation but one can reduce
this equation to a second order linear differential equation. This
second order differential equation is related to the equation of
motion for transverse graviton (gauge invariant excitations).
Therefore
we can use this equation to study causality violation in Gauss-Bonnet
gravity. In \cite{myers1,myers2} it was found that to preserve
causality of a conformal
fluid there exists a bound on second order transport coefficients,
\be
\tau_{\pi}T-2 {\eta\over s}\geq 0\ .
\ee
In Fig \ref{fig3} we plot $\tau_{\pi}T-2 {\eta\over s}$ for our result
and find the following bound on $\lambda_{gb}$
which is in agreement with \cite{myers1}.
\be
-0.711\leq \lambda_{gb}\leq 0.113\ .
\ee

\lfig{Bound on $\lambda_{gb}$.}
{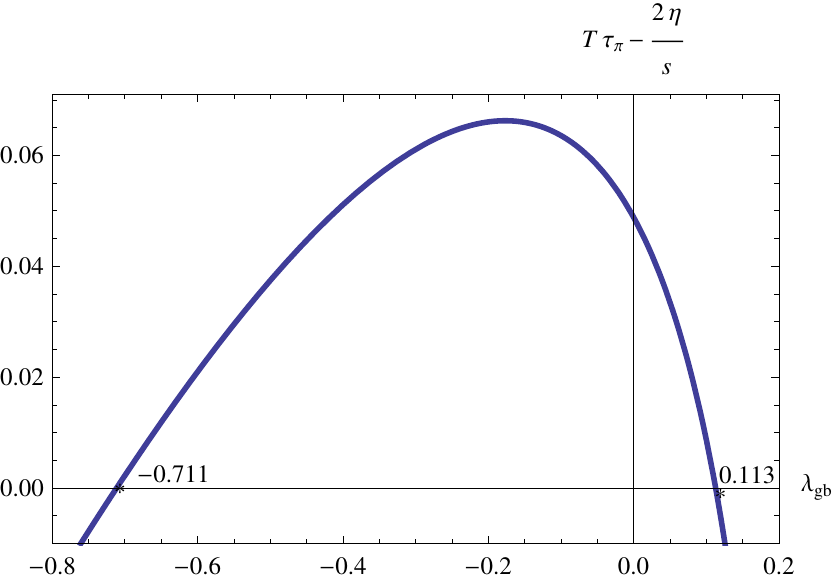}{6.5cm}{fig3}


\section{Flow equation for charged black holes}\label{cbh}

Electrically charged black holes in five dimensions have drawn a lot
of interest in the
context of AdS/CFT. The electric charge of these black holes are
mapped to the global
R-charge of the dual field theory. Because of the presence of the
electric charges, the
thermodynamics and the phase structure of these black holes are rather
complicated and also
interesting at the same time. There have been a lot of study of
thermodynamics and phase
transitions of these charged black hole with different horizon
topologies (see \cite{Cvetic} and
references therein).

The goal of the present section is to apply the AdS/CFT correspondence
to understand
how non-vanishing chemical potentials effect the hydrodynamic behavior
of strongly coupled gauge
theories.
We study the second order hydrodynamics in two cases:
(a) Generic R-charge black holes and (b) Charge black holes in
higher-derivative gravity.

\subsection{R-charged black holes}

We consider a conformal field theory with conserved charge (density)
in addition to energy and
momentum. This is especially an interesting extension of the
hydrodynamics of the uncharged fluids.

The second order hydrodynamics of charged fluid has been studied in
\cite{chargehydro,haack}.
They consider Reissner-Nordstrom black hole in five dimensions and
found the effect of chemical
potential on second order transport coefficients in some limits of
chemical potential. One important outcome of their analysis was to find
a new non-dissipative contribution to the charge current. However, we consider
generic parity preserving $R$-charged black holes with three (unequal) 
charges (chemical potentials) and find
the exact expressions for second order transport coefficients in
presence of three chemical potentials. As we
have mentioned in the introduction that solving the flow equation (of
retarded Green's function
of energy momentum tensor) we can only find two second
order transport coefficients whereas in \cite{chargehydro, haack} all
other second order transport
coefficients have been reported.


We consider R-charged black holes in five dimensions. A consistent
truncation of ${\cal N}=8$, $D = 5$ gauged supergravity
with $SO(6)$ Yang-Mills gauge group, which can be obtained by $S^5$
reduction of type $IIB$
supergravity, gives rise to ${\cal N} = 2$, $D = 5$ gauge supergravity
with $U(1)^3$ gauge group. The
same theory can also be obtained by compactifying eleven dimensional
supergravity, low
energy theory of $M$ theory, on a Calabi-Yau three folds. The bosonic
part of the action of
${\cal N} = 2$, $D = 5$ gauged supergravity is given by \cite{Cvetic}.
We follow the notation of \cite{son5}.
\ben
{\cal I}_{\rm sugra} &=& \nt \int d^5x \sqrt{-g} \bigg[R +V(X)-
{1 \over 2}G_{IJ}(X) F^I_{\mu\nu}F^{\mu\nu J}-G_{IJ}(X)
\partial_{\mu} X_I \partial^{\mu}X_J \bigg]  \nonumber \\
&& \quad +{{\zeta\over 3} \over 16\pi G_5} \int  d^5x\
\epsilon^{\mu\nu\rho\sigma\gamma}A_{\mu} F_{\nu\rho} F_{\sigma\gamma}
\een
where, $X^I$'s are three real scalar fields, subject to the constraint
$X^1 X^2 X^3 =1$. $F^I$'s, which are field strengths of three Abelian
gauge
fields (I,J=1,2,3), and the scalar potential $V(X)$ is given by,
\ben
F_{\mu\nu}^I &=& 2 \partial_{[\mu}A^I_{\nu]}, \ \ \
G_{IJ}={1\over 2} \rm{diag} \ltb
(X^1)^{-2},(X^2)^{-2},(X^3)^{-2}\rtb, \ \ \
V(X)=2\sum_I{1\over X_I}
\een

The three-charge non-extremal STU solution is specified by
the following background values of the metric
\ben
ds^2 &=& - {\cal H}^{-2/3}\, f_k \, dt^2
+ {\cal H}^{1/3} \left( f_k^{-1} dr^2 + r^2 d
\Omega_{3,k}^2\right)\,,
\label{metric_k}
\een
\ben
f_k = k - {m_k\over r^2} + {r^2}{\cal H} \,, \qquad
 H_i = 1 + {q_i\over r^2}\,, \qquad
  {\cal H} = H_1 H_2 H_3 \,,
\een
 as well as the scalar and the gauge fields
\begin{equation}
X^i = {{\cal H}^{1/3}\over H_i} \,,  \qquad
A^i_t = \sqrt{ {k q_i + m_k\over q_i}} \left( 1 - H_i^{-1}\right)\,.
\end{equation}
The parameter $k$ determines the spatial curvature of  $d
\Omega_{3,k}^2$:
$k=1$ corresponds to the metric on the three-sphere of unit radius,
$k=0$ - to the metric on $R^3$.
as the
Hydrodynamic approximation is valid only in the
case of a translational-invariant horizon, in our case we set
$k=0$
and
$$
d \Omega_{3,0}^2 \rightarrow  \left( dx^2 + dy^2 + dz^2\right)\,.
$$
Replacing the radial coordinate $r \ra r_0/\sqrt{r}$,
where $r_0$ is the largest root of the equation $f(r)=0$,
the background solution in this new coordinate is given by,
\ben
ds^2_5 &=& - {\cal H}^{-2/3}{(\pi T_0)^2 \over r}\,f \, dt^2 + {\cal
H}^{1/3}{1 \over 4 f r^2} dr^2+  {\cal H}^{1/3}{(\pi T_0 )^2
\over r}\, \left( dx^2 + dy^2 + dz^2\right)
\label{metric_u_3}
\een
\ben
f(r) &=& {\cal H} (r) - r^2 \prod\limits_{i=1}^3 (1+\kappa_i)\,,
\;\;\;\;\; H_i = 1 + \kappa_i r \,,\hspace{1cm} \kappa_i \equiv
{q_i\over r_0^2}\, .
\label{identif}
\een
where $\kappa_i's$ are chemical potentials and
\be
 T_0 = r_0/\pi\,.
\ee
The scalar fields and the
gauge fields are given by
\begin{equation}
X^i = {{\cal H}^{1/3}\over H_i(u)} \,, \qquad
A^i_t = {\tilde{\kappa}_i \sqrt{2} u\over L H_i(u)}
\label{scal_gauge_u_3}
\end{equation}
where,
\be
\tilde{\kappa}_i = {\sqrt{q_i}}
 \prod\limits_{i=1}^3 (1+\kappa_i)^{1/2}
\,.
\ee
The Hawking temperature of the background
(\ref{metric_u_3}) is given by
\begin{equation}
T_H =
{2 + \kappa_1 + \kappa_2 + \kappa_3 - \kappa_1 \kappa_2  \kappa_3\over
2\sqrt{(1+\kappa_1)(1+\kappa_2) (1+\kappa_3)}}\, T_0\,.
\end{equation}

We perturb the $xy$ component of background metric and the action
for transverse graviton is given by,
\ben
S_{\rm eff} &=& \nt \int {d^4 k \over (2 \pi)^4} dr \bigg[ {\cal
A}_1^{Q}(r,k)
 \phi'(r,k)\phi'(r,-k) + {\cal A}_0^{Q}(r,k)
\phi(r,k)\phi(r,-k)\bigg]
\een
where,
\be
{\cal A}_1^{\rm Q}= -{r_0^4 f(r) \over r}
\ee
and
\ben
{\cal A}_0^{\rm Q}   &=& {r_0^2\over 4r^2} \lb {H_1 H_2 H_3\over f(r)}
-q^2\rb\ .
\een

Therefore the flow equation is given by,
\be\label{Rflow}
\partial_r \bar \chi^{\rm Q}(k_{\mu},r)= i \omega \sqrt{- {g_{rr}
\over g_{tt}}}
\Bigg[{\bar \chi^{\rm Q}(k_{\mu},r)^2 \over \Sigma^{\rm Q}(r,k)}-
{\Upsilon^{\rm Q}(r,k) \over \omega^2}\Bigg].
\ee

Solving this equation perturbatively in $\omega$ and $q$ we get
\ben
\bar \chi^{\rm
Q}(k_{\mu},r)&=&-\frac{r_0^3\prod_i(1+\kappa_i)^{1/2}}{16\pi
G_5}+{i r_0^2\over 2 \omega}\frac{(q^2-\omega^2)}{16\pi
G_5}\lb1-{1\over r}\rb+{i \omega r_0^2 \prod_i(1+\kappa_i) \over
16\pi G_5\sqrt{4 P_{\kappa} +(1+S_{\kappa})^2}}\nn
&&\bigg({1\over 2} \ln \ltb \frac{1+ S_{\kappa} -2 P_{\kappa}-\sqrt{4
P_{\kappa} +(1+S_{\kappa})^2}}{1+ S_{\kappa} -2 r S_{\kappa}-\sqrt{4
P_{\kappa} +(1+S_{\kappa})^2}}\rtb\nn
&&+{1\over 2} \ln \ltb \frac{1+ S_{\kappa} -2 r P_{\kappa}+\sqrt{4
P_{\kappa} +(1+S_{\kappa})^2}}{1+ S_{\kappa} -2 S_{\kappa}+\sqrt{4
P_{\kappa} +(1+S_{\kappa})^2}}\rtb\bigg) + {\cal O}(q\omega^2,
\omega q^2, q^3, \omega^3)
\een
where,
\ben
S_{\kappa}&=&\sum_i \kappa_i, \ \ \
P_{\kappa}=\prod_i\kappa_i.
\een

Computing the response function at the boundary (throwing away the
divergent piece)
 we get the following transport coefficients,
\ben
\eta &=& \frac{r_0^3}{16\pi G_5}\prod_i(1+\kappa_i)^{1/2}
\een
and
\ben
\kappa &=& {\eta\over \pi T} {1 +S_{\kappa}/2-P_{\kappa}/2 \over
\prod_i (1+\kappa_i)}\nn
\tau_{\pi}T&=& {2 +S_{\kappa}-P_{\kappa}\over 4 \pi
\prod_i(1+\kappa_i)} \bigg[ 2- {\prod_i(1+\kappa_i)\over \sqrt{4
P_{\kappa}+(1+S_{\kappa})^2}} \ln\lb {3+S_{\kappa} + \sqrt{4
P_{\kappa}+(1+S_{\kappa})^2}}\over 3+S_{\kappa} - \sqrt{4
P_{\kappa}+(1+S_{\kappa})^2}\rb\bigg].
\een
These are the new results in this paper. It is easy to check that for
$\kappa_i \ra 0$ limit
we recover the results in section \ref{prob2}.

To complete the discussion on the second order transport coefficients
for
$R$-charged black holes one should find the flow of Green's functions
for two
point correlation functions of $R$-currents. As we will mention in
section \ref{rcflow}
that in presence of finite charges (or chemical potentials) it is very
hard to
solve the $Riccati$ equation even perturbatively in $\omega$ and $q$.
We find
it very difficult to get any analytic solution for $R$-current Green's
function.
However we consider a simple model in section \ref{rcflow} and study
the flow of
$R$-current Green's function numerically.

\subsection{Charged black holes in higher derivative gravity}

In this section, we will study five-dimensional gravity in presence of a negative
cosmological constant and coupled to $U(1)$ gauge field. The model has been studied
in \cite{mps,Maeda,sl}, the action is given as,
\ben\label{hdcbacn}
S&=&\frac{1}{16 \pi G_5} \int d^5x \sqrt{-g}\bigg [R +12 -\frac{1}{4}F^2  +
\frac{\zeta}{3} \epsilon^{abcde}A_{a}F_{bc}F_{de} + \alpha' \bigg( c_1 R_{abcd}R^{abcd} \nn
&& \,\,\,\,\,\,\,\,\,\,\,\,\ + c_2 R_{abcd}F^{ab}F^{cd} +c_3(F^2)^2+c_4 F^4
+ c_5 \epsilon^{abcde}A_{a}R_{bcfg}R_{de}^{fg}\bigg) \bigg].
\een
Here, $F^2= F_{ab}F^{ab}, F^4=F_{ab}F^{bc}F_{cd}F^{da}$, and the AdS radius is set to unity.
The action includes the Chern-Simon term and also a generic set of four derivative terms.
All the four derivative terms will be treated perturbatively in our computation and here
$\alpha' << 1$ is the perturbation parameter. In \cite{mps}, it was shown that, within
perturbative approach, after using field-redefinition, this is the most generic four
derivative action that one can write down. In this section, we will closely follow their
work. The background metric and the gauge field in presence of these higher derivative
terms have the following form,
\ben\label{chargehdsol}
ds^2= - r^2 f(r) dt^2 + \frac{1}{r^2 g(r)}dr^2 + r^2 (dx^2+ dy^2+dz^2),\ \ \
A = h(r) dt,
\een
where,
\ben\label{hds}
f(r)&=&f_0(r)(1+ \alpha' F(r)), \ \ \ \ \
g(r)= f_0(r)(1+\alpha' (F(r)+ G(r))), \nn
h(r)&=&h_0(r)+\alpha'H(r).
\een
Here $f_0(r),g_0(r)$ and $h_0(r)$ are the solution of the background in absence of
the higher-derivative terms in the action and they are given as,
\ben
f_0=g_0=\bigg(1-\frac{r_0^2}{r^2}\bigg)\bigg(1+\frac{r_0^2}{r^2}
-\frac{Q^2}{r_0^2 r^4} \bigg), \ \
h_0={\sqrt{3}Q}\bigg(\frac{1}{r_0^2}-\frac{1}{r^2}\bigg).
\een
Here, $Q$ is related to the physical charge of the system and $r_0$ is the
 position of the horizon. From (\ref{hds}), it is clear that even in presence
 of the higher-derivative terms, the horizon remains at $r_0$. The higher-derivative
 corrections to this background are given by the functions $F(r),G(r)$ and $ H(r)$.
 The form of these functions are given in \cite{mps}. We would not write those
 expressions and refer the reader to that paper.

Using the flow equation,
 we will study the higher order transport coefficient of the plasma theory dual to
 this gravity model. For this, we will write the effective action for the metric
 fluctuation in (\ref{petmet}), as we have done in previous sections,
\ben
S_{\rm eff} &=& \nt \int {d^4 k \over (2 \pi)^4} dr \bigg[ {\cal A}_1^{CB}(r,k)
 \phi'(r,k)\phi'(r,-k)+ {\cal A}_0^{CB}(r,k) \phi(r,k)\phi(r,-k)\bigg]
\een
where, ${\cal A}_1^{CB}$ and ${\cal A}_0^{CB}$ are given in appendix \ref{hdcb}. The
corresponding the flow equation (\ref{flow}) for this case with the coefficients
${\cal A}_1^{CB}$ and ${\cal A}_0^{CB}$ is,
\be\label{CBflow}
\partial_r \bar \chi^{\rm CB}(k_{\mu},r)= i \omega \sqrt{- {g_{rr} \over g_{tt}}}
\Bigg[{\bar \chi^{\rm CB}(k_{\mu},r)^2 \over \Sigma^{\rm CB}(r,k)}- {\Upsilon^{\rm CB}(r,k) \over \omega^2}\Bigg],
\ee
where we define
\ben
\Sigma^{\rm CB}(r,k)&=& - 2 {\cal A}_1^{\rm CB}(r,k_\mu)\sqrt{-{g_{rr} \over g_{tt}}}, \ \ \
\Upsilon^{\rm CB}(r,k)= 2 {\cal A}_0^{\rm CB}(r,k_\mu)\sqrt{-{g_{tt} \over g_{rr}}}.
\een

We solve this flow equation to find the effect of higher derivative terms and chemical
potential (or charge)
on transport coefficient. Here, we present our result for small $Q$ only, though it is possible
to find the results for any $Q$.

The boundary condition (\ref{bocon}) will take the following form\footnote{Note that we are
working in a different coordinate where $r\ra \infty$ is the boundary, therefore we choose
the positive branch of the boundary condition (\ref{bocon})},
\be
\bar \chi(k_\mu,r_0)=\frac{r_0^3}{16 \pi G_5}\bigg[1- \alpha'\frac{24 c_1 Q^2}{r_0^6}\bigg].
\ee
In this case, the horizon value of the response function is independent of
momenta. With this boundary condition, we can solve the flow equation
(\ref{CBflow}), and the solution is given by,
\ben
i \omega \bar \chi(k_\mu,\infty)&=& i \omega \frac{r_0^3}{16
\pi G_5}\bigg[1- \alpha'\frac{24 c_1 Q^2}{r_0^6}\bigg]
- \omega^2\frac{r_0^2}{32 \pi G_5} \bigg [(1-\ln2) -\frac{Q^2}{2 r_0^6}(3-\ln2)\nn
&& + \frac{\alpha'}{6}\bigg(c_1(2-5 \ln2)
-\frac{Q^2}{2 r_0^6}(c_1(35 -58 \ln2 - 48 c_2(2-\ln2))\bigg)\bigg] \nn
&&+\frac{q^2 r_0^2}{32 \pi G_5}\bigg[1-
\frac{\alpha'}{3}\bigg(25 c_1 +\frac{Q^2}{r_0^6}(32 c_1+24 c_2)\bigg)\bigg]
+ \higho + {\cal O}(Q^4)\ .\nn
\een

It is easy to read off the transport coefficients from this expressions.
\ben
\eta&=&\frac{r_0^3}{16 \pi G_5}\bigg[1- \alpha'\frac{24 c_1 Q^2}{r_0^6}\bigg]
+ {\cal O}(Q^4)\nn
\kappa&=& \frac{\eta}{\pi T}\bigg[\lb 1-\frac{Q^2}{2 r_0^6}\rb -
\alpha'\lb 10 c_1 - \frac{Q^2}{3 r_0^6}(37 c_1 -48 c_2)\rb\bigg]+ {\cal O}(Q^4)\nn
\tau_{\pi} T &=&{2 - \ln 2\over 2\pi} -{Q^2(5-3 \ln 2)\over 4 \pi r_0^6}
 + \app \ltb -{11 c_1 \over 2 \pi}
 +{Q^2 \over 4 \pi r_0^6}(-16 c_2 + 5 c_1(11-4\ln 2))\rtb+ {\cal O}(Q^4)\nn
\een
where, the temperature $T$ of the system is given by,
\be
T=\frac{r_0}{\pi}\bigg[\lb 1-{Q^2 \over 2 r_0^6}\rb-{\alpha' \over 3}\bigg(5 c_1 +\frac{Q^2}{2 r_0^6}(31 c_1 + 48 c_2)\bigg)\bigg]+ {\cal O}(Q^4).
\ee

We see that the first order as well as the second order transport coefficients coming
from retarded Green's function of energy momentum tensor\footnote{It would be
interesting to study the flow of retarded Green's function for boundary $R$-current.
We found it to be difficult to get any analytical solution for response function
in presence of finite chemical potential and higher derivative terms. However, it would be
nice to know the higher derivative corrections to other second order transport coefficients
appear in $R$ current\cite{chargehydro, haack}.} only depends on two coefficients
$c_1, c_2$. This feature was observed in \cite{mps} for entropy density $s$ and first-order
transport coefficients $\eta$. The coefficients $c_3, c_4$, which parameterize
couplings in the four point function
of the dual $U(1)$ current does not play any role in these hydrodynamic coefficients. They
should be important for the computation of conductivity, which
comes from the Green's function of the boundary $R$-current. Two
other coefficients $\zeta, c_5$ also do not appear in the expressions. One can find a magnetic
brane solution of the action (\ref{hdcbacn}) like \cite{D'Hoker:2009bc}. In that case it would be interesting to
find the effect of magnetic field on transport coefficients.

So far we have discussed the flow equation for energy-momentum tensor. 
However, in general the fluid can have other conserved currents like $R$ current, 
if the theory has a global $R$ symmetry. Therefore, in the same spirit, 
one can also discuss the flow of $R$ current in the context of fluid/gravity correspondence. 
This is the final topic of our discussion.



\section{Flow of retarded Green's function of boundary $R$
current}\label{rcflow}

Finally, in this section, we study the flow of retarded Green's
function
of boundary $R$-current,
\be
G^R_{i,j}(k) =-i \int dt d^3x e^{ik\cdot x}
\langle[J_i(x),J_j(0)]\rangle
\ee
where $J_{\mu}(x)$ is the $CFT$ current dual to a bulk gauge field
$A_{\mu}$.

In hydrodynamic approximation one can express the current in powers of
boundary derivatives.
Up to first order in derivative expansion it has the following form,
\be
J_{\nu}= - \tilde{\kappa} P^{\alpha}_{\nu} \partial_{\alpha}{\mu\over
T} + \Omega l_{\nu} + {\cal O}(\partial^2)
\ee
where, $\tilde{\kappa}$ and $\Omega$ are two first order transport
coefficients, $\mu$ is chemical
potential, $T$ is temperature and
\ben
P_{\mu\nu}&=&u_{\mu}u_{\nu} + \eta_{\mu\nu} \nn
l_{\mu}&=&
\epsilon_{\mu}^{\alpha\beta\gamma}u_{\alpha}\partial_{\beta}u_{\gamma}
\ .
\een
The expression of $J_{\mu}$ up to second order in derivative expansion
can be found in \cite{chargehydro,haack}.
From conformal invariance of the theory it is possible to write
all possible second order transport coefficients appear in the
expression of $J_{\mu}$. However,
 like energy momentum tensor, from the expression of retarded Green's
function it is not possible
to compute all the transport coefficients that appear in different
order of derivative expansion.

In this section we study the flow equation of retarded Green's
function of boundary $R$-current.
Unfortunately we find it difficult
to solve the flow equation analytically to extract any transport
coefficient. We present our calculation
how to write the flow equation for retarded Green's function of
$R$-currents in presence of generic higher derivative
terms in bulk Lagrangian and some numerical results.

We start with Einstein-Maxwell action
\be
S= \nt \int d^5x \sqrt{-g}\lb R + 12 -\frac{1}{4}F^2\rb.
\ee
Solution is given by equation (\ref{chargehdsol}) with $\app=0$.
The temperature of the black hole is given by,
\be
T={r_0\over \pi}\lb 1-\frac{Q^2}{2r_0^6}\rb
\ee
and the chemical potential is given by,
\be
\mu=\frac{\sqrt{3} Q}{r_0^2}.
\ee

For technical advantage we write the metric and gauge field in
a different coordinate. We change
the radial coordinate $r\ra {r_0 \over \sqrt{r}}$. In this coordinate
the metric
and gauge field is given by,
\ben
ds^2 &=& - {r_0^2 U(r)\over r} dt^2 + {dr^2 \over 4 r^2 U(r)}+
{r_0^2\over r}(d\vec{x}^2)\nn
A_t(r) &=& E(r)
\een
where,
\ben
U(r)=(1-r)(1+r-{Q^2 r^2\over r_0^6}), \ \ \
E(r)={\sqrt{3}Q\over r_0^2}(1-r)\ .
\een

We turn on small fluctuations for $x$ component of gauge fields.
Since the $A_t$ component of the bulk vector is non-vanishing
in this background, the perturbations $A_x$ can couple to the $tx$
component of graviton
Therefore we also need to consider small
metric fluctuations for components $g_{tx}$.
Writing them in momentum space
\ben
A_{x}(r,x)&=&\int{d^4k\over (2\pi)^4} e^{i k.x} A_{1}(r,k)\nn
g_t^{x}(r,x)&=&\int{d^4k\over (2\pi)^4} e^{i k.x} \Phi(r,k)\ .
\een
However, there exists a constraint relation between $A_x$ and
$g_{tx}$. We
use this relation to replace $g_{tx}$ from equation of motion of
$A_x$.

 The on-shell action for gauge field fluctuations
are given by,
\ben\label{Axacn}
S_{A} &=& \nt \int {d^4k\over (2 \pi)^4} \bigg[ -r_0^2 U(r)
A_x'^2(r,k) +  \lb {\omega^2 \over 4 r U(r)} - {q^2\over 4 r}\rb
A_x^2\bigg].
\een
The current corresponding to $A_x$ fluctuation is given by,
\ben
J_x(r,k)&=&{\delta S_{A} \over \delta A_x'(r,k)}
=-2 r_0^2 U(r) A_x'(r,k).
\een
The equation of motion for $A_x(r,k)$ is given by,
\ben
(U(r) A_x'(r,k))' &=& -{1\over 4 r_0^2} \lb{\omega^2 \over r U(r)} -
{q^2\over r}\rb A_x(r,k) -E'(r) \phi'(r,k)\nn
 &=& -{1\over 4 r_0^2} \lb{\omega^2 \over r U(r)} - {q^2\over r}\rb
A_x(r,k) + {r E'(r)^2 \over r_0^2}A_x(r,k).
\een
Using the constraint relation (coming from $rx$ component of Einstein
equations),
\be
\phi'(r,k) =- {r E'(r) A_x(r,k) \over r_0^2}.
\ee
and the equations of motion one gets,
\ben
J_x'(r,k) &=& {1\over 2} \lb{\omega^2 \over r U(r)} - {q^2\over r}\rb
A_x(r,k)  - {2r E'(r)^2} A_x(r,k).
\een

Next we define a response function
\be
\sigma(r,k)= {J_x(r,k) \over i \omega A_x(r,k)}.
\ee
Taking the derivative with respect to $r$ and using the equation of
motion we
find the flow is given by,
\ben\label{Jfloweq}
\sigma'(r,k) &=& {i \omega\over 2 r_0 U(r)} \bigg[ \sigma(r,k)^2
- \bigg({r_0\over  r} \bigg( 1 - {q^2 U(r) \over \omega^2} \bigg) - {4
r_0 r E'(r)^2 U(r) \over \omega^2} \bigg)\bigg].
\een
From the regularity of the response function at the horizon we find
the boundary condition
is given by,
\be
\sigma(1,k)^2 = 1.
\ee

With this boundary condition one can integrate this nonlinear equation
to find finite frequency response of boundary Green's function.

\lfig{Flow of R current correlation function.}
{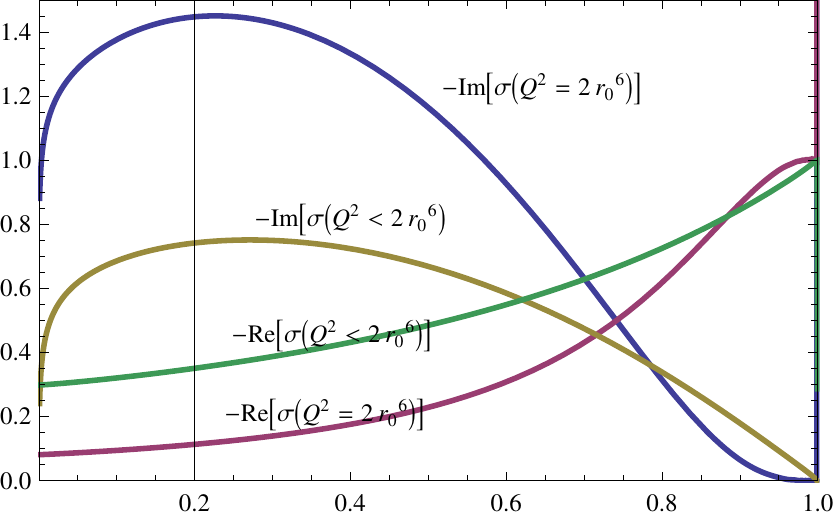}{6.5cm}{currentplot}

In presence of generic higher derivative terms in the Lagrangian the
on-shell
action for fluctuation $A_x$ may not have canonical form like
(\ref{Axacn}).
In that case one has to write an effective action for $A_x$ like
transverse
graviton. The effective will have the same form as (\ref{Axacn}) only
the
coefficients will depend on coupling constant of higher derivative
terms.

We conclude this section by presenting some numerical solutions of the
flow equation (\ref{Jfloweq}) for both non-extremal extremal black
holes in
fig. \ref{currentplot}.

\section{Conclusion and future directions} \label{conclusion}

In this review we have discussed the hydrodynamic behavior of
strongly
coupled systems in the context of the gauge/gravity correspondence. We
have learnt how we can apply the holographic methods to compute
different transport coefficients of boundary fluid systems. We have
also elaborated the effect of higher derivative terms, which come from
string theory, on these coefficients.

First, we have shown how the shear-viscosity coefficient can be viewed as the effective 
coupling of transverse graviton in a two derivative theory. Next, 
we generalized this statement to any arbitrary
 higher derivative theory. We have developed a procedure to construct an effective action for
transverse
graviton in canonical form in presence of any higher derivative terms
in
bulk and showed that the horizon value of the effective
coupling
obtained from the effective action gives the shear viscosity
coefficient of
boundary fluid. Our results are valid up to first
order in $\mu$ (coefficient of higher derivative term). We discussed
two non trivial examples to check the method. We have considered four
derivative and eight derivative ($Weyl^4$) Lagrangian and calculated
the
correction to the shear viscosity using our method. We found complete
agreement between our result and the results obtained using other
methods.

In section \ref{nhsection} we have discussed that the shear viscosity to
entropy density ratio is controlled by the near horizon geometry of the dual
black hole spacetime. Therefore the knowledge of the near-horizon
behavior of different bulk fields are enough to compute the ratio.
That is why the computations are not very complicated
even in the presence of higher derivative terms. We would like to
emphasize that there is no ambiguity in defining the overall coefficient
of the effective action and the results are consistent.
We apply this idea to compute $\eta/s$ in section \ref{etabys}
and see that the ratio is controlled by the
horizon values of the scalars and so, in the non-extremal case, an
operator deformation in the QFT will produce an interpolating non-trivial
flow in which the moduli approach the (IR) black hole
horizon. In the extremal case, though, the horizon moduli
values are fixed and so the shear viscosity to entropy density
ratio does not depend on the asymptotic values of the scalars. Therefore,
QFTs with different UV fixed points can flow to
the same IR fixed point.

One important point to mention here is that the radial independence
of retarded Green's function (response function) depends on the
massless properties of transverse graviton $h_{xy}$ at $k\ra 0$
limit. We have seen that the mass term is proportional to $k^2$ and it
vanishes in low frequency limit. However, in general this observation
is not true. If we have other matter fields in the Lagrangian (like
magnetic field considered in section \ref{model}) then all the components of
transverse graviton may not be massless. Please look at appendix \ref{hxy}.
The isotropy (SO(3) symmetry) in spatial direction is lost in presence of
a magnetic field. Therefore though $xy$ component of transverse
graviton ($h_{xy}$) is massless but other components $h_{xz}, h_{yz}$
are not decoupled form gauge field perturbations and hence not
massless even in low frequency limit. Therefore the retarded Green's
function (response function) corresponding to these massive
fluctuations are not independent of radial direction.
Hence, horizon value and asymptotic values are different in these
case. As a result the isotropy of viscosity tensor is also
lost\footnote{Since, $[T_{xy},T_{xy}]$ correlator is different than
$[T_{xz},T_{xz}]$ or $[T_{yz},T_{yz}]$ correlators.}.

We have also studied the flow equations for
retarded Green's function of boundary theory analytically and found
higher order transport coefficients of the boundary plasma solving
this equation. We have generalized the analysis for generic higher
derivative gravity theory. The flow equation for Green's function is a
first order non-linear differential equation of Riccati type. Because
of its non-linear nature it is hard to solve this equation exactly.
After a change of variable one can reduce this non-linear equation to
a second order linear homogeneous differential equation. But to solve
this we need to specify two boundary conditions. In this article we
have dealt with the non-linear equation and specify the boundary
condition at the horizon. Therefore the hydrodynamic characteristic of
the field theory at $UV$ fix point is determined by $IR$ boundary
condition. In this way of computing the transport coefficients has an
advantage over usual Kubo approach. In Kubo approach, one has to first
find the transverse graviton by solving a second order differential
equation and then compute regarded Green's function. Instead, the flow
equation is a first order differential equation (although
non-linear). As we want a perturbative expansion of Green's function
in powers of $\omega$ and $q$ the equation turns out to be a linear
first order differential equation. Thus, technically, it is simpler to
get results for causal hydrodynamics, particularly when the dual bulk
theory is complicated.

It would be interesting to use $renormalization-group$ $flow$
techniques to understand how the response function evolves
as a function of the radial direction. One can also apply
Blackfold techniques of \cite{emparan} and the idea of
Wilsonian RG flow explained in \cite{mukund-liu}.
Blackfolds are higher dimensional black holes with two
separate length scales (mass and angular-momentum) and
their equations of motions are obtained from fluid correspondence
(in leading Einstein gravity). Generalizing this technique
in reverse, one should obtain equations of boundary plasma
from the Blackfold equations in a generic gravity theory. It would be
really nice to check this explicitly.

There are other methods to compute first and second order transport 
coefficients holographically. We have discussed two of them in little 
details in appendix \ref{other methods}. In 
\ref{revshiraz} we have briefed the method developed in \cite{causa1}. 
This is an elegant method to understand the properties of boundary fluid
from boosted black brane geometry. This method shows that, 
all the constituent equations of fluid system are captured in the 
Einstein equations. In \cite{sd}, this method has been extended to higher 
derivative gravity (Gauss-Bonnet gravity) and shown that 
the above mentioned statement is true 
even in presence of higher derivative terms. Although this method captures 
almost all transport coefficients (except $\kappa$, the second order 
transport coefficient) up to second order in derivative expansion, but 
it is technically very hard to extend this idea to study higher derivative 
corrections. One problem is that the exact form of boundary stress tensor 
is not known for any generic higher derivative gravity, for example Weyl$^4$
term. 

In \ref{revpaolus} we have discussed another method to compute first 
order transport coefficients developed in \cite{Paulos}. In \cite{ramy1},
the authors proposed a Wald like formula for shear viscosity coefficient
in generic higher derivative gravity. Though the formula works for generic four 
derivative term in bulk action, but unfortunately it fails to produce the correct result 
for Wely$^4$ term \cite{ns1}. Motivated by this formula the author in \cite{Paulos}
proposed a new Wald like formula for the first order transport coefficients.
The author has shown that the
  transport coefficients can simply be obtained by evaluating the
  residue of the pole of the quadratic action near the horizon. It
  only assumes the massless behavior of low energy perturbation and
  regularity of the field at horizon. The result is independent of the
  boundary terms and thus one can neglect them.

We conclude this review with some discussion on the currents trends on this subject.

In this article we have mainly considered the characteristic of conformal
fluid which has zero bulk viscosity at first order. Bulk viscosity measures the
reaction of fluid against any bulk stress. The holographic computation of bulk viscosity
is also an interesting topic in this area of research. In \cite{buchel-bulkvisco} the author
has proposed a holographic bound on bulk viscosity to entropy density ratio of
strongly coupled gauge theory plasma,
\be\label{bulkviscobound}
\frac{\zeta}{s}\geq \frac1{2\pi} \lb \frac1p - c_s^2 \rb,
\ee
where $c_s$ is the speed of sound and $p$ is the spatial dimension of the system.
This bound is dynamical, i.e. the right hand side depends on temperature unlike
the bound on shear viscosity to entropy density ratio. They observed that the bound is
saturated by the $p + 1$ space-time dimensional gauge theory plasma holographically
dual to a stack of near-extremal flat $Dp$-branes at leading order (i.e. infinitely large
't Hooft coupling). The bulk viscosity
bound \ref{bulkviscobound} is also saturated in toroidal compactifications of conformal
theories \cite{buchel-bulkvisco,skenderis-nonconformal}. It has been found that in various
examples of string theory embedding of the holographic gauge-gravity correspondence the
bound is saturated. For example, ${\cal N}=2^*$ gauge theory plasmas, gauge theories
with adjoint $R$ charges, cascading gauge theories etc. There are further evidences
in favor of saturation of this bound in
\cite{buchel-pagnutti,Gursoy-Kiritsis-Michalogiorgakis-Nitti} for
phenomenological models of fluid-gravity correspondence. However
\cite{gubser-nellore,Gubser-Pufu-Rocha} considered some other phenomenological models
where the bound is violated. Recently in \cite{buchel-bulkvisko-violation} the author
addressed the question if the violation of bulk viscosity bound is only limited to
the phenomenological models of gauge/gravity correspondence. They considered
strongly coupled ${\cal N} = 4$ supersymmetric Yang-Mills plasma compactified on a
two-manifold of constant curvature and found that the resulting (1 + 1)-dimensional
hydrodynamic system can have bulk viscosity coefficients which violates the bound
if the curvature of the compact manifold is negative. It would be also 
interesting to check this bound explicitly in presence of some higher derivative 
terms in bulk.

Non-relativistic generalization of the original AdS/CFT
correspondence has opened up new directions in current
research after a series of beautiful experiments on cold
atoms at unitarity. The correspondence has been extended
to explore the holographic duals of strongly coupled
non-relativistic conformal field theories. Cold atoms are
system of fermions interacting through a short-range
potential which can be fine-tuned to obtain a massless
bound state. Since the theory is scale invariant, it can
be described as a non-relativistic conformal field theory
with Schr\"odinger symmetry. This symmetry consists of the
usual Galilean invariance, the scaling symmetry as well as
the particle number symmetry. There has been some activity
along this direction where solution generating techniques
have been used to obtain bulk geometries with asymptotic
Schr\"odinger symmetry.

At the first order in derivative expansion,
we have one interesting transport coefficient, which is shear viscosity
coefficient. It turns out that for non-relativistic fluid the shear viscosity to entropy
density ratio saturates the $KSS$ bound \cite{rangamani-ross}. However,
\cite{rangamani-ross} has also computed other transport coefficients like,
thermal conductivity kinematic viscosity, thermal diffusivity, Prandtl number holographically,
using the dual geometry proposed in \cite{mukund-nonrel,maldacena-nonrel,bala-nonrel}.

As mentioned earlier, fluid dynamics in general can be thought of as an effective
field theory with an infinite number of irrelevant terms obtained as usual in a
derivative expansion. In relativistic case consideration of second order hydrodynamics
is essential because the first order formalism is inconsistent with causality issues.
However in non-relativistic setup one is not forced to consider the second order
(beyond ideal and viscous terms) terms in stress tensor. But it is interesting to study
those terms if we want to view hydrodynamics as a derivative expansion of its local
variables.

One interesting problem in this direction is to study the light-cone reduction of relativistic
conformal/non-conformal hydrodynamic stress tensor
\cite{causa1,causa2,causa3,haack,chargehydro}. In second order we
encounter different transport coefficients in case of
relativistic theory (see section \ref{prob2}). It is
possible to reduce the theory along one of
the light-cone coordinates and understand how the higher
dimensional transport coefficients descend down to new transport
coefficients for the lower dimensional non-relativistic theory.
The reduction of first order stress tensor has been done in \cite{mukund-nonrel}.
Following the same procedure one can write the second order corrections to the stress tensor
of a charged fluid with Schr\"odinger symmetry.
Once we have all possible second order transport coefficient of
a non-relativistic fluid we can evaluate their holographic
values using the non-relativistic gauge/gravity correspondence.


To compute the second order non-relativistic charged fluid's transport 
coefficients holographically, one has to take the bulk spacetime
to be five dimensional Reissner-Nordstr\"om black hole
(embedded in 10 dimensional geometry) and perform $TsT$
transformation to get asymptotically Schr\"odinger spacetime.
Then one should turn on hydrodynamic fluctuations in this
background and study the flow of corresponding response
function (retarded Green's function) like \cite{nsflow}.
Evaluating the response function at asymptotic boundary
it is possible to compute the non-relativistic second
order transport coefficients of 2 (spatial)-dimensional
dual fluid system. However, in this way of computing the
transport coefficients (Kubo method) may not capture all
possible coefficient (like relativistic case). Applying
the method described in \cite{causa1} (\cite{mukund-nonrel} has
already applied this method to evaluate first order transport
coefficient for non-relativistic fluid) would be a good
idea to compute the second order transport coefficients
holographically.\\

\vspace{1cm}

\noindent
{\bf \large Acknowledgement}

It is a great pleasure to thank various people who encouraged and 
helped us to write this review. 
We would like to give a special thank to our collaborators Dumitru 
Astefanesei, Jyotirmoy Bhattacharya, Sayantani Bhattacharyya, R Loganayagam, P Surowka for interesting 
and useful discussion during projects. We would also like to thank Shiraz Minwalla for many useful 
insights that helped us understanding the subject better. NB acknowledges IISc, Bangalore and TIFR, 
Mumbai for hospitality during this program. N.B. is supported by a Veni grant of the `Nederlandse 
Organisatie voor Wetenschappelijk Onderzoek (NWO)'. Finally, we would like to thank people of India for 
their unconditional and generous support to researches in string theory.\\


\noindent
{\Large \bf Appendix}

\appendix


\section{Second order transport coefficients from Kubo
formula}

In this appendix we re-derive second order transport coefficients
$\kappa$ and $\tau_{\pi}$
 from usual Kubo approach.
Let us now consider the action with solution given in section
\ref{prob2},
\be\label{action}
S={1 \over 16 \pi G_5}\int d^5x\sqrt{-g}  [R + 12 ]\, .
\ee
For well defined variation of this action we need to add a
Gibbons-Hawking
boundary term. Also, requiring the on-shell action
being finite at boundary, we have to add counter-terms following usual
approach
of holographic renormalization. They are as follows:
\ben
S_{GH}&=&{1 \over 8 \pi G_5}\int d^4x\sqrt{-\gamma} {\cal K},\nn
S_{CT}&=& {1 \over 16 \pi G_5}\int d^4x\sqrt{-\gamma}[6+ {1 \over 2}
{\cal R}]
\een
where $\gamma$ and ${\cal R}$ are boundary metric and Ricci scalar
(constructed out of $\gamma$)
respectively.

We consider the following metric perturbation,
\be
g_{xy}=g^{(0)}_{xy}+ h_{xy}(r,x)=g^{(0)}_{xy}(1+\ep \Phi(r,x)),
\ee
where $\epsilon$ is an order counting parameter. We are interested in
quadratic on-shell
action for this transverse graviton  $\Phi(r,x)$. Let us define the
Fourier transform,
\be \label{phifu}
\ph = \int {d^4x \over (2 \pi)^4} e^{-i k.x} \Phi(r,x)\ ,
\ee
and $k=\{-\omega,\vec k\}$. Substituting this fluctuation in action
(\ref{action}),
we get (\ref{fullaction})\footnote{$a_3,a_6$ are zero here}. Now, we
rewrite this
action (\ref{action}) as equation of motion piece (which vanishes
on-shell) and
boundary term \footnote{We ignore contribution from horizon as
\cite{son4}}.
Thus, on-shell $S, S_{GH}, S_{CT}$  become,
\ben
S&=&\frac{1}{16 \pi G_5}\int_{r=\delta}\frac{d^4k}{(2
\pi)^4}L(\phi(r,k)), \nn
S_{GH}&=&\frac{1}{8 \pi G_5}\int_{r=\delta}\frac{d^4k}{(2
\pi)^4}L_{GH}(\phi(r,k)), \nn
S_{CT}&=&\frac{1}{16 \pi G_5}\int_{r=\delta}\frac{d^4k}{(2
\pi)^4}L_{CT}(\phi(r,k)),
\een
where we  define,
\ben
L(\phi(r,k))&=& a2(r) \phi'(r,k) \phi^*(r,k) + {a4(r)-a5'(r)
\over 2}\phi(r,k) \phi^*(r,k), \\
L_{GH}(\phi(r,k))&=&2(g1) \phi(r,k) \phi^*(r,k) + 2(g2) \phi'(r,k)
\phi^*(r,k), \\
L_{CT}(\phi(r,k))&=&(  c_0 + c_1 \omega^2 + c_2 q^2) \phi(r,k)
\phi^*(r,k).
\een
The coefficients $a_2,a_4,a_5$ are given in (\ref{boeff}) and other
coefficients are,
\ben
g_1&=&\frac{2-r^2}{r^2}, \,\,\ g_2= -2\frac{1-r^2}{r}, \ \
\
c_0=-3 \frac{\sqrt{1-r^2}}{r^2}, \,\,\ c_1=\frac{1}{4 r
\sqrt{1-r^2}}, \ \ \
c_2=- \frac{\sqrt{1-r^2}}{4 r}.
\een
The retarded Green's function is defined at asymptotic infinity as,
\be
G^R(k)=2\lim_{r \rightarrow
0}\frac{(L+L_{GH}+L_{CT})|_{on-shell}}{\phi_0(k)\phi_0(-k)}.
\ee
Here, $\phi_0(k)$ is the boundary value of the fluctuation
(\ref{phifu}).
The retarded Green's function $G^R$ is a function of boundary momenta
$k_{\mu}=(\omega,0,0,q)$.
Now the leading action and the Gibbons-Hawking action get divergences
from $\phi'(r,k) \phi^*(r,k)$ and $\phi(r)\phi^*(r,k)$ parts. Both
these divergences
get canceled by the counter-term action which is always proportional
to only
$\phi(r) \phi^*(r,k)$. In this case of leading Einstein's gravity, it
is even
more simplified. Divergences coming from $\phi(r)\phi^*(r,k)$ piece of
leading and Gibbons-Hawking action gets canceled by momentum
independent
piece of Counter-term action. It turns out that there is a
cancellation among
the corresponding coefficients as,
\be
\lim_{r \rightarrow 0}({1\over 2}(a4(r)-a5'(r))+2 g1 + c_0)={1 \over
2},
\ee
i.e. the final contribution from $\phi(r)\phi^*(r,k)$ piece is only a
finite number
${1 \over 2}$. As the graviton fluctuation $\phi(r,\omega, \vec k)=
\phi_0(1+ F(r,\omega,\vec k))$
and moreover $\lim_{r \rightarrow 0}F(r,\omega,\vec k)=0$, we see that
the
$\phi(r)\phi^*(r)$ term above only contribute to pressure (the
$\omega$
independent piece of $G^R$). It would never contribute to any
transport coefficient.

Also, the divergences coming from $\phi'(r,k) \phi^*(r,k)$ piece of
original and
Gibbons-Hawking action, get canceled with the piece of the
counter-term
proportional to $\omega^2, q^2$ ($c_1\,\,\ and \,\,\ c_2$ are purely
divergent at boundary). Here, the situation is more subtle, as there
is no cancellation
among the coefficients. One actually needs to put the solution of
$\phi(r,\omega, \vec k)$ to see the cancellation.

The overall lesson from this detailed analysis is that counter-term
only cancel the
UV divergences in usual holographic renormalization process and at
most contribute
to pressure of the boundary plasma. It has no effects on any transport
coefficients.
In \cite{causa2}, the author have computed second order transport
coefficients for
the plasmas dual to leading Einstein's gravity following this usual
approach. The
results are as follows,
\be\label{HOTC}
\tau_{\pi}={2- \ln 2 \over 2 \pi T}, \,\,\,\,\,\,\,\,\ \kappa={\eta
\over \pi T}.
\ee
These results match with the one we obtained in (\ref{htf}) by solving
the flow equations.


\section{Equivalence of Boundary Terms}\label{BOUN}

In this appendix we will show explicitly why the transport
coefficients
computed from the original action and the effective action are same,
even
for any higher derivative theory. It was already noticed \cite{ns1},
that the
two would give same first-order transport coefficient $\eta$ with a
suitable choice of the overall normalization constant. Here, we show
that, not
just the first order transport coefficients, rather any higher order
transport
coefficients computed from the original action and the effective
action are same.

We consider a general class of action for $\phi$ which appears when
the higher
derivative terms are made of different contraction of Ricci tensor,
Riemann
tensor, Weyl tensor, Ricci scalar etc. or their different powers.
Since, all
these tensors involve two derivatives of metric they can only have
terms like
$\partial_a \partial_b \Phi(r,x)$ and its lower derivatives. Therefor
the most generic
quadratic (in $\Phi(r,x)$,
in linear response theory) action for this kind of higher derivative
gravity
has the following form (in
momentum space)\footnote{In all the expressions we have omitted $k$
dependence
  of $\phi$.}
\ben\label{fullaction}
S&=&\nt \intk dr \bigg[
 a1(r) \phi (r)^2+ a2(r)  \phi ' (r)^2 \hspace{0.5 cm} + a4(r) \phi(r)
   \phi'(r)
+ \alpha' \ a6(r) \phi''(r) \phi'(r)  \nn
&& \hspace{0.5 cm} + \alpha' \ a3(r) \phi''(r)^2 + a5(r) \phi(r)
\phi''(r)\bigg]
\een
where,
\ben\label{boeff}
a1(r)&=&
\frac {-8 r^2 + \omega ^2 r + 8} {4 r^3 - 4 r^5} + \alpha' \
f2(r), \ \ \ a2(r)=
-3 r + \frac {3} {r} + \alpha' \ h2 (r)  \nonumber \\
a4(r)&=&
- \frac {6} {r^2} - 2 + \alpha' \ g2 (r) \ \ \
a5(r)=
-4 r + \frac {4} {r}  +\alpha' \ j2 (r)
\een
and $a3(r), a6(r), j2(r), g2(r), h2(r)$ and $f2(r)$ depends on
higher derivative terms in the action and hence are computed purely
from
the background solution with $\app \rightarrow 0$. Among these
coefficients
$a3$ is special, as, it couples to $\phi''^2$. All four derivatives
act on the
graviton fluctuation and thus $a3$ only depends on background metric functions
and there r-derivatives. It is easy to convince ourselves that
$a3 \propto r (r^2-1)^2 f(r, \app)$, where $f(r,\app)$ is a function
that
depend on the higher derivative terms and finite (constant or 0) at
the
boundary $r \rightarrow 0$.
Now let us write the effective Lagrangian as follows,
\ben
S_{eff}&=& {1 + \alpha'
\Gamma \over 16 \pi G_5} \intk dr \bigg[
\frac {4 r \left (r^2 - 1 \right)^2
\phi'(r)^2 - \omega^2 \phi(r)^2} {4 r^2 \left (r^2 - 1 \right)}
 + \alpha' \bigg (
b2
(r) \phi(r)^2 + \ b1 (r) \phi'(r)^2 \bigg )\bigg] \ .\nn
\een
Demanding that the equation of motion (up to order $\alpha'$) of
$\phi$ derived from
the original action and the above action are same we get,
\ben
b1(r)&=&{1 \over 2 r \left (r^2 - 1 \right)^2} [ (-4 r^3 - 12 r +
\omega
^2) a3(r) +  (r^2 - 1)
(2 \kappa
r^4 - a6'(r) r^3 - 4 \kappa r^2 + 2 a3'(r) r^2\nn
&& + 2 (r^2 -
1 ) h2(r) r - 2  (r^2 - 1 ) j2(r) r +
a6' (r) r  + 2 \kappa + 2 a3' (r))]
\een
\ben
b2(r)&=&
-{1\over 16 r^2 \left (r^2 -
      1 \right)^4} \bigg [ (\omega ^4 + 144 r^3 \omega^2 )
a3(r)
  + 4  (r^2 - 1 ) \bigg (-4 r^2 f2(r) (r^2 - 1)^3  \nn
&&  + ((\omega^2 \kappa - 2 r^2  (r^2 - 1 ) j2'' (r) ) (r^2 - 1)
 + 2 r^2 g2' (r)  (r^2 - 1)^2 + r \omega^2 a3'' (r) )  (r^2 -1 ) \nn
&& +
(1 - 11 r^2 )\omega^2 a3' (r)\bigg) \bigg]\ .
\een
The boundary terms coming from the original action (after adding {\it
Gibbons-Hawking} boundary terms) are given by
\footnote{There was a sign error in \cite{ns1}},
\ben
S^{{\cal B}}&=& \nt \intk\bigg[
-\frac {\phi (r)^2} {r^2} + \phi (r)^2 +
 r \phi ' (r) \phi (r) - \frac {\phi ' (r)
        \phi (r)} {r}\nn
&& + \alpha' \bigg (\frac {1} {2} g2 (r) \phi (r)^2 - \frac {1} {2}
         j2' (r) \phi (r)^2+( h2 (r)  - j2 (r)
         - \frac {
a6' (r)}{2})\phi ' (r) \phi
        (r) \nonumber \\
  &&+\frac {
a3' (r) \left (\phi (r) \omega^2 +
          4 \left (r^4 - 1 \right)
               \phi ' (r) \right) \phi (r)} {4 r \left (r^2 -
          1 \right)^2} - \frac {a3 (r)
           (6 r \phi (r)\phi ' (r) \omega^2)} {4 r
           \left (r^2 - 1 \right)^3} \nonumber \\
           &&- \frac {a3 (r)
           \left ( \left (r^2 -
              1 \right) \left (8 r^3 + 24 r -
\omega^2 \right) \phi ' (r) \right) \phi (r)} {4 r
           \left (r^2 - 1 \right)^3}- \frac
{a3 (r) \phi ' (r) \left (\phi (r)
              \omega^2 +
         4 \left (r^4 - 1 \right) \phi ' (r) \right)} {4 r
           \left (r^2 - 1 \right)^2} \nn
  &&- a3 (r) \phi ' (r) \left (-\frac {\phi (r)
              \omega^2} {2 r \left (r^2 -
             1 \right)^2} - \frac {\left (r^4 - 1 \right) \phi
               ' (r)} {r \left (r^2 - 1 \right)^2} \right) \bigg
)\bigg]\ .
\een
And the boundary terms coming from the effective action are given by,
\ben
S_{seff}^{{\cal B}}&=& {1 \over 16 \pi G_5} \intk \bigg[
\left (r - \frac {1} {r}
\right) \phi (r) \phi ' (r)   + {\alpha' \over 2 r \left (r^2 - 1
\right)^2} \bigg (\ \phi (r) (2
            \Gamma  \left (r^2 -
           1 \right)^3
+ (- a6' (r) r^3
\nonumber \\
&&  + 2 a3' (r) r^2 +
           2 \left (r^2 - 1 \right) h2 (r) r - 2
                \left (r^2 - 1 \right) j2 (r) r
 + a6' (r) r  + 2
               a3' (r) ) \left (r^2 -
          1 \right) \nn
          &&  + \left (-4 r^3 - 12 r +
\omega^2 \right) a3 (r)
) \phi ' (r)\bigg ) \bigg]\ .
\een
 Now, it is interesting to compute the difference between these two
boundary terms and the result
is\footnote{It has been shown in \cite{ns1} that $\Gamma=0$.},
\ben\label{diff}
S^{\cal B}- S^{\cal B}_{eff}&=& \nt \intk\bigg[
-\frac {\phi (r)^2} {r^2} + \phi (r)^2   +
 \alpha' \bigg (\frac {1} {2} g2 (r) \phi (r)^2 - \frac {1} {2}
         j2' (r) \phi (r)^2 \nonumber \\
&& \ \ \ \ \ \  \ \,\,\,\,\,\ +\frac {
a3' (r) \omega^2 \phi (r)^2} {4 r \left (r^2 -
          1 \right)}  - \frac {a3 (r)
           \left (6 r \omega^2 \right) \phi (r)^2} {4 r
           \left (r^2 - 1 \right)^3}\bigg )
\een
The term proportional to $a3$ in the parenthesis of (\ref{diff})
vanishes
at the boundary whereas the term proportional to $a3'$ gives a pure UV
divergent
piece and a vanishing piece, due to the property of $a3$ mentioned
above. This
is true irrespective of the choice of the higher derivative terms.
Thus, we see
that the two boundary terms differ only by terms which are either
purely divergent
or of the form $g(r)\phi^2$, where $g(r)$ is any function of $r$. The
divergent terms would get canceled once appropriate
boundary terms are added (which has been discussed in sections
\ref{prob2} and \ref{modelhd}). The terms
proportional to $\phi^2$ can only contribute to pressure of the
boundary theory
and are not important for the computation of transport coefficients of
the boundary
plasma. Thus we see that, it is obvious that the transport
coefficients coming from
the original action and the boundary action are same.

Here we have considered only $R^{(n)}$ gravity theory. A more rigorous
proof is required
for theories involving covariant derivatives of curvature tensors and
scalars.


\section{Functions appeared in four derivative action}\label{gbapp}

\ben
{\cal A}_0^{\rm GB}(r,k)&=& -\frac{q^2
   \left(r^2-1\right)+\omega ^2}{4 r^2 \left(r^2-1\right)}
+ {\alpha ' \over 12 r^2 \left(r^2-1\right)} \bigg(q^2 (2 \beta _3
(13 r^2-3 r
   \omega ^2-13) +130 (r^2-1) \beta _1\nn
&&+(-36
   r^4+25 r^2+11) \beta _2)+\omega ^2 ((6
   r^2-11) \beta _2+130 \beta _1+26 \beta
   _3)\bigg)
\een
and
\ben\label{a1gb}
{\cal A}_1^{\rm GB}(r,k)&=& r-\frac{1}{r}
-\frac{\left(r^2-1\right) \left(\left(18 r^2-13\right) \beta
   _2+110 \beta _1+22 \beta _3\right) \alpha '}{3
   r} .
\een


\section{Coefficients in expansion (\ref{nonextremal}) at ${\cal O}(r-r_h)^2$}
\label{hocoeff}

For completeness, in this appendix we present the other coefficients that
appear in the near horizon expansion
of different fields in eq. (\ref{nonextremal}). They are
\ben
\label{BB}
u_2 &=& e^{\alpha  \varphi _h} \left(\frac{5 B^2}{3 r_h^4}+\frac{7
   q^2}{3}\right)+\frac{3}{4} z_1^2 r_h^2-2 \nonumber
\een
\ben
v_2&=&{1\over 288 \pi ^2 T^2}\bigg(e^{-2 \alpha  \phi _h}
\bigg(24 B^4 \zeta ^2 e^{\alpha  \phi _h}+24 \bigg(5 B^2+2 q^2\bigg)
e^{3 \alpha  \phi _h}-72 B^2 \zeta ^2\nn
&&+\bigg(6
   B^4 \bigg(\alpha ^2-4\bigg)-B^2 q^2 \bigg(9 \alpha ^2+8\bigg)+q^4
\bigg(3 \alpha ^2-4\bigg)\bigg) e^{4 \alpha  \phi _h}-144 e^{2 \alpha
   \phi _h}\bigg)\bigg)\nonumber
\een
\ben
w_2 &=& {e^{-4 \alpha  \phi _h}\over 288 \pi ^2 T^2}
\bigg(-36 B^4 \zeta ^4-24 B^2 \zeta ^2 \bigg(B^2+q^2\bigg)
e^{3 \alpha  \phi _h}+48 \bigg(q^2-2 B^2\bigg) e^{5
   \alpha  \phi _h}+288 B^2 \zeta ^2 e^{2 \alpha  \phi _h} \nn
&& +\bigg(-3 B^4 \bigg(\alpha ^2-4\bigg)-8 B^2 q^2+q^4 \bigg(3 \alpha ^2-4\bigg)\bigg)
   e^{6 \alpha  \phi _h}-144 e^{4 \alpha  \phi _h}\bigg)\nonumber
\een
\ben
z_2&=&-\frac{B \zeta  \left(B^2 \left(9 \zeta ^2
e^{-3 \alpha  \phi _h}+1\right)+5 \left(q^2-6
e^{-\alpha  \phi _h}\right)\right)}{6 \pi  T}
\nonumber
\een
\ben
p_1&=&\frac{B q \zeta  e^{-3 \alpha  \phi _h}
\left(\left(3 \alpha ^2+4\right) \left(B^2-q^2\right)
e^{3 \alpha  \phi _h}-12 B^2 \zeta ^2+24 e^{2
   \alpha  \phi _h}\right)}{96 \pi ^2 T^2}.
\een

\section{Decoupling of $h_{xy}$ mode}
\label{hxy}
In what follows, we provide a detailed derivation of the decoupling
of the dual gravitational mode $(4.1)$. We have explicitly checked that
the $h_{xy} = e^{i t \omega +2 V(r)} \epsilon  \Phi (r)$ mode does not
couple with any other field when the momentum vanishes.

For two derivative gravity theory, this can be easily seen from the
equations of motion (\ref{einstein}, \ref{scalar}, \ref{gauge}). However,
we are interested in the most general four-derivative action (\ref{hdaction}).
In this case, instead of writing the equations of motion in the
presence of higher derivative terms, we will explicitly compute
the action up to order $\epsilon^2$. In this way, it can be explicitly
checked that there is no coupling between $h_{xy}$ and the other fields.

Let us turn on the following perturbations of the metric
\ben
g_{\alpha\beta}&=&g^{(0)}_{\alpha\beta}+ \epsilon h_{\alpha\beta} \nn
&=&\left(
\begin{array}{ccccc}
 \frac{1}{U}+e^{i t \omega } \epsilon  \xi_1 & e^{i t \omega }
\epsilon  \xi_2 & e^{i t \omega } \epsilon  \xi_3 & e^{i t
\omega } \epsilon  \xi_4 & e^{i t \omega } \epsilon  \xi_5 \\
 e^{i t \omega } \epsilon  \xi_2 & e^{2 W} Z^2-U & e^{i t \omega } 
\epsilon  \Upsilon & 0 & e^{2 W} Z+e^{i t \omega } \epsilon  \upsilon \\
 e^{i t \omega } \epsilon  \xi_3 & e^{i t \omega } \epsilon  
\Upsilon & e^{2 V} & e^{i t \omega +2 V} \epsilon  
\Phi & e^{i t \omega } \epsilon  \chi \\
e^{i t \omega } \epsilon  \xi_4  & 0 & e^{i t \omega +2 V} 
\epsilon  \Phi & e^{2 V} & 0 \\
 e^{i t \omega } \epsilon  \xi_5 & e^{2 W} Z+e^{i t \omega } 
\epsilon  \upsilon & e^{i t \omega } \epsilon  \chi  & 0 & e^{2 W}
\end{array}\right)
\een
and gauge gauge fields
\ben
A_{\alpha}&=&A^{(0)}_{\alpha} + \epsilon f_{\alpha} \nn
&=& \left(e^{i t \omega } \epsilon  a_r(r) , e^{i t \omega } \epsilon  a_t(r)-E(r)
, \frac{B y}{2}+e^{i t \omega } \epsilon  a_x(r), -\frac{B x}{2}+e^{i t \omega }
\epsilon  a_y(r),e^{i t \omega } \epsilon  a_z(r)+P(r)\right)\nn
\een
Here $g^{(0)}_{\alpha\beta}$ and $A^{(0)}_{\alpha}$ are the background
metric (\ref{anz2}) and the background gauge field; $\epsilon$ dependent terms
are the perturbations.

Using these field excitations, we can now compute the action.\footnote{
We use the Mathematica notebook for this computation. We emphasize that our
system is symmetric in $x-$ and $y-$directions.}  The result is complicated,
but for our purpose it is enough to pick up the $\Phi(r)$-dependent
terms --- for concreteness, let us write the $\Phi(r)$-dependent part
of the action:

\ben
S_{\Phi, \Phi}&=&
\Phi(r)^2     \bigg[ {1 \over 4 U(r)} ((e^{-2 V(r)-W(r)} (-U(r) e^{2 W(r)}
(4 B^2 e^{\alpha  \varphi (r)}+4 Q(r)^2 e^{4 V(r)+\alpha  \varphi (r)} \nn
&&-2 e^{4 V(r)} U''(r)
-4
   e^{4 V(r)} U'(r) \left(2 V'(r)+W'(r)\right)+e^{4 V(r)+2 W(r)} Z'(r)^2+24 e^{4 V(r)}) \nn
&&+2 U(r)^2 e^{4 V(r)} (2 P'(r)^2 e^{\alpha
   \varphi (r)}+e^{2 W(r)} (4 V''(r)+4 V'(r) W'(r)+6 V'(r)^2  \nn
&&+2 W''(r)+2 W'(r)^2-\varphi '(r)^2))+14 \omega ^2 e^{4 V(r)+2
   W(r)})))\bigg]  \nn
&& +
\Phi(r) \Phi'(r) \bigg[  (2 e^{2 V(r)+W(r)} \left(U'(r)+U(r)
\left(3 V'(r)+W'(r)\right)\right)  \nn
&&-(2 \alpha' e^{W(r)-2 V(r)} (-2 U(r) V'(r) (U(r) \left(B^2 c_2+c_1
   e^{4 V(r)+2 W(r)} Z'(r)^2\right)-2 c_1 U(r)^2 e^{4 V(r)}  \nn
&& \left(2 V''(r)+3 V'(r)^2+W'(r)^2\right)+2 c_1 \omega ^2 e^{4 V(r)})+2 c_1 U(r)
   e^{4 V(r)} U'(r)^2 V'(r) \nn
&& +c_1 e^{4 V(r)} U'(r) \left(2 U(r)^2
\left(V''(r)+3 V'(r)^2\right)+\omega ^2\right)))/U(r))\bigg]\nonumber\\
&&+
\Phi(r)\Phi''(r)  \bigg[(2 U(r) e^{2 V(r)+W(r)}-4 c_1
\alpha' U(r) e^{2 V(r)+W(r)}(U'(r) V'(r)
+2 U(r) \left(V''(r)+V'(r)^2\right)))\bigg] \nn
&&+\Phi'(r)^2  \bigg[( \alpha' e^{W(r)-2 V(r)} (U(r)
(B^2 c_2  +2 c_1 e^{4 V(r)} U'(r) V'(r)-c_1 e^{4 V(r)+2 W(r)} Z'(r)^2)  \nn
&& +c_1 e^{4 V(r)} \left(U'(r)^2+4
   \omega ^2\right)+2 c_1 U(r)^2 e^{4 V(r)} \left(-2 V''(r)+V'(r)^2+W'(r)^2\right))
+\frac{3}{2} U(r) e^{2 V(r)+W(r)})\bigg] \nn
&& + \Phi'(r)   \Phi''(r) \bigg[ ( 2 c_1 \alpha' U(r) e^{2 V(r)+W(r)}
\left(U'(r)+4 U(r) V'(r)\right))\bigg]
+ \Phi''(r)^2  \bigg[2 c_1 \alpha' U(r)^2 e^{2 V(r)+W(r)}\bigg]\nn
\een

Since there is no mixing between $\Phi$ and the other modes in the zero momentum
limit, this mode remains massless and a computation of the related Green's
function from the near-horizon data is still possible.

However, the other metric fluctuations namely $h_{xz}$ or $h_{yz}$ are not decoupled
from the gauge field perturbations in the zero momentum limit (even in the absence of higher
derivative terms) --- these coupled terms are proportional to B or $\zeta$ (CS term).
Therefore, these modes are not massless in the zero momentum limit and
the near horizon geometry is not sufficient to compute the related two point
correlation functions, i.e. $<[T_{xz},T_{xz}]>$ or $<[T_{yz},T_{yz}]>$ .


\section{Functions appeared in Higher-derivative Charged black-hole action}\label{hdcb}

\ben
{\cal A}_1^{CB}&=&\frac{r^5(\omega ^2-q^2)+q^2 r_0^4 r}{2 (r^4-r_0^4)}
+\frac{c_1 \alpha' (11 r^8 (\omega ^2-q^2)-r_0^4 r^4(25 q^2+6 \omega ^2)+36 q^2 r_0^8)}{6 r^3
   (r^4-r_0^4)} \nonumber\\
&+& \frac{Q^2}{2 r_0^2 (r^2-r_0^2)
   (r^2+r_0^2)^2} \bigg[r^3 \omega ^2+\frac{\alpha'}{3 r^5}
\bigg(c_1 [28 q^2 r_0^6 r^2 -r^8 (36 q^2+127 \omega ^2)\nn
&&-4 r_0^2 r^6 (7 q^2-3 \omega ^2)
-8 q^2 r_0^8
   +4 r_0^4 r^4 (11 q^2+3
   \omega ^2)]-24 c_2 [r_0^2 r^6 (2 q^2-3 \omega ^2)-2 q^2 r_0^6
   r^2 \nn
   &+&r_0^4 r^4 (2 q^2-3 \omega ^2)-2 q^2 r_0^8+4 r^8 \omega ^2]\bigg) \bigg] +  {\cal O}(Q^4)
\een
and
\ben
{\cal A}_1^{CB}&=&\frac{1}{2} \left(r r_0^4-r^5\right)
-\frac{c_1 \alpha' \left(13 r^4-18 r_0^4\right) \left(r^4-r_0^4\right)}{6 r^3}
 +  \frac{Q^2 (r^2-r_0^2)}{2 r r_0^2 } \bigg[1-\frac{\alpha'}{3
   r^4} \bigg(24 c_2 (4 r^4-3 r_0^2 r^2-3 r_0^4) \nn
   &&+ c_1 (101 r^4-156 r_0^2 r^2-120 r_0^4)\bigg)\bigg] + {\cal O}(Q^4).
\een


\section{Other methods of holographic hydrodynamics}\label{other methods}

In this appendix we outline two other methods of computing various 
transport coefficients very briefly. First we discuss about the boosted black 
brane method by \cite{causa1}. Next, we talk about the pole 
methods by \cite{Paulos}. 

\subsection{Boosted black branes and fluid dynamics} \label{revshiraz}

In this section we will briefly sketch the working procedure of
\cite{causa1}. For detailed discussion readers are referred to the
original paper. We will also skip the technical details in this
section.

\bi
\ii
Consider the Einstein-Hilbert action with negative cosmological
constant
\be
I = -{1 \over \Ncg} \int d^5x \sqrt{-g}\lb R + {12 \over L^2} \rb
\ee
where $L$ is the radius of $AdS$ space.
\ii
The equation of motions are given by \footnote{$x^M = \{v, r,
\vec x\}$.}
\be
E_{MN} = R_{MN} -{1\over 2} R g_{MN} - {6\over L^2} g_{MN}=0.
\ee
\ii
There exists a class of solutions to these equations of motion given
by the ``boosted black branes'' \footnote{$x^{\mu} = \{v,\vec x
  \}$.},
\be
ds^2 = -2 u_{\mu}dx^{\mu}dr -\rl f(b r) u_{\mu} u_{\nu}
dx^{\mu}dx^{\nu} + \rl \lb u_{\mu} u_{\nu} + \eta_{\mn} \rb dx^{\mu}
dx^{\nu}
\ee
with,
\ben
f(r)&=& 1- {1 \over r^4}, \nn
u_v&=& - \ga \nn
and \ \ u_i&=& \ga \beta_i
\een
where, $\ga = 1/\sqrt{1-\vec \beta^2}$.
\ii
Putting the values of $u_{\mu}$'s the metric can also be written as,
\ben \label{boostmetric}
ds^2 = && 2 \ga dv dr - {r^2\over L^2} \ga^2 f(br) dv^2 + \rl dx^i dx^i \nn
&+& \rl (\ga^2-1) dv^2
- 2 \ga \beta_i dx^i dr - 2 \rl \ga^2 (1- f(br)) \beta_i dx^i dv \nn
&+&
\rl \ga^2 (1-
f(br)) \beta_i \beta_j dx^i dx^j.
\een
The solution is parametrized by four constant parameters $b$ and
$\beta_i$'s.
\ii
The black brane horizon is located at $r_H=1/b$ and the temperature of
this black brane is given by,
\be \label{temp0}
T= {1 \over \pi b L^2}.
\ee
\ii
Consider the metric (\ref{boostmetric}) and replace the constant
parameters $b$ and $\beta_i$'s by slowly varying functions $b(x^{\mu})$
and $\beta_i(x^{\mu})$'s of boundary coordinates $x^{\mu}$
\ben \label{boostmetric2}
ds^2 =&& 2\ga dv dr - {r^2\over L^2} \ga^2 f(b(x^{\alpha})r) dv^2
+ \rl dx^i dx^i \nn
&+& \rl (\ga^2 -1)dv^2 - 2 \ga \beta_i(x^{\alpha}) dx^i dr - 2
\rl\ga^2 (1-
f(b(x^{\alpha})r)) \beta_i(x^{\alpha}) dx^i dv \nn
&+&\rl \ga^2 (1-
f(b(x^{\alpha})r)) \beta_i(x^{\alpha}) \beta_j(x^{\alpha}) dx^i dx^j.
\een
We will call this metric $g^{(0)}(b(x^{\alpha}),\beta_i(x^{\alpha}))$.
\ii
In general the metric (\ref{boostmetric2}) is not a solution to Einstein
equations unless one adds some corrections to the metric and also the
parameters $b(x^{\alpha}),\beta_i(x^{\alpha})$ satisfy some set of
equations, which turn out to be the equations of boundary fluid
mechanics.
\ii
Write the parameters $b(x^{\alpha})$ and $\beta_i(x^{\alpha})$ and the
metric as a derivative expansion of the parameters. Up to first order
in derivative expansion,
\be \label{metexpn1st}
g= g^{(0)}(b(x^{\alpha}),\beta_i(x^{\alpha})) + \epsilon
g^{(1)}(b(x^{\alpha}),\beta_i(x^{\alpha})),
\ee
\be \label{bexpan}
b(x^{\alpha}) = b^{(0)}(x^{\alpha})
\ee
and
\be \label{betaexpn}
\beta_i(x^{\alpha}) = \beta^{(0)}_i(x^{\alpha})
\ee
where $\epsilon$ is a dimension less parameter whose power counts the
number of (boundary)spacetime derivatives acting on the
parameters. Since $b^{(1)}(x^{\alpha})$ and
$\beta^{(1)}_i(x^{\alpha})$ do not enter in to the first order
equation of motions, we have kept the expansion for $b$ and $\beta_i$'s
up to leading order.
\ii
In general one can write the metric and parameters as power series of
$\epsilon$. Then plug the metric in Einstein equations and solve the
metric and the parameters order by order (in $\epsilon$).  For example
in our case since we are interested up to first order, we will plug the
metric in Einstein equations and solve for $g^{(1)}$ and the constraint
equations imply some relations between the zero$^{th}$ order
parameters. We will work in a particular gauge,
\be
Tr ((g^{(0)})^{-1} g^{(1)})=0.
\ee
\ii
After finding the metric with first order fluctuations one can find
the boundary stress tensor (using the
definition given in \cite{bala-kraus}). The form of the
boundary stress up to first order in derivative expansion is given by,
\be
16 \pi G_5 T_{\mn} = {T^4
 \pi^4L^3 } \lb4 u_{\mu} u_{\nu} +
\eta_{\mn}\rb - {2 T^3 \pi^3 L^3} \sigma_{\mn},
\ee
where $\sigma_{\mn}$ is given by,
\be \label{sigmamn}
\sigma_{\mn} = P_{\mu}^{\alpha} P_{\nu}^{\beta} \partial_{( \alpha}
u_{\beta )} -{1\over 3} P_{\mn} \partial_{\alpha}u^{\alpha}
\ee
and $P_{\mn} = u_{\mu}u_{\nu}+ \eta_{\mn}$.
\ei

One can follow this procedure to second order in derivative expansion.
We are not discussing that lengthy algebra here. Rather we are going to present the
final answer for stress tensor up to second order in derivative expansion.

\begin{equation} \label{fmst} \begin{split}
T^{\mu\nu}=& (\pi \,T)^4
\left( \eta^{\mu \nu} +4\, u^\mu u^\nu \right) -2\, (\pi\, T )^3 \,\sigma^{\mu \nu} \\
& + (\pi T)^2 \,\left( \left(\ln 2\right) \, T_{2a}^{\mu \nu} +2\, T_{2b}^{\mu\nu} +
\left( 2- \ln2 \right)  \left[ \frac{1}{3} \, T_{2c}^{\mu \nu}
+ T_{2d}^{\mu\nu} + T_{2e}^{\mu\nu} \right] \right) \\
\end{split}
\end{equation}
where
\begin{equation}\label{defst}
\begin{split}
\sigma^{\mu\nu}&= P^{\mu \alpha} P^{\nu \beta} \,
\, \partial_{(\alpha} u_{\beta)}
-\frac{1}{3} \, P^{\mu \nu} \, \partial_\alpha u^\alpha \\
T_{2a}^{\mu\nu}&= \epsilon^{\alpha \beta \gamma (\mu} \, \sigma_{\;\;\gamma}^{\nu)} \, u_\alpha \, \l_\beta \\
T_{2b}^{\mu\nu}&= \sigma^{\mu\alpha} \sigma_{\;\alpha}^{\nu} - \frac{1}{3}\,P^{\mu\nu}\, \sigma^{\alpha \beta } \sigma_{\alpha \beta} \\
T_{2c}^{\mu\nu}&=\partial_\alpha u^\alpha\,\sigma^{\mu\nu}\\
T_{2d}^{\mu\nu}&=  \CD u^\mu \,   \CD
 u^\nu  - \frac{1}{3}\, P^{\mu\nu}\, \CD u^\alpha  \, \CD u_\alpha  \\
  T_{2e}^{\mu \nu}&= P^{\mu\alpha} \, P^{\nu\beta}\, \CD \left(\partial_{(\alpha} u_{\beta)}\right)
 - \frac{1} {3}  \,P^{\mu \nu}\,P^{\alpha \beta} \, \CD \left(\partial_\alpha u_\beta  \right) \\
 \l_\mu &=\epsilon_{\alpha \beta \gamma \mu} \, u^\alpha \partial^\beta u^\gamma .
 \end{split}
 \end{equation}
 Our conventions are  $\epsilon_{0123} = -\epsilon^{0123} =  1$ and
$\CD \equiv u^\alpha \partial_\alpha$ and the brackets $()$ around the indices to  denote
symmetrization, \ie, $a^{(\alpha} b^{\beta)}=
(a^\alpha b^\beta + a^\beta b^\alpha)/2$.

From this expression of second order stress tensor, one can read-off the transport coefficients,
\begin{equation}
\eta = \frac{\pi}{8}\,N^2 \, T^3, \qquad
{\tau}_\Pi = \frac{2 - \ln 2}{\pi\, T} \ , \qquad {\lambda}_1 = \frac{2\,\eta}{\pi \,T}
 \ , \qquad {\lambda}_2 = \frac{2\,\eta\, \ln 2}{\pi\,T}\ , \qquad \lambda_3=0.
\label{}
\end{equation}	
As the coefficient $\kappa$ does not enter the equations of fluid dynamics in
flat space (see equation (\ref{T2nd})), this analysis leaves this coefficient undetermined.

\subsection{Pole Method} \label{revpaolus}

The method is very well explained in the original paper \cite{Paulos}.
Here we outline the working formula in brief.  Let us consider
a Lagrangian containing arbitrary higher-derivative terms. As usual,
we treat the higher-derivative terms perturbatively small and the
equations of motions can still be treated as quadratic. The most
generic form of the quadratic action in Fourier space (at zero spatial
momenta) is given as follows:
\be
S= \int \Pi _{i=1}^{d-1} dx^i \int \frac{d\omega}{2 \pi} ({S_z +S_t + S_B}[\phi(z,x^i,\omega)] ),
\ee
Where $S_z$ is radial action, containing at least two radial $(z)$
derivative and no time derivative. $S_t$ contains time derivatives and
are proportional to $\omega^2$. $S_B$ is the boundary Lagrangian. The
only constraint that has been put to write the above Lagrangian
density is that the field $\phi$ is massless. Studying the generic
structure of the boundary terms, it can logically be shown that these
boundary terms do not contribute to the computation of first order
transport coefficients. 

The next step is to compute the quadratic action near the
horizon. Now, by demanding the regularity at the horizon, it can be
shown that the metric perturbation $\phi$ has following solution
(independent of the details of the action), 
\be
\phi_{sol}=\phi_0 \exp{-i \frac{\omega}{4 \pi T} \log z},
\ee
where, $T$ is the temperature of the background.
The Lagrangian density computed on this solution will always have a
pole and other regular terms in the radial variable $z$. The first
order transport coefficient is then given by a simple formula as, 
\be
\Xi=8 \pi T \frac{Residue{z=0}{\cal L}_2}{\omega^2},
\ee
where, ${\cal L}_2$ is the quadratic Lagrangian density evaluated at
$\phi_{sol}$. This method is really simple and easily usable for
theories with arbitrary  higher-derivative couplings (powers of
curvature tensors as well as their covariant derivatives) .  But its
extension to higher order transport coefficient is rather unclear.




\begin{thebibliography}{100}


\bibitem{landau} L.D.Landau  \& E.M.Lifshitz
(1997). \emph{Fluid mechanics}, Pergamon Press.

\bibitem{rhic}
  M.~Gyulassy and L.~McLerran,
  Nucl.\ Phys.\  A {\bf 750}, 30 (2005)



\bibitem{maldacena} J.M.Maldacena,
  Adv.\ Theor.\ Math.\ Phys.\  {\bf 2}, 231 (1998)
  [Int.\ J.\ Theor.\ Phys.\  {\bf 38}, 1113 (1999)].


\bibitem{lattice}
  H.~B.~Meyer,
  Phys.\ Rev.\  D {\bf 76}, 101701 (2007)



\bibitem{maldarev}
  O.~Aharony, S.~S.~Gubser, J.~M.~Maldacena, H.~Ooguri and Y.~Oz,
  Phys.\ Rept.\  {\bf 323}, 183 (2000)



\bibitem{witten} E.Witten,
  Adv.\ Theor.\ Math.\ Phys.\  {\bf 2}, 253 (1998).

\bibitem{son1}
  G.~Policastro, D.~T.~Son and A.~O.~Starinets,
  Phys.\ Rev.\ Lett.\  {\bf 87}, 081601 (2001).

\bibitem{son2}
  D.~T.~Son and A.~O.~Starinets,
  JHEP {\bf 0209}, 042 (2002).


\bibitem{liu}
  N.~Iqbal and H.~Liu,
  Phys.\ Rev.\  D {\bf 79}, 025023 (2009).


\bibitem{ns1}
  N.~Banerjee and S.~Dutta,
  JHEP {\bf 0903}, 116 (2009).


\bibitem{nsnh}
  N.~Banerjee and S.~Dutta,
  Nucl.\ Phys.\  B {\bf 845}, 165 (2011).


\bibitem{ns2}
  N.~Banerjee and S.~Dutta,
  JHEP {\bf 0907}, 024 (2009)
  [arXiv:0903.3925 [hep-th]].

\bibitem{Kats}
  Y.~Kats and P.~Petrov,
  arXiv:0712.0743 [hep-th].

\bibitem{Liu:2010sa}
  JHEP {\bf 0810}, 072 (2008)
  [arXiv:0807.1100 [hep-th]];\\
  D.~Cassani, G.~Dall'Agata and A.~F.~Faedo,
  JHEP {\bf 1005}, 094 (2010)
  [arXiv:1003.4283 [hep-th]];\\
  J.~T.~Liu, P.~Szepietowski and Z.~Zhao,
  Phys.\ Rev.\  D {\bf 81}, 124028 (2010)
  [arXiv:1003.5374 [hep-th]] ;\\
  J.~P.~Gauntlett and O.~Varela,
  JHEP {\bf 1006}, 081 (2010)
  [arXiv:1003.5642 [hep-th]].


\bibitem{Taylor:2000xw}
  M.~Taylor,
  arXiv:hep-th/0002125.

\bibitem{Skenderis:2002wp}
  K.~Skenderis,
  Class.\ Quant.\ Grav.\  {\bf 19}, 5849 (2002)
  [arXiv:hep-th/0209067].

\bibitem{Gibbons:1996af}
  G.~W.~Gibbons, R.~Kallosh and B.~Kol,
  Phys.\ Rev.\ Lett.\  {\bf 77}, 4992 (1996)
  [arXiv:hep-th/9607108];\\
  D.~Astefanesei, K.~Goldstein and S.~Mahapatra,
  Gen.\ Rel.\ Grav.\  {\bf 40}, 2069 (2008)
  [arXiv:hep-th/0611140].



\bibitem{ABD}
  D.~Astefanesei, N.~Banerjee and S.~Dutta,
  JHEP {\bf 1102}, 021 (2011)
  [arXiv:1008.3852 [hep-th]].


\bibitem{D'Hoker:2009bc}
  E.~D'Hoker and P.~Kraus,
  arXiv:0911.4518 [hep-th];\\
  E.~D'Hoker and P.~Kraus,
  JHEP {\bf 0910}, 088 (2009)
  [arXiv:0908.3875 [hep-th]].


\bibitem{Astefanesei:2008wz}
  D.~Astefanesei, N.~Banerjee and S.~Dutta,
  JHEP {\bf 0811}, 070 (2008)
  [arXiv:0806.1334 [hep-th]].



\bibitem{cai1}
  R.~G.~Cai, Z.~Y.~Nie and Y.~W.~Sun,
  Phys.\ Rev.\  D {\bf 78}, 126007 (2008)
  [arXiv:0811.1665 [hep-th]].



\bibitem{ramy1}
  R.~Brustein and A.~J.~M.~Medved,
  Phys.\ Rev.\  D {\bf 79}, 021901 (2009)
  [arXiv:0808.3498 [hep-th]].



\bibitem{mps}
  R.~C.~Myers, M.~F.~Paulos and A.~Sinha,
  JHEP {\bf 0906}, 006 (2009)
  [arXiv:0903.2834 [hep-th]].

\bibitem{Hanaki:2006pj}
  K.~Hanaki, K.~Ohashi and Y.~Tachikawa,
  Prog.\ Theor.\ Phys.\  {\bf 117}, 533 (2007)
  [arXiv:hep-th/0611329];\\
  S.~Cremonini, K.~Hanaki, J.~T.~Liu and P.~Szepietowski,
  Phys.\ Rev.\  D {\bf 80}, 025002 (2009)
  [arXiv:0903.3244 [hep-th]].




\bibitem{Wald:1993nt}
  R.~M.~Wald,
  Phys.\ Rev.\  D {\bf 48}, 3427 (1993)
  [arXiv:gr-qc/9307038].



\bibitem{dg}
  S.~Dutta and R.~Gopakumar,
  Phys.\ Rev.\  D {\bf 74}, 044007 (2006)
  [arXiv:hep-th/0604070].
























\bibitem{loga}
  R.~Loganayagam,
  JHEP {\bf 0805}, 087 (2008)
  [arXiv:0801.3701 [hep-th]].

\bibitem{causa1}
  S.~Bhattacharyya, V.~E.~Hubeny, S.~Minwalla and M.~Rangamani,
  JHEP {\bf 0802}, 045 (2008).

\bibitem{causa2}
  R.~Baier, P.~Romatschke, D.~T.~Son, A.~O.~Starinets and
M.~A.~Stephanov,
  JHEP {\bf 0804}, 100 (2008).

\bibitem{causa3}
  M.~Natsuume and T.~Okamura,
  Phys.\ Rev.\  D {\bf 77}, 066014 (2008) [Erratum-ibid.\  D {\bf 78},
089902 (2008)].

\bibitem{nsflow}
  N.~Banerjee and S.~Dutta,
  JHEP {\bf 1008}, 041 (2010).

\bibitem{skende}
  M.~Henningson and K.~Skenderis,
  JHEP {\bf 9807}, 023 (1998).




\bibitem{gks}
  S.~S.~Gubser, I.~R.~Klebanov and A.~A.~Tseytlin,
  Nucl.\ Phys.\  B {\bf 534}, 202 (1998).





\bibitem{buchel-paulos1}
  A.~Buchel and M.~Paulos,
  Nucl.\ Phys.\  B {\bf 805}, 59 (2008).



\bibitem{Berg}
  E.~A.~Bergshoeff and M.~de Roo,
  Nucl.\ Phys.\  B {\bf 328}, 439 (1989).

\bibitem{panda}
  W.~A.~Chemissany, M.~de Roo and S.~Panda,
  JHEP {\bf 0708}, 037 (2007)
  [arXiv:0706.3636 [hep-th]].




\bibitem{Blau}
  M.~Blau, K.~S.~Narain and E.~Gava,
  JHEP {\bf 9909}, 018 (1999)
  [arXiv:hep-th/9904179].

\bibitem{BD2}
  N.~Banerjee and S.~Dutta,
  JHEP {\bf 0907}, 024 (2009)
  [arXiv:0903.3925 [hep-th]].




\bibitem{sd}
  S.~Dutta,
  JHEP {\bf 0805}, 082 (2008)
  [arXiv:0804.2453 [hep-th]].



\bibitem{myers-shenker}
  M.~Brigante, H.~Liu, R.~C.~Myers, S.~Shenker and S.~Yaida,
  Phys.\ Rev.\  D {\bf 77}, 126006 (2008)
  [arXiv:0712.0805 [hep-th]].

\bibitem{myers1}
  A.~Buchel, J.~Escobedo, R.~C.~Myers, M.~F.~Paulos, A.~Sinha and
M.~Smolkin,
  JHEP {\bf 1003}, 111 (2010)
  [arXiv:0911.4257 [Unknown]].




\bibitem{kss}
  P.~Kovtun, D.~T.~Son and A.~O.~Starinets,
  Phys.\ Rev.\ Lett.\  {\bf 94}, 111601 (2005)
  [arXiv:hep-th/0405231].


\bibitem{myers2}
  A.~Buchel and R.~C.~Myers,
  JHEP {\bf 0908}, 016 (2009)
  [arXiv:0906.2922 [hep-th]].


\bibitem{Cvetic}
  M.~Cvetic and S.~S.~Gubser,
  JHEP {\bf 9904}, 024 (1999)
  [arXiv:hep-th/9902195].

  A.~Chamblin, R.~Emparan, C.~V.~Johnson and R.~C.~Myers,
  Phys.\ Rev.\  D {\bf 60}, 104026 (1999)
  [arXiv:hep-th/9904197].

  R.~G.~Cai and A.~Wang,
  Phys.\ Rev.\  D {\bf 70}, 064013 (2004)
  [arXiv:hep-th/0406057].

  S.~S.~Gubser,
  Nucl.\ Phys.\  B {\bf 551}, 667 (1999)
  [arXiv:hep-th/9810225].

  N.~Banerjee and S.~Dutta,
  JHEP {\bf 0707}, 047 (2007)
  [arXiv:0705.2682 [hep-th]].


\bibitem{chargehydro}
 N.~Banerjee, J.~Bhattacharya, S.~Bhattacharyya, S.~Dutta, R.~Loganayagam and P.~Surowka,
  JHEP {\bf 1101}, 094 (2011)
  [arXiv:0809.2596 [hep-th]].

\bibitem{haack}
  J.~Erdmenger, M.~Haack, M.~Kaminski and A.~Yarom,
  JHEP {\bf 0901}, 055 (2009)
  [arXiv:0809.2488 [hep-th]].


\bibitem{son5}
  D.~T.~Son and A.~O.~Starinets,
  JHEP {\bf 0603}, 052 (2006)
  [arXiv:hep-th/0601157].

\bibitem{Maeda}
  K.~Maeda, M.~Natsuume and T.~Okamura,
  Phys.\ Rev.\  D {\bf 73}, 066013 (2006)
  [arXiv:hep-th/0602010].

\bibitem{sl}
  S.~Cremonini, K.~Hanaki, J.~T.~Liu and P.~Szepietowski,
  Phys.\ Rev.\  D {\bf 80}, 025002 (2009)
  [arXiv:0903.3244 [hep-th]].





\bibitem{son4}
  D.~T.~Son and A.~O.~Starinets,
  JHEP {\bf 0209}, 042 (2002)
  [arXiv:hep-th/0205051].

\bibitem{Buchbinder:2008nf}
  E.~I.~Buchbinder and A.~Buchel,
  Phys.\ Rev.\  D {\bf 79}, 046006 (2009)
  [arXiv:0811.4325 [hep-th]].







\bibitem{emparan}
  R.~Emparan, T.~Harmark, V.~Niarchos and N.~A.~Obers,
  JHEP {\bf 1003}, 063 (2010)
  [arXiv:0910.1601 [hep-th]].


\bibitem{mukund-liu}
  T.~Faulkner, H.~Liu and M.~Rangamani,
  JHEP {\bf 1108}, 051 (2011)
  [arXiv:1010.4036 [hep-th]].

\bibitem{Paulos}
  M.~F.~Paulos,
  JHEP {\bf 1002}, 067 (2010)
  [arXiv:0910.4602 [hep-th]].

\bibitem{buchel-bulkvisco}
  A.~Buchel,
  Phys.\ Lett.\  B {\bf 663}, 286 (2008)
  [arXiv:0708.3459 [hep-th]].


\bibitem{skenderis-nonconformal}
  I.~Kanitscheider and K.~Skenderis,
  JHEP {\bf 0904}, 062 (2009)
  [arXiv:0901.1487 [hep-th]].


\bibitem{buchel-pagnutti}
  A.~Buchel and C.~Pagnutti,
  Nucl.\ Phys.\  B {\bf 834}, 222 (2010)
  [arXiv:0912.3212 [hep-th]].



\bibitem{Gursoy-Kiritsis-Michalogiorgakis-Nitti}
  U.~Gursoy, E.~Kiritsis, G.~Michalogiorgakis and F.~Nitti,
  JHEP {\bf 0912}, 056 (2009)
  [arXiv:0906.1890 [hep-ph]].

\bibitem{gubser-nellore}
  S.~S.~Gubser and A.~Nellore,
  Phys.\ Rev.\  D {\bf 78}, 086007 (2008)
  [arXiv:0804.0434 [hep-th]].

\bibitem{Gubser-Pufu-Rocha}
  S.~S.~Gubser, S.~S.~Pufu and F.~D.~Rocha,
  JHEP {\bf 0808}, 085 (2008)
  [arXiv:0806.0407 [hep-th]].


\bibitem{buchel-bulkvisko-violation}
  A.~Buchel,
  arXiv:1110.0063 [hep-th].



\bibitem{rangamani-ross}
  M.~Rangamani, S.~F.~Ross, D.~T.~Son and E.~G.~Thompson,
  JHEP {\bf 0901}, 075 (2009)
  [arXiv:0811.2049 [hep-th]].


\bibitem{mukund-nonrel}
  C.~P.~Herzog, M.~Rangamani and S.~F.~Ross,
  JHEP {\bf 0811}, 080 (2008)
  [arXiv:0807.1099 [hep-th]].

\bibitem{maldacena-nonrel}
  J.~Maldacena, D.~Martelli and Y.~Tachikawa,
  JHEP {\bf 0810}, 072 (2008)
  [arXiv:0807.1100 [hep-th]].

D.~T.~Son,
  Phys.\ Rev.\  D {\bf 78}, 046003 (2008)
  [arXiv:0804.3972 [hep-th]].





\bibitem{bala-nonrel}
  A.~Adams, K.~Balasubramanian and J.~McGreevy,
  JHEP {\bf 0811}, 059 (2008)
  [arXiv:0807.1111 [hep-th]].

\bibitem{bala-kraus}
  V.~Balasubramanian and P.~Kraus,
  Commun.\ Math.\ Phys.\  {\bf 208}, 413 (1999)
  [arXiv:hep-th/9902121].




\end{thebibliography}
\end{document}